\documentclass[a4paper,notitlepage,twocolumn,superscriptaddress,longbibliography,nofootinbib]{revtex4-1}
\usepackage[utf8]{inputenc}
\usepackage{braket}
\usepackage{amsmath}
\usepackage{mathtools}
\usepackage{amssymb}
\usepackage{amsfonts}
\usepackage{graphicx}
\usepackage{hyperref}
\usepackage{bbold}
\usepackage{color}
\usepackage{makecell}
\usepackage{enumitem}
\usepackage{upgreek}
\usepackage{makecell}

\usepackage{amsmath,amsthm,amssymb}
\usepackage[dvipsnames]{xcolor}
\usepackage{braket}
\usepackage{bbm}
\usepackage{bm}
\usepackage{longtable}
\usepackage{adjustbox}
\usepackage{array}
\usepackage{multirow}
\usepackage{tabularx}

\usepackage{float}

\usepackage{tikz}
\usetikzlibrary{shapes,arrows,patterns}

\newcommand{\X}{\mathrm{x}}
\newcommand{\Y}{\mathrm{y}}
\newcommand{\Z}{\mathrm{z}}

\newcommand{\ii}{\mathrm{i}}
\newcommand{\ie}{{\it i.e.},\ }
\newcommand{\eg}{{\it e.g.},\ }

\newcommand{\id}{\mathbb{1}} 
\newcommand{\fe}{\hat{f}}

\newcommand{\fd}{\hat{f}^\dagger}

\newcommand{\tr}{\operatorname{tr}}
\newcommand{\Tr}{\operatorname{Tr}}
\newcommand{\m}{\operatorname{\gamma}}
\newcommand{\mt}{\operatorname{\tilde{\gamma}}}

\usetikzlibrary{decorations.markings,arrows,decorations.pathreplacing,calc,intersections,through,backgrounds}
\newcommand\manifold[3][]{
	\draw[every to/.style={out=-20,in=160,relative},#1] (#2) 
	to ($(#2 -| #3)!0.2!(#2 |- #3)$)
	to (#3)
	to ($(#2 -| #3)!0.8!(#2 |- #3)$)
	to cycle;
}
\tikzset{->-/.style={decoration={markings,mark=at position #1 with {\arrow{>}}},postaction={decorate}}}
\tikzset{
	partial ellipse/.style args={#1:#2:#3}{
		insert path={+ (#1:#3) arc (#1:#2:#3)}
	}
}

\begin{document}

\title{Entanglement and Complexity of Purification\texorpdfstring{\\}{}in (1+1)-dimensional free Conformal Field Theories}

\author{Hugo A. Camargo}
\email{hugo.camargo@aei.mpg.de}
\affiliation{Max-Planck-Institut für Gravitationsphysik, \\
Am Mühlenberg 1, 14476 Potsdam-Golm, Germany}
\affiliation{Dahlem Center for Complex Quantum Systems, Freie Universität Berlin, Arnimallee 14, 14195 Berlin, Germany}
\author{Lucas Hackl}
\email{lucas.hackl@unimelb.edu.au}
\affiliation{School of Physics, The University of Melbourne, Parkville, VIC 3010, Australia}
\affiliation{QMATH, Department of Mathematical Sciences, University of Copenhagen, Universitetsparken 5, 2100 Copenhagen, Denmark}
\affiliation{Max-Planck-Institut für Quantenoptik, Hans-Kopfermann-Str.~1, 85748 Garching, Germany}
\affiliation{Munich Center for Quantum Science and Technology, Schellingstr.~4, 80799 München, Germany}
\author{Michal P. Heller}
\email{michal.p.heller@aei.mpg.de}
\altaffiliation[\emph{On leave of absence from:}]{ National Centre for Nuclear Research, Pasteura 7, 02-093 Warsaw, Poland}
\affiliation{Max-Planck-Institut für Gravitationsphysik, \\
Am Mühlenberg 1, 14476 Potsdam-Golm, Germany}
\author{Alexander Jahn}
\email{a.jahn@fu-berlin.de}
\affiliation{Dahlem Center for Complex Quantum Systems, Freie Universität Berlin, Arnimallee 14, 14195 Berlin, Germany}
\author{Tadashi Takayanagi}
\email{takayana@yukawa.kyoto-u.ac.jp}
\affiliation{Center for Gravitational Physics, Yukawa Institute for Theoretical Physics,\\ 
Kyoto University, Kyoto 606-8502, Japan}
\affiliation{Inamori Research Institute for Science,
620 Suiginya-cho, Shimogyo-ku,
Kyoto 600-8411 Japan}
\affiliation{Kavli Institute for the Physics and Mathematics of the Universe,
University of Tokyo, Kashiwa, Chiba 277-8582, Japan}
\author{Bennet Windt}
\email{bennet.windt17@imperial.ac.uk}
\affiliation{Blackett Laboratory, Imperial College London, Prince Consort Road, SW7 2AZ, UK}

\begin{abstract}
Finding pure states in an enlarged Hilbert space that encode the mixed state of a quantum field theory as a partial trace is necessarily a challenging task. Nevertheless, such purifications play the key role in characterizing quantum information-theoretic properties of mixed states via entanglement and complexity of purifications. In this article, we analyze these quantities for two intervals in the vacuum of free bosonic and Ising conformal field theories using, for the first time, the~most general Gaussian purifications. We provide a comprehensive comparison with existing results and identify universal properties. We further discuss important subtleties in our setup: the massless limit of the free bosonic theory and the corresponding behaviour of the mutual information, as well as the Hilbert space structure under the Jordan-Wigner mapping in the spin chain model of the Ising conformal field theory.
\end{abstract}

\maketitle


\section{Introduction}
Understanding quantum information theoretic properties of quantum field theories (QFTs) and, via holography, also of quantum gravity has been an enormously fruitful research front of the past two decades (as seen, for example, in ~\cite{Casini:2009sr,Harlow:2014yka,Rangamani:2016dms,Susskind:2018pmk,Headrick:2019eth}).

The main player in this endeavour has been the notion of entanglement and its entropy~$S$. Starting with a pure state $|\Psi\rangle$ and a subsystem $A$ (its complement denoted by $\bar{A}$), the entanglement entropy is defined as the von Neumann entropy of the reduced density matrix\footnote{This approach assumes factorization of the Hilbert space between $A$ and $\bar{A}$. This is not the case for gauge theories, where more refined approaches need to be invoked, see for example~\cite{Buividovich:2008yv,Donnelly:2011hn,Casini:2013rba,Ghosh:2015iwa,Aoki:2015bsa}.} $\rho_{A} = \Tr_{\bar{A}} |\Psi \rangle \langle \Psi |$ associated with~$A$, specifically
\begin{equation}
S(A) \equiv - \Tr_{A} \rho_{A} \log{\rho_{A}} \ .
\end{equation}

While entanglement entropy is very hard to calculate in a generic QFT, by now many results exist for free quantum fields, conformal field theories (primarily in two spatial dimensions) and strongly coupled QFTs with a holographic description. In the latter case, the entanglement entropy acquires a natural geometric description in terms of the Bekenstein-Hawking entropy of certain codimension-2 surfaces penetrating anti-de Sitter (AdS) geometries~\cite{Ryu:2006bv,Hubeny:2007xt,Lewkowycz:2013nqa,Dong:2016hjy} and led to a wealth of results on quantum gravity in this setting.

Complexity is another quantum information-theoretic notion that made its appearance in the context of QFTs only recently and is directly motivated by holography. To this end, it was observed in~\cite{Susskind:2014moa,Susskind:2014rva,Stanford:2014jda,Brown:2015bva,Brown:2015lvg,Couch:2016exn} that codimension-one boundary-anchored maximal volumes and codimension-zero boundary-anchored causal diamonds have properties expected from the hardness of preparing states using tensor networks~\cite{Orus:2014poa} in chaotic quantum many-body systems.

Subsequent articles starting with~\cite{Chapman:2017rqy,Jefferson:2017sdb} saw in this conjecture a strong motivation to define the notion of complexity in the realm of QFTs in a similar spirit in which pioneering works~\cite{Bombelli:1986rw,Srednicki:1993im} introduced the notion of entanglement entropy in the same context. The~articles~\cite{Chapman:2017rqy,Jefferson:2017sdb} were largely inspired by the continuous tensor network of cMERA~\cite{Haegeman:2011uy} and viewed preparation of a pure target state $| \psi_{T} \rangle$ in QFT as a unitary transformation from some pure reference state $| \psi_{R} \rangle$
\begin{equation}
\label{eq.TUR}
| \psi_{T} \rangle = U | \psi_{R} \rangle ,
\end{equation}
where the unitary $U$ is obtained as a sequence of layers constructed by exponentiation of more elementary Hermitian operators~${\cal O}_{I}$
\begin{equation}
U = {\cal P} e^{- i \int_{0}^{1} d\tau \sum_{I} {\cal O}_{I} \, Y^{I}(\tau)}.
\end{equation}
Following the approach of~\cite{Nielsen1133}, which was originally devised to bound complexity of discrete quantum circuits, one can associate the cost of invocations of different gates generated by~${\cal O}_{I}$ as related to the infinitesimal parameter $Y^{I}(\tau) \, d\tau$ in the exponent. Translating this literally into a mathematical formula would lead to
\begin{equation}
\label{eq.costL1}
\mathrm{cost}_{L^1} = \int_{0}^{1} d\tau \sum_{I} |Y^{I}|,
\end{equation}
which is an integral over the circuit of a $L^1$ norm of a formal vector formed from the parameters $Y^{I}$. Complexity $\cal C$ arises then as the minimum of~\eqref{eq.costL1} subject to the condition~\eqref{eq.TUR}
\begin{equation}
\label{eq.defC}
{\cal C} = \min\left[\mathrm{cost}\right].
\end{equation}
As anticipated already in~\cite{Jefferson:2017sdb}, cost functions based on $L^1$ norms such as~\eqref{eq.defC} lead to challenging minimization problems. In the present work our rigorous results on complexity will be based on a particular choice of a $L^{2}$ norm
\begin{equation}
\label{eq.costL2}
\mathrm{cost}_{L^2} = \int_{0}^{1} d\tau \sqrt{\sum_{I} \upeta_{IJ} Y^{I}\,Y^{J}},
\end{equation}
where, following~\cite{Hackl:2018ptj,Chapman:2018hou}, $\upeta_{IJ}$ is going to be a particular non-negative definite constant matrix and $O_{I}$ are going to be normalized accordingly. This choice of $\upeta_{IJ}$ is naturally induced from the reference state~$|\psi_{R}\rangle$ (bosonic systems) or from Lie algebra (fermionic systems).

The essence of recent progress on defining complexity in QFT using broadly defined approach of~\cite{Nielsen1133} lies in making educated choices for ${\cal O}_{I}$ and $|\psi_{R}\rangle$, which allow one to perform minimization encapsulated by Eq.~\eqref{eq.defC}. In the vast majority of cases, it was achieved by focusing on free QFTs and utilizing powerful toolkit of Gaussian states and transformations~\cite{Weedbrook2012}.

The discussion so far concerned pure states, \ie von Neumann entropy as an entanglement measure between a subregion and a complement in pure states and complexity as a way of quantifying hardness of preparation of pure states. Much less understood in the QFT context are quantum information properties of mixed states and the present paper concerns precisely this important subject. Of interest to us will be entanglement of purification (EoP)~\cite{Takayanagi:2017knl,Nguyen:2017yqw} and complexity of purification (CoP)~\cite{Agon:2018zso}. We will introduce these quantities in more detail in, respectively, sections~\ref{sec:EoP} and~\ref{sec:CoP}. Here want to stress instead that the key motivating feature behind our work stems from both of these quantities involving in their definition scanning over purifications of mixed many-body states\footnote{Otherwise, EoP and CoP use regular notions of, respectively, entanglement entropy and pure state complexity, which is the reason why they already made an appearance in the text.}.

Such purifications, \ie embedding a mixed state in an enlarged Hilbert space in which it arises as a reduced density matrix, in the context we are interested in, \ie QFT physics, are clearly challenging to operate with. Earlier works on EoP and CoP in high-energy physics include respectively~\cite{Bhattacharyya:2018sbw,Bhattacharyya:2019tsi} and~\cite{Caceres:2019pgf}\footnote{One should also mention in this context~\cite{Camargo:2018eof}, which, motivated by holographic complexity proposals, explored properties of CoP in the setting of a single harmonic oscillator.} and focus on free QFTs in which mixed states of interest, such as vacuum reduced density matrices or thermal states, are Gaussian. Gaussian mixed states allow for purification to pure Gaussian states, which underlay strategies employed in the aforementioned references. However, even purifications within the Gaussian manifold of states for large subsystems can be challenging to operate with and the above works made additional choices in this respect.

This is where the key novelty of our present work appears, which is to consider the most general Gaussian purifications. To this end, we will consider free QFTs on a lattice and, whenever possible, encode reduced density matrices in terms of corresponding quadratic correlations represented by covariance matrices. Considering the most general Gaussian purifications amounts then to embedding mixed state covariance matrices as parts of larger covariance matrices corresponding to pure states. Utilizing efficient Gaussian techniques allows us to minimize the two quantities of interest, EoP and CoP for a judicious choice of a definition of pure state complexity~\cite{Chapman:2018hou}, over general purifications to a given number of bosonic or fermionic modes.

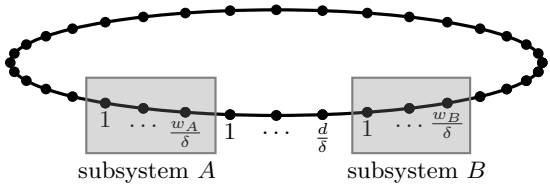
\begin{figure}
    \centering
    \begin{tikzpicture}
    \draw[very thick] (0,0) ellipse (3.5 and .7);
    \fill (-80:3.5 and .7) node[below]{$\tfrac{d}{\delta}$} circle (2pt);
    \fill (-90:3.5 and .7) node[below,yshift=-1mm]{$\ldots$} circle (2pt);
    \fill (-100:3.5 and .7) node[below]{$1$} circle (2pt);
    
    \fill (-70:3.5 and .7) node[below]{$1$} circle (2pt);
    \fill (-60:3.5 and .7) node[below,yshift=-1mm]{$\ldots$} circle (2pt);
    \fill (-50:3.5 and .7) node[below]{$\tfrac{w_{B}}{\delta}$} circle (2pt);
    
    \fill (-110:3.5 and .7) node[below]{$\tfrac{w_{A}}{\delta}$} circle (2pt);
    \fill (-120:3.5 and .7) node[below,yshift=-1mm]{$\ldots$} circle (2pt);
    \fill (-130:3.5 and .7) node[below]{$1$} circle (2pt);
    
    \foreach \x in {0,...,40}
    \fill (-10*\x:3.5 and .7) circle (2pt);
    
    \filldraw[thick,gray,fill opacity=.3] (-2.5,-1.2) rectangle (-.8,-.2);
    \draw (-1.7,-1.45) node{subsystem $A$};
    \draw (1.85,-1.45) node{subsystem $B$};
    \filldraw[thick,gray,fill opacity=.3] (2.55,-1.2) rectangle (1.0,-.2);
    \end{tikzpicture}
    \caption{The subsystem that defines reduced density matrices for our discretized bosonic and fermionic models in their vacuum state consists of two intervals of a width of $w_{A}/\delta$ and $w_{B}/\delta$ sites and separated by a distance of $d/\delta$ sites, where $\delta$ is the lattice spacing. When $d = 0$, we will keep $w_{A}$ and $w_{B}$ generic. When $d > 0$, we will set for simplicity $w_{A} = w_{B} \equiv w$ and the natural continuum combination is $w/d$. We will see that numerically determined MI and EoP approach in the continuum limit functions of $w/d$. With CoP the situation is more complicated, as it turns out to be ultraviolet divergent and brings in an additional dimensionful scale through the class of reference states of interest $|\psi_R\rangle$.}
    \label{fig:twointervals}
\end{figure}

Our primary focus is on a particularly simple yet revealing setup of two-intervals vacuum reduced density matrices in free QFTs with a vanishing or very small mass. In quantum information context, such setup arose first in studies of mutual information (MI) defined using two subsystems $A$ and $B$ as
\begin{equation}
\label{MIde}
I(A:B) = S(A) + S(B) - S(A \cup B)
\end{equation}
in (1+1)-dimensional conformal field theories (CFTs), where $A$ and $B$ are two disjoint or adjacent intervals on a flat spatial slice as depicted by figure~\ref{fig:twointervals}. MI will play an important role in our studies providing us with a guidance regarding both the behaviour of EoP, as in~\cite{Bhattacharyya:2018sbw}, as well as will help us to understand subtleties underlying our models. Our studies will mostly concern scaling of MI, EoP and CoP with control parameters such as interval size, separation and, when present, system size and the mass. While EoP turns out be such a ultraviolet finite quantity by itself, for CoP we will consider a combination of single and two interval CoP results akin to~\eqref{MIde} for which the leading ultraviolet divergences cancel and only milder divergences remain.

Our paper is structured as follows. In section~\ref{sec:models}, we review the two models we consider, the Klein-Gordon field in the massless limit and the critical transverse field Ising model, on a lattice paying a particular attention to description of their ground states in terms of covariance matrices. In section~\ref{sec:mutualinformation}, we benchmark our abilities to reach continuum limit in lattice calculations by comparing the results of our numerics with existing analytic formulas for MI in the aforementioned two interval case. In section~\ref{sec:purifications}, we discuss briefly the mathematics of purifications of Gaussian states as seen by covariance matrices, which is the working horse behind most of the results reported in the present article. Subsequently, we use this machinery to study EoP and CoP in the two-interval case of figure~\ref{fig:twointervals}, respectively, in sections~\ref{sec:EoP} and~\ref{sec:CoP}. In section~\ref{sec.comment}, we comment on two subtleties relevant for our model, namely the zero mode when taking the massless limit for a bosonic theory and the different notions of locality in the spin vs. fermion picture of the Ising model. We summarize our results and present an outlook in section~\ref{sec.discussion}. We also provide an extensive appendix that provides further details regarding our methods. 

\begin{table*}
\centering
\renewcommand{\arraystretch}{1.45}
\begin{tabular}{@{} c @{\hspace{0.4cm}} c @{$\quad$} c @{$\quad$} c p{1.2cm} c @{$\quad$} c @{$\quad$} c @{}}
\toprule
& \multicolumn{7}{c}{\bf Decompactified free boson CFT ($c = 1$)} \\
\cline{2-8}
& & Analytical predictions & & & & {Gaussian numerics}  \\
 \cline{2-4} \cline{6-8}
 & $S(A)$ & $I(A:B)$ & $E_P$ & & $S(A)$ & $I(A:B)$ & $E_P$ \\
 \cline{2-4} \cline{6-8}
$d=0$ & & $\frac{c}{3} \log \frac{w_A w_B}{(w_A + w_B)\delta}$& $\frac{c}{6} \log \frac{2 w_A w_B}{(w_A + w_B)\delta}$\cite{Caputa:2018xuf} &  &
& $\frac{1}{3} \log \frac{w_A w_B}{(w_A + w_B)\delta}$& $\frac{1}{6} \log \frac{2 w_A w_B}{(w_A + w_B)\delta}$ 
\\
$d\ll w$ & \multirow{2}{*}{$\simeq \frac{c}{3} \log\frac{w}{\delta}$\cite{Holzhey:1994we}} & $\simeq \frac{c}{3} \log\frac{w}{d}$\cite{Calabrese:2004eu} & $\simeq \frac{c}{6} \log\frac{w}{d}$\cite{Caputa:2018xuf}  &  &
\multirow{2}{*}{$\simeq \frac{1}{3} \log\frac{w}{\delta}$\;} & $\simeq \binom{{<}0.40}{>0.27} \log\frac{w}{d}$ & $\simeq \frac{1}{6} \log\frac{w}{d}$
\\
$d\gg w$ & & $\propto \left(\frac{w}{d}\right)^{0}$\cite{Cardy:2013nua,Ugajin:2016opf} & $\propto \left(\frac{w}{d}\right)^{0}\;$ * &
 & & $< \left(\frac{w}{d}\right)^{0.15}$ & $< \left(\frac{w}{d}\right)^{0.15}$ 
\\
\colrule

& \multicolumn{7}{c}{\bf Ising CFT ($c = \frac{1}{2}$)} \\
\cline{2-8}
& & Analytical predictions & & & \multicolumn{3}{c}{Gaussian numerics (fermions)} \\
 \cline{2-4} \cline{6-8}
 & $S(A)$ & $I(A:B)$ & $E_P$ & & $S(A)$ & $I(A:B)$ & $E_P$ \\
 \cline{2-4} \cline{6-8}
$d=0$ & & $\frac{c}{3} \log \frac{w_A w_B}{(w_A + w_B)\delta}$& $\frac{c}{6} \log \frac{2 w_A w_B}{(w_A + w_B)\delta}$~\cite{Caputa:2018xuf}  & &
& $\frac{1}{6} \log \frac{w_A w_B}{(w_A + w_B)\delta}$& $\frac{1}{12} \log \frac{2 w_A w_B}{(w_A + w_B)\delta}$
\\
$d\ll w$ & \multirow{2}{*}{$\simeq \frac{c}{3} \log\frac{w}{\delta}$} & $\simeq \frac{c}{3} \log\frac{w}{d}$ & $\simeq \frac{c}{6} \log\frac{w}{d}$\cite{Caputa:2018xuf}  & &
\multirow{2}{*}{$\simeq \frac{1}{6} \log\frac{w}{\delta}$\;} & \multicolumn{2}{c}{\multirow{2}{*}{(non-Gaussian setting)}}
\\
$d\gg w$ &  & $\propto\left(\frac{w}{d}\right)^{1/2}$\cite{Cardy:2013nua,Ugajin:2016opf} & $\propto\left(\frac{w}{d}\right)^{1/2}\;$ *  & & \multicolumn{2}{c}{}
& \\
\botrule
\end{tabular}
\caption{Overview over known analytical results and numerical fits with approximate coefficients for entanglement entropy $S(A)$, MI $I(A:B)$, and EoP $E_P$, all for an infinite-size system. Analytical entries marked with a star (*) are guesses based on analogous behavior between MI and EoP. Twice the free fermion result in the $d = 0$ Ising is the $c = 1$ free Dirac fermion CFT quantity.
}
\label{tab:EoP-MI}
\end{table*}

\begin{table*}
\centering
\renewcommand{\arraystretch}{1.75}
\begin{tabular}{@{} l @{\hspace{0.7cm}} c @{\hspace{0.5cm}} c @{}}
\toprule
 {\bf CFT and complexity definition}
& {\bf Single interval complexity / CoP}
& {\bf Mutual complexity} \\

 \colrule
 
Hologr.\ CFTs: subregion-${\cal C}_{V}$ & $\propto \frac{w}{\delta} -\pi$ & const.\ \\

Hologr.\ CFTs: subregion-${\cal C}_{V\,2.0}$ & $\propto \frac{w}{\delta }-2 \log \left(\frac{w}{\delta }\right)-\frac{\pi^2}{4}$ & $\propto \log{\frac{w_{A} \, w_{B}}{(w_{A}+w_{B})\delta}} + \frac{\pi^2}{8}$ \\

Hologr.\ CFTs: subregion-${\cal C}_{A}$ &  $\propto \log\left(\frac{\ell_{CT}}{{\cal L}}\right)\frac{w}{2\delta}-\log\left(2\frac{\ell_{CT}}{{\cal L}}\right)\log\left(\frac{w}{\delta}\right)+\frac{\pi^{2}}{8}$ & $\propto \log{\left( 2 \frac{\ell_{CT}}{\cal L}\right)} \times \log{\frac{w_{A} \, w_{B}}{(w_{A}+w_{B})\delta}} - \frac{\pi^2}{8}$\\

Decomp.\ free boson CFT ($c = 1$) & $\left( f_2(\mu\, \delta) \frac{w}{\delta} + f_1(\frac{m}{\mu}, \mu\, \delta) \log\frac{w}{\delta} + f_0(\frac{m}{\mu}, \mu\, \delta) \right)^{\frac{1}{2}}$   & 
\makecell{$f_1(\frac{m}{\mu}, \mu\, \delta) \, \log{\frac{w_{A}w_{B}}{(w_{A}+w_{B})\delta}}$ 
$+f_0(\frac{m}{\mu}, \mu\, \delta)$} \\

Ising CFT ($c = \frac{1}{2}$) & $\left(0.103 \, \frac{w}{\delta} + 0.0544 \log{\frac{w}{\delta}} + 0.0894\right)^{\frac{1}{2}}$ & $0.0544 \, \log{\frac{w_{A}w_{B}}{(w_{A}+w_{B})\,\delta}} + 0.0894$ \\

\botrule
\end{tabular}
\caption{Summary of mixed states complexity results consisting of predictions of holographic complexity proposals collected from~\cite{Abt:2017pmf,Abt:2018ywl,Caceres:2019pgf,Auzzi:2019vyh} and our numerical CoP results in free CFTs. The latter are obtained using Gaussian bosonic and fermionic states, the $L^{2}$ norm circuit complexity encapsulated by~\eqref{eq.defourCoP} and spatially disentangled reference states. The mutual complexity is defined differently for the holographic complexity proposals, see~\eqref{eq:mutualCholo}, and for our implementation of CoP, see~\eqref{eq:UVMichalReg}. In the case of the ${\cal C}_{A}$ proposal, $\cal L$ is the AdS curvature radius and $\ell_{CT}$ is an arbitrary length scale arising from counter-terms~\cite{Lehner:2016vdi,Reynolds:2016rvl}. In the case of the bosonic calculation, $\mu$ is the reference state scale and functions $f_{0}$, $f_{1}$ and $f_{2}$ are defined in~\eqref{eq.CoPf0f1f2}.}
\label{TAB_ANALYTIC_RES_COMP}
\end{table*}

\section{Setup}\label{sec:models}
In the present work we focus on two paradigmatic models: the \emph{Klein-Gordon field in the massless limit} and the \emph{critical transverse field Ising model} in 1+1 dimensions. For our numerical calculations, we discretize both theories either on a lattice with $N$ sites and periodic boundary conditions (\ie we identify the sites $N+j\equiv j$) or on an infinite lattice. We will consider subsystems consisting of intervals of width $w/\delta$ sites and separated by a distance of $d/\delta$ sites, where $\delta$ is the lattice spacing (see figure~\ref{fig:twointervals}). Both theories describe CFTs in the respective limits with central charge $c=1$ (Klein-Gordon) and $c=\frac{1}{2}$ (Ising model). We will review the Hamiltonians of both models and their ground states with a particular focus on the covariance matrix formulation. The latter for free bosons will allow for an efficient calculation of EoP and CoP using Gaussian techniques. For the Ising model, we will discuss in detail how there are two distinct notions of locality associated to the spin and fermion formulation, respectively.

\subsection{Klein-Gordon field \label{sec:models.KG}}
We consider the well-known Klein-Gordon scalar field with a mass $m$ that we will later take to zero. Its discretized Hamiltonian on a lattice with $N$ sites is
\begin{align}
    \hat{H}=\frac{\delta}{2}\sum^N_{i=1}\left(\hat{\pi}_i^2+\frac{m^2}{\delta^2}\hat{\varphi}^2_i+\frac{1}{\delta^4}(\hat{\varphi}_i-\hat{\varphi}_{i+1})^2\right)\,,\label{eq:H-KG}
\end{align}
where $\delta$ represents the lattice spacing. We thus have a circumference 
\begin{equation}
L=N\,\delta.   
\end{equation}
We define canonical variables
\begin{equation}
\hat{\xi}^a_i\equiv(\hat{\varphi}_i,\hat{\pi}_i)\,,
\end{equation}
where $a=1,2$. It is well-known that the Hamiltonian can be diagonalized via Fourier transformations leading to $N$ decoupled harmonic oscillators with frequencies
\begin{equation}
\label{eq.dispersionrelation}
\omega_k=\sqrt{m^2+\frac{4}{\delta^2}\sin^2\frac{\pi k}{N}}.
\end{equation}
The ground state $\ket{0}$ is Gaussian and fully characterized by its covariance matrix
\begin{align}
\begin{split}
    G_{ij}^{ab}&=\braket{0|\hat{\xi}^a_i\hat{\xi}^b_j+\hat{\xi}^b_j\hat{\xi}^a_i|0}\\
    &=\frac{1}{N}\sum^N_{k=1}e^{\ii\frac{2\pi k}{N}(i-j)}\begin{pmatrix}
    \omega_k & 0\\
    0 & \frac{1}{\omega_k}
    \end{pmatrix}\,,
\end{split}\label{eq:KG-cov}
\end{align}
where $a$ and $b$ label the entries of the matrix. Continuum limit on a circle requires taking $N \rightarrow \infty$ while keeping the product of meaningful continuum quantities $m\,L = m \, \delta \, N$ fixed. Each value of this product corresponds to a different QFT as a continuum limit within the class of free Klein-Gordon theories. Furthermore, when considering subsystems, as depicted in figure~\ref{fig:twointervals}, continuum limit requires keeping ratios of $w \, \delta$ and $d \, \delta$ to $L$ fixed as $N \rightarrow \infty$. In practice, one takes $N$ to be large but finite and requires that as $N$ is increased well-defined quantities, for example the MI~\eqref{MIde}, stop changing significantly with~$N$ and stabilize in the vicinity of their QFT values.

When $\frac{w \, \delta}{L} \ll 1$, $\frac{d\, \delta}{L} \ll 1$, then the results of the calculations should be effectively indistinguishable from the situation in which the spatial direction is a line. The mass $m \ll \frac{1}{\delta}$ becomes then the only dimensionful parameter in the continuum theory. Also, in this case the number $k$ associated with discrete momenta in~\eqref{eq.dispersionrelation} gets incorporated into a continuum variable and a sum in~\eqref{eq:KG-cov} needs to be replaced by an appropriate integral, see for example~\cite{Chapman:2018hou}.

We are particularly interested in the massless limit $m\to0$, where the Klein-Gordon field describes the CFT with central charge $c=1$. More precisely, the $c = 1$ CFT with the periodic boundary conditions we imposed should be regarded as a 1-parameter family of theories arising in the path integral language from the compactification of the bosonic field $\varphi$ (\ie periodically identified):
\begin{equation}
\label{eq.defcompactboson}
\varphi + 2 \pi R \equiv \varphi.
\end{equation}
The dimensionless parameter~$R$ is the radius of compactification in the field space and plays the role of a moduli specifying a particular $c = 1$ CFT. The scaling dimension of the lightest operator is then given by
\begin{equation}
\label{eq.Deltaminboson}
\Delta_{\mathrm{min}} = \mathrm{min}\left(\frac{1}{R^2},\frac{R^2}{4}\right). \end{equation}
The above formula is a hint of an underlying duality between theories with field compactification radia of $R$ and $\frac{2}{R}$~\cite{DiFrancesco:1997nk}. The massless limit of~\eqref{eq:H-KG} corresponds to the decompactification limit of compact free boson CFTs ($R\rightarrow \infty$), which is a subtle limit since in light of~\eqref{eq.Deltaminboson} the gap in the operator spectrum approaches $0$. While this limit leads to correct correlation functions of vertex operators or a single interval entanglement entropy, for other quantities the situation is more complicated. In particular, the modular invariant thermal partition function for the free boson reads~\cite{DiFrancesco:1997nk}
\begin{equation}
\label{eq.Zmodinv}
Z_{\mathrm{mod-inv}} \sim \frac{1}{(\beta/L)^{1/2} \, \eta(i \, \beta/L)^2},
\end{equation}
whereas the free massive boson calculation for $m \, L \ll 1$ and upon keeping the regularized zero point energy is
\begin{equation}
\label{eq.Zmassive}
Z_{m \, L \ll 1} \sim \frac{1}{(\beta \, m) \, \eta(i \, \beta/L)^2} \, .
\end{equation}
In both expressions $\eta$ is the Dedekind eta function defined as
\begin{equation}
\eta(i \, \beta/L) = e^{- \frac{\pi}{12} \beta/L} \, \Pi_{n = 1}^{\infty} (1-e^{- 2\pi \, n \, \beta/L }).
\end{equation}
The mismatch between the two calculations can be understood using the representation of the partition function on a circle as an Euclidean path integral on a torus. In the case of~\eqref{eq.Zmodinv}, the zero mode contribution is neglected, as its inclusion would lead to an infinite volume term coming from the integration over the field space. In the case of~\eqref{eq.Zmodinv}, the zero mode $\phi$ contribution to the path integral is included and is finite, as it originates in the path integral language from
\begin{equation}
\label{eq.pathintegralzeromodefactor}
\int_{-\infty}^{\infty} d\phi \, e^{-\frac{1}{2} \, \beta \, L \, m^2 \, \phi^2} \sim \frac{1}{m \sqrt{\beta \, L}},
\end{equation}
where the product $\beta \, L$ is the torus spatial volume. Multiplying the partition function~\eqref{eq.Zmodinv} by the factor~\eqref{eq.pathintegralzeromodefactor} leads  to~\eqref{eq.Zmassive}, which explains the relation between the two partition functions. We will come back to these calculations in section~\ref{sec:zero-mode}, where we discuss the influence of the zero mode on MIdecay with separation between two intervals.

In our studies, we will be using the free massive boson setup to extract the properties of the modular invariant $c = 1$ free boson CFT in the decompactification limit $R \rightarrow \infty$. From this perspective, the partition function of interest, \ie~\eqref{eq.Zmodinv}, can be indeed recovered from the massive boson Gaussian calculation~\eqref{eq.Zmassive} by dividing it by the known zero mode contribution~\eqref{eq.pathintegralzeromodefactor}. However, in the case of other quantities calculated using Gaussian techniques at non-vanishing mass the effect of the zero mode is not straightforward to isolate. As we already mentioned, numerical studies showed that Gaussian calculations with a small mass reproduce the universal entanglement entropy result for a single interval~\cite{Casini:2009sr}. Furthermore, one may expect the two interval case at small separations to be reliably described by the massive free boson calculation, as the zero mode affects primarily the long distance physics. As a result, these will be the mixed state setups that we will consider in our EoP and CoP explorations. On the other hand, the two interval case at large separations is delicate and we will return to it in the case of MI in section~\ref{sec:zero-mode}. 

Another subtlety that originates in the massless limit is that the ground state is only defined distributionally. The issue is best understood by diagonalizing the Hamiltonian~\eqref{eq:H-KG} by transforming to momentum modes $\tilde{\pi}_k=\frac{1}{\sqrt{N}}\sum^N_{j=1}e^{-\ii\frac{2\pi k j}{N}}\pi_j$ and $\tilde{\varphi}_k=\frac{1}{\sqrt{N}}\sum^N_{j=1}e^{\ii \frac{2\pi k j}{N}}\varphi_j$ leading to
\begin{align}
    \hat{H}=\frac{1}{2}\sum^N_{k=1}\left(\delta \, |\tilde{\pi}_k|^2+\tfrac{\omega_k^2}{\delta}\,|\tilde{\varphi}_k|^2\right)\,,
\end{align}
where we find $N$ decoupled harmonic oscillators. For the oscillator with $k=0$ (zero momentum mode), we have $\omega_0=m$, which vanishes in the massless limit. Consequently, the ground state of this mode approaches a delta distribution, which does not lie in Hilbert space. This leads to the divergence of certain terms in the covariance matrix~\eqref{eq:KG-cov}. However, we are still able to define expectation values of observables and entanglement measures, such as the entanglement entropy, by computing those quantities for finite $m$ and generating numerically results for values of $m$ gradually approaching $0$. In section~\ref{sec:zero-mode}, we will discuss the role of the zero mode for such calculations in more detail using MI as an example.

\subsection{Critical transverse field Ising model \label{sec.Ising}}
We consider the transverse field Ising model~\cite{katsura1962statistical,pfeuty1970one}
\begin{align}
\label{eq.HIsing}
\hat{H}=-\sum^N_{i=1}(2J\,\hat{S}^{\X}_i \hat{S}^{\X}_{i+1}+h\,\hat{S}^{\Z}_i)
\end{align}
in the critical limit $J=h$, where $\hat{S}^{\X,\Z}_{i}$ are spin-$\frac{1}{2}$ operators in the direction $\X$ or $\Z$ at position~$i$ in the chain, \ie related $\hat{S}^{\X,\Z}_{i}=\frac{1}{2}\sigma^{\X,\Z}_{i}$ to the well-known Pauli matrices. The system consists of $N$ spin-$\frac{1}{2}$ degrees of freedom arranged in a circle, \ie we assume periodic boundary conditions with $N+i\equiv i$.

The transverse field Ising model can be solved analytically by employing the Jordan-Wigner transform~\cite{jordan1993paulische}, \ie eigenvalues and eigenvectors of the Hamiltonian $\hat{H}$ can be constructed in closed form. The transformation is based on introducing fermionic creation and annihilation operators $\hat{f}_i^\dagger$ and $\hat{f}_i$. For the transformation, we write $S^\pm_i=S^{\X}_i\pm\ii S^{\Y}_i$ as
\begin{align}
    S_i^+=\hat{f}_i^\dagger \,\exp\left(\ii \pi\sum^{i-1}_{j=1}\hat{f}^\dagger_j\hat{f}_j\right),\label{eq:jordan-wigner}
\end{align}
which leads to the \emph{almost} quadratic Hamiltonian
\begin{align}
\label{eq.HIsingfermions}
	\begin{split}
	\hat{H}&=-\sum^N_{i=1}\left(\frac{J}{2}\left[\hat{f}_i^\dagger(\hat{f}_{i+1}+\hat{f}_{i+1}^\dagger)+\text{h.c.}\right]+h\hat{f}_i^\dagger \hat{f}_i\right)\\
	&\qquad+\frac{J}{2}[\hat{f}_N^\dagger(\hat{f}_{1}+ \hat{f}_{1}^\dagger)+\text{h.c.}](\hat{P}+1)+\frac{Nh}{2}\,,
\end{split}
\end{align}
where h.~c. stands form Hermitian conjugation and $\hat{P}=\exp{(\ii \pi\sum^N_{j=1}\hat{f}^\dagger_j \hat{f}_j)}$ is the parity operator.

In this picture, the operators $\hat{f}_i^\dagger$ and $\hat{f}_i$ are fermionic creation and annihilation operators, but with a different notion of locality than the spin operators appearing in~\eqref{eq.HIsing}. From the Jordan-Wigner transformation~\eqref{eq:jordan-wigner}, it is clear that the fermionic operator on site $i$ are local to the whole region from site $1$ to $i$ in the fermionic picture, and vice versa. This ensures that bipartite entanglement of a connected region of sites is equivalent in the spin and the fermionic picture, because we can use translational invariance to identify this region with the sites $\{1,\dots,w\}$, for which the spin and the fermionic pictures are isomorphic, \ie the density operators are unitarily equivalent leading to the same entanglement entropy.

It is well-known~\cite{vidmar2016generalized} that the fermionic Hamiltonian~\eqref{eq.HIsingfermions} can be written as a sum of two quadratic Hamiltonians $\hat{H}_{\pm}$ of the form
\begin{align}
	\hat{H}&=\hat{H}_+\mathbb{P}_++\hat{H}_-\mathbb{P}_-\,,
\end{align}
where $\mathbb{P}_\pm$ represent orthogonal projectors onto the Hilbert subspaces $\mathcal{H}_{\pm}$ of even and odd number of excitations, respectively, \ie
\begin{align}
    \mathcal{H}_\pm=\mathrm{span}\left\{\ket{n_1,\dots,n_L}\,\big|\,e^{\ii\pi\sum^N_{i=1}n_i}=\pm 1\right\}
\end{align}
with $\hat{f}_i^\dagger \hat{f}_i\ket{n_1,\dots,n_n}=n_i\ket{n_1,\dots,n_n}$.
We can equivalently describe the Hamiltonian in terms of Majorana modes $\hat{\xi}^a_i\equiv(\hat{q}_i,\hat{p}_i)$ with
\begin{align}
    \hat{q}_i:=\frac{1}{\sqrt{2}}(\hat{f}_i^\dagger+\hat{f}_i)\quad\text{and}\quad\hat{p}_i:=\frac{\ii}{\sqrt{2}}(\hat{f}^\dagger_i-\hat{f}_i)\,,\label{eq:def-qp-ferm}
\end{align}
which leads to the Hamiltonian given by
\begin{align}
    \hspace{-2mm}\hat{H}=\ii\sum^{N}_{i=1}\left(J\hat{p}_{i}\hat{q}_{i+1}-h \hat{q}_i\hat{p}_i\right)-J\ii \hat{p}_N\hat{q}_{1}(\hat{P}+1)\,.
\end{align}
We are particularly interested in the ground state $\ket{0}$ of the critical model with $J=h>0$, which is completely characterized by its covariance matrix
\begin{align}
\begin{split}
    \hspace{-2mm}\Omega^{ab}_{ij}&=\braket{0|\hat{\xi}^a_i\hat{\xi}^b_j-\hat{\xi}^b_j\hat{\xi}^a_i|0}\\
    &=\frac{1}{N}\sum_{\kappa}\begin{pmatrix}
    0 & -\cos{\kappa(\tfrac{1}{2}+i-j)}\\
    \cos{\kappa(\tfrac{1}{2}+j-i)} & 0
    \end{pmatrix}\,,
\end{split}
\end{align}
where the variables $\hat{\xi}^{a}_{i}$ are defined using~\eqref{eq:def-qp-ferm} as
\begin{equation}
    \hat{\xi}^a_i\equiv(\hat{q}_i,\hat{p}_i),
\end{equation}
and $\kappa=\frac{\pi}{N} (2k+1)$ with $k=-\tfrac{N}{2},\dots,\tfrac{N}{2}-1$ for even $N$.

An important subtlety arises if we compute bipartite entanglement of disconnected regions, because in this case the entanglement entropies are different in the spin and fermion picture. This subtle fact has been recognized numerous times in the literature~\cite{caban2005entanglement,banuls2007entanglement,friis2013fermionic, Radicevic:2016tlt, Lin:2018bud} and plays an important role when relating the lattice model with the continuum CFT~\cite{Radicevic:2019mle}. The key observation is that the canonical anticommutation relations induce a different notion of tensor product and partial trace for fermions~\cite{coser2016spin}. Interestingly, this different notion only affects the bipartite entanglement entropy of disjoint regions, \ie the reduced state in a subsystem consisting of two non-adjacent intervals on the circle will be different if we compute it using the spin vs. fermion picture. We comment on this in section~\ref{sec:fermion-subsystem} and review the respective literature in appendix~\ref{app:spin-vs-fermion}. In practical terms, this fact will lead us to apply our Gaussian numerics based on purifications only to the case when the two intervals are adjacent, \ie $d = 0$ in figure~\ref{fig:twointervals}.

Finally, note that the $c = 1$ free Dirac fermion CFT can be obtained from two copies of Ising model (\ie Majorana fermion CFT) by imposing a different GSO projection~\cite{DiFrancesco:1997nk}. As a result, the discussion in the present section about spatial locality and Gaussianity applies also to the free Dirac fermion CFT. In effect, our fermionic Gaussian methods reproduce the properties of the free Dirac fermion CFT only for a single subregion or adjacent subregions and the answers in these cases are simply given by twice the answers for the corresponding Ising calculation. It is well-known that the Dirac fermion CFT is equivalent to the free compactified boson CFT at the compactification radius $R=1$ (or equally $R=2$) via the bosonization procedure~\cite{Elitzur:1986ye,Ginsparg:1988ui}. Let us also emphasize on this occasion that modular invariance is a property that is not always imposed in free fermion calculation available in the literature (see~\cite{Lokhande:2015zma} for a discussion of the modular invariance in the context of entanglement entropy in CFTs). This sometimes leads to apparent tensions between CFT expectations and free fermion results. We will come back to this point in the next section.

\section{Mutual information \label{sec:mutualinformation}}

MI defined in~\eqref{MIde} provides an important correlation measure between two subsystems $A$ and $B$ and below we summarize some of its properties. One reason to do this is to test our ability to reproduce them using our numerics before we apply it to a much less understood case of EoP and CoP. Another one is to explore what kind of behaviour to expect from EoP and CoP.

MI is generically a non-universal quantity in CFTs, as it is related to a four-point function of twist operators and the latter is spectrum-dependent~\cite{Calabrese:2004eu}.

At large distances between the intervals, i.e.\ for $d \gg w$ in the notation of figure~\ref{fig:twointervals}, the operator product expansion analysis predicts the following behaviour of MI
\begin{eqnarray}
I(A:B) \sim| \langle O^{\min}_AO^{\min}_B\rangle|^2\sim  \left(\frac{w}{d}\right)^{-4\Delta_{\min}},  \label{CCDd}
\end{eqnarray}
where $O^{\min}$ is the operator with lowest (but non-zero) conformal dimension and where $\Delta_{\min}=h_{\min}+\bar{h}_{\min}$~\cite{Cardy:2013nua,Ugajin:2016opf}.

At short separations, $d \ll w$,  one expects the following universal result~\cite{Calabrese:2009qy,Calabrese:2004eu}
\begin{eqnarray}
I(A:B)\simeq \frac{c}{3}\log\frac{w}{2d}. \label{CCQq}
\end{eqnarray}
For $d = 0$, one can use a universal, \ie only $c$-dependent, formula for a single interval entanglement entropy in the vacuum to arrive at a variant of~\eqref{CCQq} with $d = \delta$. In table~\ref{tab:EoP-MI} we provided the form of~\eqref{CCQq} when the two intervals have arbitrary lengths.

Moving on to the two models we are considering: for the free massless scalar QFT one has continuous and gapless spectrum of primary operators. As a result, the formula~\eqref{CCDd} does not apply as such and in section~\ref{sec:zero-mode} we comment on a possible generalization. However, for a compactified free boson CFT, see~\eqref{eq.defcompactboson}, the spectrum of operators develops a gap~\eqref{eq.Deltaminboson} and calculations of entanglement entropy in~\cite{Calabrese:2009ez} do reproduce this behaviour.

For the $c=1/2$ Ising CFT in the limit $d\gg w$, we expect  $I(A:B)\sim (w/d)^{1/2}$, because $O^{\min}$ is the spin operator $\sigma \sim \hat{S}^{\X}_i$, see~\eqref{eq.HIsing}, which has $h=\bar{h}=\frac{1}{16}$, \ie $\Delta_{\min}=\frac{1}{8}$. Furthermore, as the free Dirac fermion CFT is dual to the free compact boson theory at $R = 1$ (or, equivalently, $R = 2$), according to~\eqref{eq.Deltaminboson} one expect MI to decay as $1/d$ governed by the $h = \bar{h} = \frac{1}{8}$ operator. Such an operator would be natural to interpret as a product of two $\sigma$ operators from each underlying Majorana fermion model.

Let us also note that existence of the following formula for MI for Dirac fermions~\cite{Casini:2005rm}
\begin{eqnarray}
I(A:B)=\frac{1}{6}\log\frac{(d+w)^2}{d\,(2w+d)}  \label{CCFf}
\end{eqnarray}
While this formula agrees at short distances with~\eqref{CCQq}, at large distances it falls off as $1/d^2$ rather than the aforementioned $1/d$ predicted by bosonization. This is related to the fact that the calculation in~\cite{Casini:2005rm} utilizes torus partition function with the anti-periodic boundary condition for fermions. However, the modular invariant partition function leading to~\eqref{CCDd} includes also contributions from sectors in which fermions satisfy period boundary conditions. This is directly related to the discussion about reduced density matrices in the fermionic formulation of the Ising model mentioned in section~\ref{sec.Ising} and expanded later in section~\ref{sec:fermion-subsystem} and appendix~\ref{app:spin-vs-fermion}.

Having discussed the analytic expectations, let us show how our bosonic and fermionic Gaussian method reproduces them. This should be regarded as a cross-check of both our numerical lattice setup and its ability to reproduce features of the continuum limit. Furthermore, it will illustrate to what extent considering a decompactified free boson with the zero mode regulated via non-vanishing mass captures long ($d \gg w$) and short distance ($d \ll w$) CFT expectations.

\begin{figure*}[htb]
\centering
\begin{tabular}{c  c p{0.5cm} c}
    & \hspace{0.5cm} {\bf Large block width $w$, small distance $d$} & & \hspace{0.5cm}  {\bf Small block width $w$, large distance $d$} \\[0.15cm]
    \rotatebox{90}{\hspace{0.05\textheight} {\bf Bosonic MI}} &  
    \includegraphics[height=0.17\textheight]{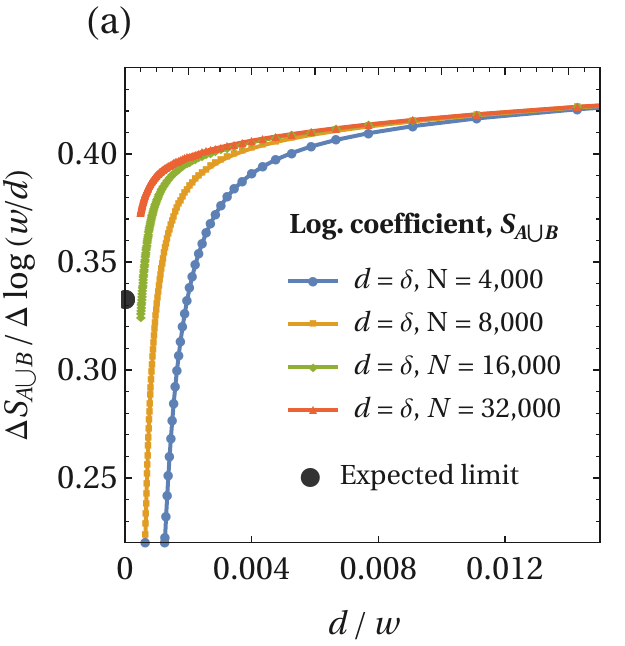}
    \includegraphics[height=0.17\textheight]{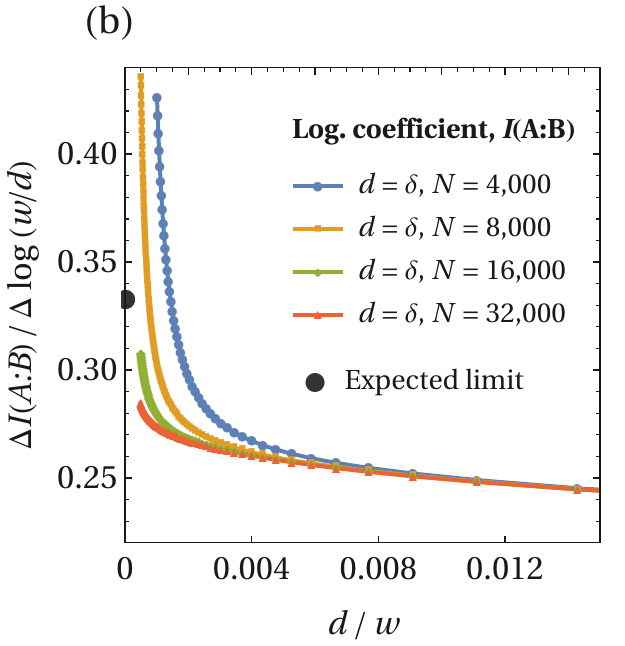} & &
    \includegraphics[height=0.17\textheight]{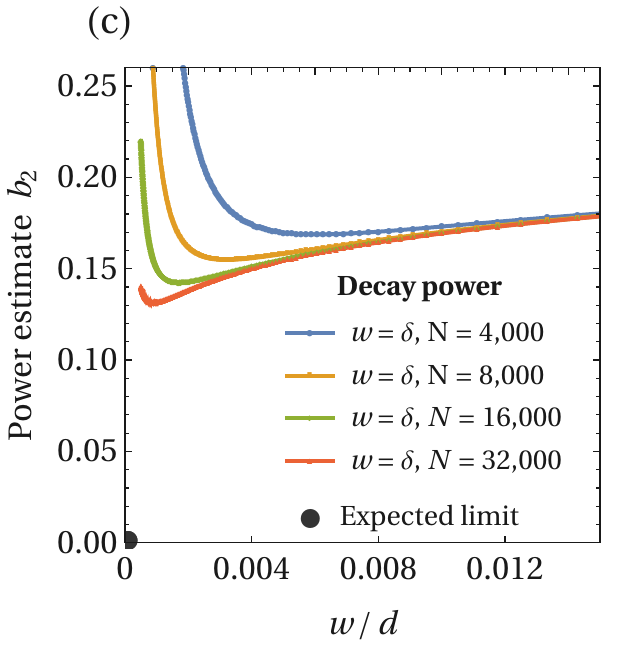}
    \includegraphics[height=0.17\textheight]{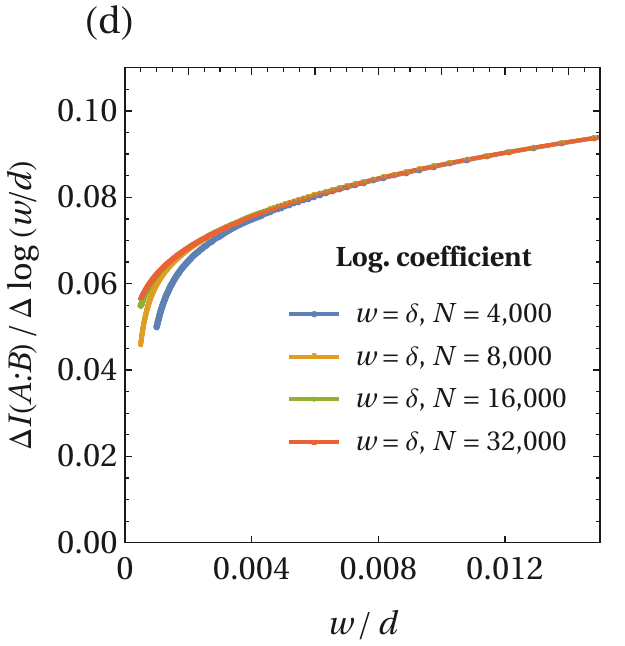} \\[0.15cm] 
    \rotatebox{90}{\hspace{0.045\textheight} {\bf Bosonic EoP}} &  \includegraphics[height=0.17\textheight]{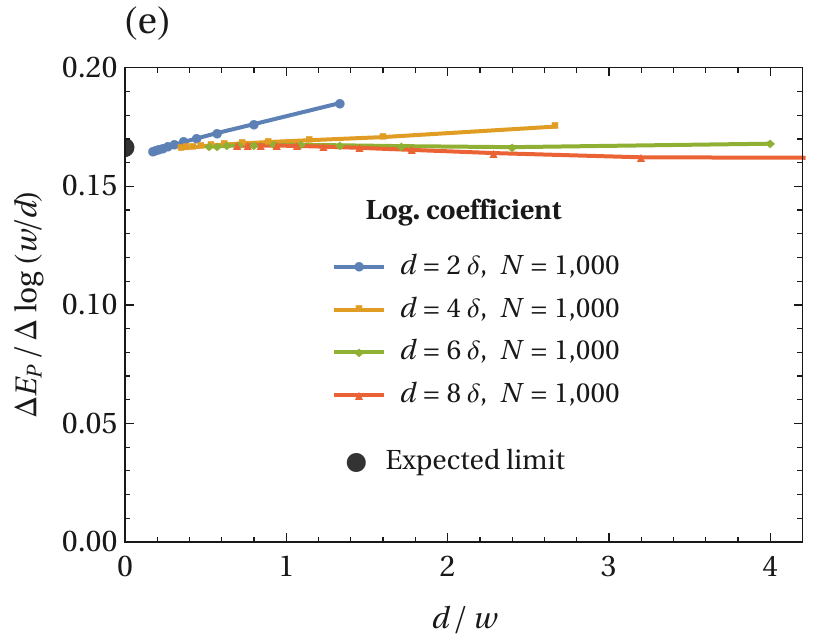} & &
    \includegraphics[height=0.17\textheight]{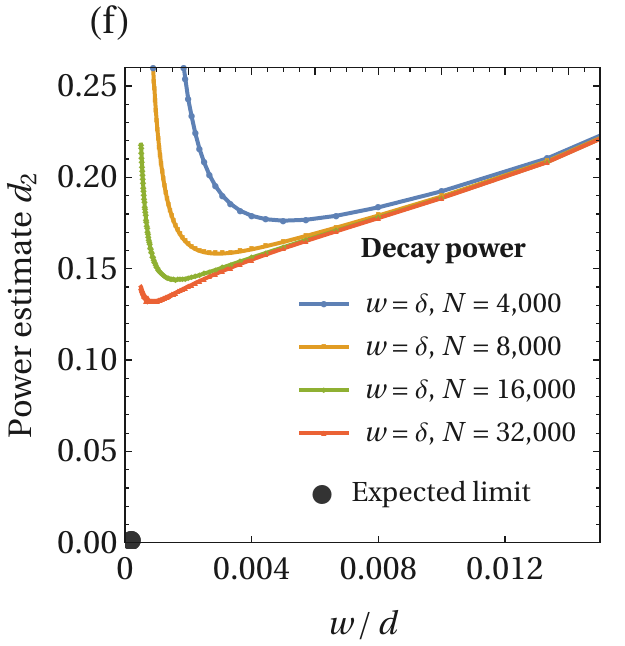}
    \includegraphics[height=0.17\textheight]{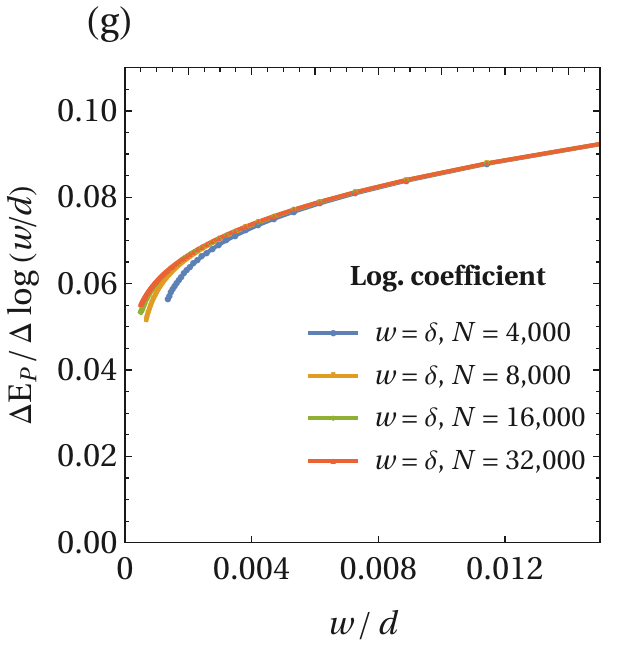}
\end{tabular}
\caption{Numerical data for bosonic MI (a-d) and EoP (e-g) in the regimes of large (a,b,e) and small (c,d,f,g) ratio of block width $w$ to distance $d$ on a periodic system of $N$ sites of bosons. (a-b) Logarithmic coefficient of $S_{A\cup B}$ and $I(A:B)$. (c-d) Decay power and logarithmic coefficient of $I(A:B)$. (e) Logarithmic coefficient of $E_P$. (f-g) Decay power and logarithmic coefficient of $E_P$. Expected limits and numerical estimates are in table~\ref{tab:EoP-MI}. The bosonic scale $m L = m N \delta$ is set to $10^{-5}$ for MI and $10^{-3}$ for EoP, as the MI computation is more stable at small values of $m L$.}
\label{fig:MI_EoP}
\end{figure*}

First, we consider the behavior of MI for a free bosonic field with central charge $c=1$, shown in the first row of figure~\ref{fig:MI_EoP}. As anticipated in section~\ref{sec:models.KG}, employing Gaussian methods restricts us to the decompactified free boson with non-vanishing mass.

In the limit of small $d/w$, we find a logarithmic dependence similar to \eqref{CCQq}, specifically
\begin{align}
I(A:B, w \gg d) &= a_0 + a_1 \log{\frac{w}{d}} - \frac{1}{2}\log(L \, m) \ .
\end{align}
Note the logarithmic $L m$ dependence was already observed in \cite{Bhattacharyya:2019tsi}. The coefficient $a_1$, expected to be $c/3\equiv1/3$ in the continuum limit, converges very slowly as the block widths $w$ and the size $L$ of the periodic system are increased. However, we can bound it by considering the behavior of $I(A:B)$ and $S(A\cup B)$ separately. Estimating $a_1$ by a discrete derivative with respect to $\log(w/d)$, we find that this estimate approaches $1/3$ from above for the $I(A:B)$ data, and from below for $S(A\cup B)$, as shown in the top-left corner of figure~\ref{fig:MI_EoP}, suggesting an asymptotic $\propto 1/3 \log{w}$ behavior identical to that of $S(A)$. Our data, extending until $N=32000$ and $w=2000\,\delta$, yields a bound $0.27 \lesssim a_1 \lesssim 0.40$, consistent with our expectations. All shown data uses a distance $\frac{d}{\delta}=1$ of one lattice site, as lattice effects on the value of the $a_1$ estimate were found to be negligible.

The behaviour at large $d/w$ can be anticipated to be subtle in light of~\eqref{CCDd} as in the present case $\Delta_{\mathrm{min}} \rightarrow 0$. In particular, assuming a power law behavior
\begin{equation}
\label{eq.IABlarged}
I(A:B, w \ll d) = b_0 + b_1 \left(\frac{w}{d}\right)^{b_2} - \frac{1}{2} \log(L m) \ ,
\end{equation}
one would expect $b_{2}$ to vanish. The power $b_2$ can be estimated by discretizing the derivatives in the expression
\begin{equation}
b_2 = 1 + \frac{\text{d}}{\text{d}\log(w/d)} \log \frac{\text{d}I(A:B)}{\text{d}(w/d)}.
\end{equation}
Indeed, we find its estimated value to gradually decrease at large $d/w$ as $N$ is increased, though MI can be well approximated by a power law with $b_2 \approx 0.2$ in the range $1 < d/w < 100$, consistent with earlier numerical studies in the $d/w < 50$ range \cite{Bhattacharyya:2019tsi}.
In the $d/w \to \infty$ limit, we can bound $b_2 \lesssim 0.15$. The coefficient of a potential logarithmic growth $I(A:B) \sim b_1^\prime \log (w/d)$ in this limit can be bounded as $b_1^\prime \lesssim 0.06$. While these results are obtained on a circle and extrapolated to a line, in section~\ref{sec:zero-mode} we will discuss the large-distance behavior of free boson directly on a line, where we will also consider possible sub-logarithmic decay functions at large $d/w$. 
Such functional dependencies, resulting from subtle large distance behavior for free, nearly massless bosons, will reappear in the context of EoP studies in section~\ref{sec:EoP}.

As we can only study the Ising CFT via a Gaussian fermionic model under the Jordan-Wigner transformation for adjacent intervals, MI computed in this approach is only relevant for the $d=0$ case, which follows directly from the entanglement entropy formula for a single interval. These formulas are included in table~\ref{tab:EoP-MI} for completeness.

\section{Gaussian purifications}\label{sec:purifications}
As discussed in section~\ref{sec:models}, we focus on free theories since their ground states are Gaussian states, so we have a powerful machinery at our disposal to analytically compute the entanglement entropy and other quantities analytically from the covariance matrix of pure Gaussian states (see appendix~\ref{app:Gaussian-entanglement}). Similar analytical formulas exist also for the $L^2$ circuit complexity of interest. The primary goal of this paper is to use this machinery to define and compute similar quantities, such as entanglement entropy and complexity, for mixed states leading to the notions of EoP \cite{Terhal:2002} and CoP~\cite{Agon:2018zso} bearing in mind the distinguishing feature that complexity is defined with respect to a reference state, whereas entanglement entropy is not. They are defined in the following way:
\begin{enumerate}
    \item We start with a function $f(\ket{\psi})$ that is defined for arbitrary pure states $\ket{\psi}$.
    \item For a mixed state $\rho_A\in\mathcal{H}_A$, we construct the purification $\ket{\psi}$ on a larger Hilbert space $\mathcal{H}=\mathcal{H}_A\otimes\mathcal{H}_{A'}$, such that $\rho_A=\Tr_{\mathcal{H}_{A'}}\ket{\psi}\bra{\psi}$. Of course, there is quite some freedom of how large the purifying Hilbert space $\mathcal{H}_{A'}$ can be.
    \item The purification $\ket{\psi}$ is not unique, but if we have found one purification $\ket{\psi}$, we can construct any other purification by acting with a unitary $U=\id_A\otimes U_{A'}$, where $U_{A'}$ is an arbitrary unitary on the purifying Hilbert space $\mathcal{H}_{A'}$.
    \item We then define a new function $F(\rho_A)$ for the mixed state to be given by
    \begin{align}
        F(\rho_A)=\min_{U=\id_A\otimes U_B} f(U\ket{\psi})\,,
    \end{align}
    \ie we minimize the original quantity $f(\ket{\psi})$ defined for pure states over all purifications of the mixed state $\rho_A$.
\end{enumerate}
Note that there are some subtleties related to the fact that the purifying Hilbert space $\mathcal{H}_{A'}$ may not have a direct physical interpretation, \eg if $\mathcal{H}_A$ represents a local subsystem (region in real space) of a QFT, it is a priori not clear what the physical meaning of $\mathcal{H}_{A'}$ is. Consequently, the function $f$ needs to be defined in an appropriate way so that it can be meaningfully applied to arbitrary extended Hilbert spaces $\mathcal{H}=\mathcal{H}_A\otimes\mathcal{H}_{A'}$. While this is relatively straightforward for entanglement entropy, one needs to be careful about circuit complexity since it is usually defined with respect to a reference state that is chosen as spatially disentangled with respect to a physical notion of locality. As explained in~\eqref{eq:CoP}, one can show that this can also be done for the purifying Hilbert space $\mathcal{H}_{A'}$ in such a way that the resulting CoP is actually independent of the notion of locality or, put differently, the outcome of the minimization procedure can even be understood as equipping $\mathcal{H}_{A'}$ with a notion of locality.

While both EoP and CoP have been introduced previously, their efficient evaluation has been an ongoing problem for practical applications. The reason is that the required minimization procedure must be generally performed numerically, while the dimension of the respective manifold over which one needs to optimize grows quickly with the number of degrees of freedom. Therefore, EoP has been only studied for small systems and often only with respect to certain subfamilies of states, while CoP was exclusively studied via purifying individual degrees of freedom~\cite{Caceres:2019pgf} rather than directly larger subsystems.

The key ingredient that enables the progress of the present paper is that we can efficiently compute EoP and CoP for the family of Gaussian states. For this, we start with a Gaussian mixed state $\rho^{\mathrm{G}}_A$ and compute a Gaussian purification $\ket{\psi^{\mathrm{G}}}$. When performing our minimization algorithm, we only sample over Gaussian states, \ie we define the new function
\begin{align}
    F^{\mathrm{G}}(\rho_A^{\mathrm{G}})=\min_{U^{\mathrm{G}}=\id_A\otimes U_{A'}^{\mathrm{G}}}f(U^{\mathrm{G}}\ket{\psi^{\mathrm{G}}})\,.
\end{align}
Clearly, we must have $F(\rho_A^{\mathrm{G}})\le F^{\mathrm{G}}(\rho_A^{\mathrm{G}})$, \ie $F^{\mathrm{G}}(\rho_A^{\mathrm{G}})$ is an upper bound for the true minimum. Moreover, it is reasonable to assume that for many quantities, such as EoP and CoP, we actually have the equality $F(\rho_A^{\mathrm{G}})= F^{\mathrm{G}}(\rho_A^{\mathrm{G}})$. This was already conjectured in~\cite{Bhattacharyya:2018sbw} and is further supported by~\cite{Windt:2020tra}. In the case of CoP, there is still limited progress in even defining circuit complexity for non-Gaussian states, which means that it is natural to only consider $F^{\mathrm{G}}(\rho_A^{\mathrm{G}})$ to start with. In both cases, it is therefore a meaningful restriction to only consider Gaussian purifications of Gaussian states.

For Gaussian states, we can use the covariance matrix and linear complex structure formalism as explained in appendix~\ref{app:Gaussian-entanglement} (see \cite{hackl2020bosonic,Windt:2020tra} for further details). Rather than working with Hilbert space vectors, which would live in an infinite dimensional Hilbert space for bosons and a $2^{N_A}$-dimensional Hilbert space for fermions, we can fully characterize the Gaussian state by a $2N_A$-by-$2N_A$ dimensional matrix, where $N_A$ represents the number of bosonic or fermionic degrees of freedom. We restrict to Gaussian states with $z^a=\tr(\hat{\xi}^a\rho_A^{\mathrm{G}})$, for which all relevant information is encoded in the so called restricted complex structure $J_A$ defined in~\eqref{eq:JA}. For a mixed Gaussian state, $J_A$ has purely imaginary eigenvalues $\pm\ii c_i$, where $c_i\in[1,\infty)$ for bosons and $c_i\in[0,1]$ for fermions. The state is pure only if all $c_i=1$. For every mixed state $\rho_A^{\mathrm{G}}$
\begin{align}\label{eq:restrictedJstandard}
    J_A\equiv\left(\begin{array}{ccc}
    \mathbb{C}_1 & 0 & 0 \\
    0 & \ddots & 0\\
    0 & 0 & \mathbb{C}_{N_A}
    \end{array}\right)\,\,\text{with}\,\, \mathbb{C}_i=c_i\begin{pmatrix}
    0 & 1\\
    -1 & 0
    \end{pmatrix}\,.
\end{align}
We can always purify such a state using a Hilbert space~$\mathcal{H}_{A'}$ with the same number of degrees of freedom as $\mathcal{H}_A$, \ie $N_{A'}=N_A$. Then, there always exists a basis in the system $A'$, such that the complex structure $J$ of the purified state $\ket{\psi^{\mathrm{G}}}$ takes the form~\cite{hackl2019minimal}
\begin{align}\label{eq:JStandard}
    J\equiv\left(\begin{array}{ccc|ccc}
    \mathbb{C}_1 & 0 & 0 & \mathbb{S}_1 & 0 & 0 \\[-2mm]
    0 & \ddots & 0 & 0 & \ddots & 0 \\
    0 & 0 & \mathbb{C}_{N_A} & 0 & 0 & \mathbb{S}_{N_A}\\[1mm]
    \hline
    \pm\mathbb{S}_1 & 0 & 0 & \mathbb{C}_1 & 0 & 0 \\[-2mm]
    0 & \ddots & 0 & 0 & 0 & 0 \\
    0 & 0 & \pm\mathbb{S}_{N_A} & 0 & 0 & \mathbb{C}_{N_A}
    \end{array}\right)\,,
\end{align}
where $(+)$ applies to bosons and $(-)$ to fermions with
\begin{align}
    \mathbb{C}_i=c_i\begin{pmatrix}
    0 & 1\\
    1 & 0
    \end{pmatrix}\quad\text{and}\quad\mathbb{S}_i=s_i\begin{pmatrix}
    0 & 1\\
    1 & 0
    \end{pmatrix}
\end{align}
and $s_i=\sqrt{c_i^2-1}$ for bosons and $s_i=\sqrt{1-c_i^2}$ for fermions\footnote{An equivalent parametrization is given by $c_i=\cosh{2r_i}$ and $s_i=\sinh{2r_i}$ for bosons and $c_i=\cos{2r_i}$ and $s_i=\sin{2r_i}$ for fermions, as used in~\cite{hackl2019minimal,Windt:2020tra}.}. From the perspective of Gaussian states, different purifications of $\rho_A^{\mathrm{G}}$ only differ in the choice of basis of the purifying system $B$, for which $J$ takes the above standard form. Consequently, we can use the action of the respective Lie group $\mathcal{G}_B$ (symplectic group $\mathrm{Sp}(2N_{A'},\mathbb{R})$ for bosons, orthogonal group $\mathrm{O}(2N_{A'},\mathbb{R})$ for fermions) to transform $J\to MJM^{-1}$ with $M=\id_A\oplus M_{A'}$, where $M_{A'}\in\mathcal{G}_{A'}$ represented as $2N_{A'}$-by-$2N_{A'}$ matrix.

As reviewed in appendix~\ref{app:algorithm}, we optimize over all Gaussian purification by taking the natural geometry (Fubini-Study metric) of the state manifold into account. Using the fact that this geometry is compatible with the group action, \ie the Fubini-Study metric on the manifold of purifications is left-invariant under the group action of $\mathcal{G}_{A'}$, we do not need to recompute the metric at every step, but can fix an orthonormal basis of Lie algebra generators equal to the dimension of the manifold. This enables us to efficiently perform a gradient descent search attuned the geometry of states, which scales polynomially in the number of degrees of freedom and enables us probe the field theory regime of our discretized models, which has not been possible previously in this setting. In particular, previous studies~\cite{Bhattacharyya:2018sbw,Bhattacharyya:2019tsi} of EoP restricted to special classes of Gaussian states (namely, real Gaussian wave functions generated by the subgroup $\mathrm{GL}(N_{A'},\mathbb{R})\subset\mathrm{Sp}(2N_{A'},\mathbb{R})$) for a small number of degrees of freedom. Similarly, CoP has been almost exclusively studied by purifying individual degrees of freedom (mode-by-mode purifications~\cite{Caceres:2019pgf}) rather than whole subsystems for larger $N_A$. 

For purifications of small subsystems, e.g.\ of $1+1$ or $2+2$ sites, this optimization only takes a few seconds on a desktop computer, and is still feasible within a few hours for $10+10$ sites, with efficiency depending on the optimization function, the accuracy threshold, and the hardware on which the computation is performed. For the particular case of CoP the optimization procedure for bosons was found to be an order of magnitude faster than the fermionic case for the same accuracy threshold, even for small subsystems. This  implied that for larger subsystems, \eg of order of $10+10$ sites, the optimization parameters such as the gradient and function tolerance were lowered without compromising the results. For instance, lowering the gradient and function tolerance by a couple of orders of magnitude resulted in changes in the final value of the optimization in the third or fourth decimal.

\section{Entanglement of purification}\label{sec:EoP}
We discuss our results for the EoP in bosonic and fermionic field theories using the purifications discussed in the previous section the algorithm described in appendix~\ref{app:algorithm}.

\subsection{Definition and existing results}
EoP is a measure of correlations, which include both classical and quantum ones, and can be regarded as a mixed state generalization of entanglement entropy~\cite{Terhal:2002}. When a mixed state $\rho_{AB}: \mathcal{H}_{AB}\to\mathcal{H}_{AB}$ is given, we first purify it into a pure state $\ket{\psi}\in\mathcal{H}$ by extending the Hilbert space $\mathcal{H}_{AB}$ according to
\begin{align}
    \mathcal{H}_{AB}=\mathcal{H}_A\otimes \mathcal{H}_B\to\mathcal{H}=\mathcal{H}_A\otimes \mathcal{H}_B\otimes \mathcal{H}_{A'}\otimes \mathcal{H}_{B'}
\end{align}
such that $\rho_{AB}=\Tr_{A'B'}(\ket{\psi}\bra{\psi})$. The EoP $E_P(\rho_{AB})$ is defined as the minimum of the entanglement entropy $S(A \cup A')=-\Tr(\rho_{AA'}\log \rho_{AA'})$ for the reduced density matrix $\rho_{AA'}=\Tr_{BB'}(\ket{\psi}\bra{\psi})$ over all possible purifications
\begin{equation}
\label{eq.EoPdef}
E_P(\rho_{AB})=\min_{\ket{\psi}}\left[S(A\cup A')\right].
\end{equation}
When $\rho_{AB}$ is pure, it simply reduces to the entanglement entropy as $E_P(\rho_{AB})=S(A)=S(B)$.

EoP is relatively new to the QFT setting and its understanding in this context is in development, which adds a strong motivation for our paper. Our knowledge about this subject is based on a conjecture in holography, results governed by local conformal transformations in CFTs and ab initio studies in free QFTs, which is the research direction the present work subscribes to. Below we briefly summarize the state of the art that sets the stage for the results of our research.

In strongly-coupled CFTs, a holographic formula which computes EoP was proposed in \cite{Takayanagi:2017knl,Nguyen:2018}. Analytical calculations of EoP, based on the idea of path-integral optimization for CFTs~\cite{Caputa:2017urj}, were given in \cite{Caputa:2018xuf}. In particular, when the subsystem $A$ and $B$ are adjacent in a CFT, both holographic and path-integral result predict the universal formula
\begin{equation}
\label{EQ_EOP_A1A2t}
    E_P = \frac{c}{6} \log \frac{2 w_A w_B}{(w_A + w_B)\delta} \ , 
\end{equation}
 where the widths of $A$ and $B$ are $w_A$ and $w_B$, respectively and $\delta$ is the lattice spacing. Exploratory numerical calculations of EoP in a lattice regularization of (1+1)-dimensional free scalar field theory have been performed in \cite{Bhattacharyya:2018sbw,Bhattacharyya:2019tsi}. Below we would like to extend such computations so that we can compare the result~\eqref{EQ_EOP_A1A2t} with our discretized numerical calculations, as well as understand better the long distance physics ($d \gg w$) in the QFT limit. The key technical difference on this front with respect to~\cite{Bhattacharyya:2018sbw,Bhattacharyya:2019tsi} is using bigger total system sizes, significantly bigger subsystems -- both of which are desired to be closer to the QFT limit -- and the most general Gaussian purifications discussed in section~\ref{sec:purifications}.

\subsection{Numerical studies using the most general Gaussian purifications}
Using the approach outlined in section~\ref{sec:purifications} and numerical techniques explained in appendix~\ref{app:algorithm}, we can now compute both bosonic and fermionic EoP for purifications on the whole Gaussian manifold. In light of the discussion of our models in section~\ref{sec:models}, for bosons we expect the Gaussian ansatz to describe well the CFT properties when the two intervals are adjacent ($d = 0$) or at small separation $d \ll w$, as in these cases we do not expect the zero mode to be a significant contribution to the calculations we perform. For fermions, we expect the Gaussian ansatz to be appropriate for CFT calculations only when the two intervals are adjacent. Otherwise, the desired starting point of our calculations, spatially reduced density matrices for the Ising and Dirac fermion CFTs are non-Gaussian and our method is not applicable. In order to complete the picture, we will nevertheless provide results of our methods for bosons and fermions in the aforementioned regimes, however, they are not supposed to be seen as CFT predictions based on lattice calculations.

\begin{figure}[tb]
\centering
\includegraphics[height=0.18\textheight]{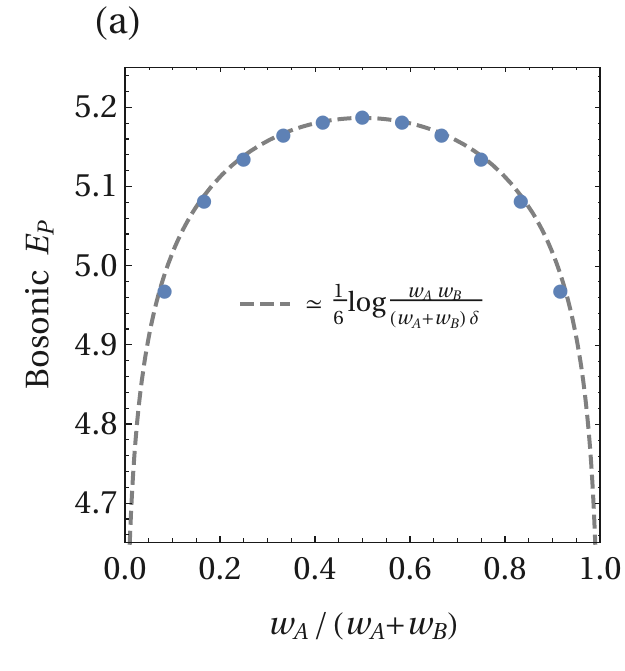}
\hspace{0.3cm}
\includegraphics[height=0.18\textheight]{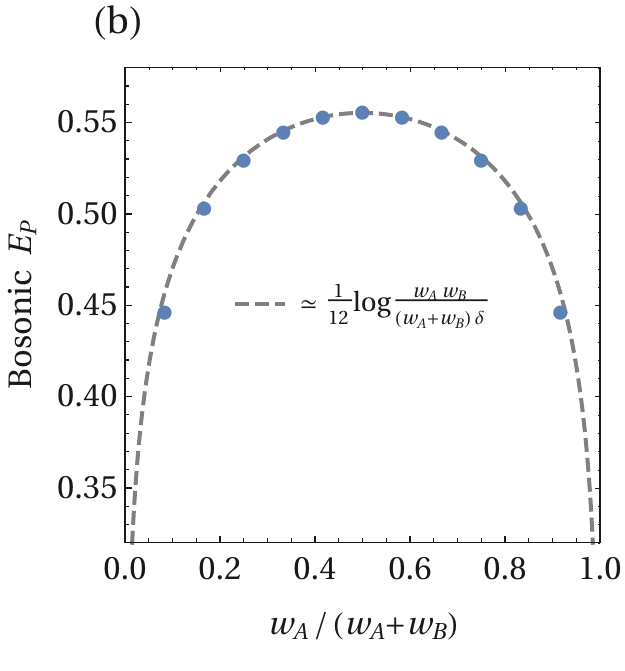}
\caption{Bosonic ($c=1$, (a)) and fermionic/Ising spin EoP ($c=\tfrac{1}{2}$, (b)) for two adjacent ($d = 0$) subsystems $A$ and $B$ on $\frac{w_A+w_B}{\delta}=12$ sites, with the continuum result~\eqref{EQ_EOP_A1A2t} for a fitted lattice spacing $\epsilon$ plotted as a dashed curve. Total system size $N=1200$. Bosonic mass scale $m \, L = 10^{-4}$. 
}
\label{FIG_EOP_A1A2t}
\end{figure}

Starting with the adjacent intervals, our Gaussian lattice calculations perfectly reproduce the behavior~\eqref{EQ_EOP_A1A2t} 
as shown in figure~\ref{FIG_EOP_A1A2t}, in both our 
bosonic and fermionic (or equivalently, Ising spin) computations up to slight lattice effects at small $w_A$ or $w_B$.

We move on to a more general case where the subsystem $A$ and $B$ are disjoint intervals in a free CFT. We again take the lengths of both intervals to be $w$ and the distance between them to be $d$. When $d\ll w$, both holographic \cite{Takayanagi:2017knl,Nguyen:2018} and path-integral approaches~\cite{Caputa:2018xuf} predict the behavior 
\begin{equation}
\label{EQ_EOP_A1A2s}
    E_P = \frac{c}{6} \log \frac{2w}{d} \ , 
\end{equation}
which agrees with \eqref{EQ_EOP_A1A2t} under the replacement $\delta=d, w_A=w_B=w$.
On the other hand, no universal results have been known for  $d\gg w$ and one possibility is a behavior similar to MI described in section~\ref{sec:mutualinformation}.

The numerical results for nearly a massless free scalar QFT are plotted in the second row of figure~\ref{fig:MI_EoP}.
As expected, we find a logarithmic dependence on $w/d$ when it becomes large, given by
\begin{equation}
E_P(w \gg d) = c_0 + c_1 \log\frac{w}{d} - \frac{1}{2} \log(L m)  \ ,
\end{equation}
with a convergence to $c_1 \approx \tfrac{1}{6}$ much faster than seen in MI. This result is consistent with the $\propto \tfrac{c}{6} \log\tfrac{w}{d}$ behavior of~\eqref{EQ_EOP_A1A2s}.

In the case of small~$w/d$, we observe that the bosonic EoP behaves extremely similar to bosonic MI. Such an observation for smaller subsystems and separations was already made in~\cite{Bhattacharyya:2019tsi} and our results should be seen as a corroboration of this earlier finding. Given this similarity and our discussion of MI in section~\ref{sec:mutualinformation}, it should not come as a surprise that a power-law fit to the bosonic EoP in the regime of small~$w/d$,
\begin{equation}
E_P(w \ll d) = d_0 + d_1 \left(\frac{w}{d}\right)^{d_2} - \frac{1}{2} \log(L m)  \ ,
\end{equation}
is unstable as $w/d \to 0$. The best we could do is to provide the upper bound on the power, $d_2 \lesssim 0.15$, which is consistent with the absence of a long-distance power behaviour. One should note that the power of such a quasi-power-law for EoP agrees well with the one extracted for MI, as can be seen by comparing the two rows of figure~\ref{fig:MI_EoP}~(right).

\section{Complexity of purification}\label{sec:CoP}
In the present section we provide a comprehensive discussion of CoP in the single and two adjacent interval case ($d = 0$ in figure~\ref{fig:twointervals}). We start by briefly reviewing the relevant results of the holographic complexity proposals, as well as the studies of pure state complexity in free QFTs. These results will guide us in the choice of a reference state and, also, in choosing the way to combine two and single interval CoPs to get a complexity analogue of MI. Subsequently, we discuss the CoP results obtained via optimization over the whole Gaussian manifold. We focus on the single and adjacent intervals to provide a clean message and we hope to report the $d$-dependence of CoP, which at least superficially seems involved, in further work. We also compare some of our results with a simplified version of a single mode purifications adopted in an earlier study of single interval CoP by~\cite{Caceres:2019pgf} to avoid the technical problem our work addresses, \ie optimizing over the full manifold of Gaussian purifications. Finally, we compare the properties of CoP with the notion of mixed state complexity discussed in~\cite{DiGiulio:2020hlz}.

\subsection{Holographic predictions}
Holographic complexity proposals relate novel gravitational observables associated with choosing a time slice on the asymptotic boundary of solutions of AdS gravity with measures of hardness of preparing corresponding pure states in dual QFTs using limited resources. The first covariant notion is the spatial volume of the boundary-anchored extremal (codimension-one) bulk time slice~(${\cal C}_{V}$)~\cite{Stanford:2014jda}. The second covariant notion is the spacetime volume (\ie a codimension-zero quantity) of the bulk causal development of such a time slice~(${\cal C}_{V2.0}$)~\cite{Couch:2016exn}. The third covariant notion is also of a codimension-zero type and is the bulk action evaluated in the causally defined region~(${\cal C}_{A}$)~\cite{Brown:2015bva,Brown:2015lvg}. The first two quantities are unique up to an overall normalization, whereas ${\cal C}_{A}$ has an additional ambiguity related to the presence of null boundaries~\cite{Lehner:2016vdi,Reynolds:2016rvl}. 

While there is also another evidence in support of the association of ${\cal C}_{V}$, ${\cal C}_{V2.0}$ and ${\cal C}_{A}$ with complexity, an important clue about the correctness of these conjectures comes from free CFT calculations of complexity of pure states along the lines of~\cite{Chapman:2017rqy,Jefferson:2017sdb}. In particular, such free CFT calculations are able to match the structure of leading divergences of holographic complexity~\cite{Chapman:2017rqy,Jefferson:2017sdb,Khan:2018rzm,Hackl:2018ptj} provided the reference state is taken to be a spatially disentangled state. Interestingly, in the case of bosonic calculations of~\cite{Chapman:2017rqy,Jefferson:2017sdb} the scale entering the definition of a spatially disentangled reference state can be linked, via the leading divergence, both to the overall normalization freedom in the case of all three proposals, as well as to an additional ambiguity appearing in the ${\cal C}_{A}$ case. Furthermore, the free boson CFT calculation in~\cite{Chapman:2018hou} explained qualitative features of the holographic complexity excess in thermofield double states as compared to the vacuum complexity reported in~\cite{Chapman:2016hwi}.

All three holographic complexity proposals acquire natural 
generalizations for mixed states represented as spatial subregions of globally pure states~\cite{Alishahiha:2015rta,Carmi:2016wjl,Ben-Ami:2016qex}. Instead of considering extremal volumes or causal developments of a full Cauchy slice in the bulk, the mixed state version of holographic complexity proposals uses the corresponding notions applied to the relevant entanglement wedge~\cite{Czech:2012bh,Wall:2012uf,Headrick:2014cta}. While there are certainly other possibilities regarding the kind of complexity the proposals~\cite{Alishahiha:2015rta,Carmi:2016wjl,Ben-Ami:2016qex} represent, see~\cite{Agon:2018zso} for a discussion of some of the available options, we will treat their properties as a guiding principle to study CoP in free CFTs.

The results of these proposals applied to a single and two interval cases of interest can be found in table~\ref{TAB_ANALYTIC_RES_COMP}. One can clearly see that the leading divergence of holographic complexity is in the volume of the combined subregions and there can be also subleading logarithmic divergences\footnote{Note also that taking $\ell_{CT} \sim \delta$ can enhance the leading divergence in the ${\cal C}_{A}$ case by a logarithm of the cut-off and change subleading divergence.}. An earlier study of divergences encountered in the case of a single interval CoP in the vacuum of a free boson theory using restricted purifications is~\cite{Caceres:2019pgf}. In the present work we lift the restriction on purifications within the Gaussian manifold of states, include also the corresponding results for fermions and carefully resolve finite contributions to CoP including their dependence on the reference state scale and residual mass for bosons. The latter we achieve by considering the two adjacent intervals case.

For two intervals it is interesting to define a better behaved (less divergent) quantity in a manner similar to the definition of MI~\eqref{MIde}. Led by the form of leading divergences, as well as simplicity, the mutual complexity~$\Delta{\cal C}$ was defined in~\cite{Alishahiha:2018lfv} as the sum of contributions for each individual intervals and subtract from it the holographic complexity of the union
\begin{equation}
\label{eq:mutualCholo}
\Delta {\cal C} = {\cal C}(A) + {\cal C}(B) - {\cal C}(A \cup B).
\end{equation}
The results in the two intervals setup at a vanishing separations are included in table~\ref{TAB_ANALYTIC_RES_COMP} and motivated us to seek for a logarithmic behaviour as a function of $\frac{w_{A} \, w_{B}}{(w_{A}+w_{B})\, \delta}$ also in the analogous setting in free CFTs. This is also reminiscent of the behaviour of MI and EoP at $d = 0$, see table~\ref{tab:EoP-MI}.

\subsection{Definition and implementation}
CoP is defined in analogy with EoP as a measure of complexity for mixed states with the use of a definition of complexity for pure states minimized with respect to all purifications~\cite{Agon:2018zso,Stoltenberg:2018ink}. This includes, in principle, purifications which contain an arbitrary number of ancilla greater or equal to number of the degrees of freedom in the subsystem.

Given a mixed state in a Hilbert Space $\mathcal{H}_A$ characterized by a density matrix $\rho_A$, we define a new Hilbert space
\begin{align}
\mathcal{H}'=\mathcal{H}_A\otimes\mathcal{H}_{A'}
\end{align}
with ancillary system $A'$. There exist many purifications $\ket{\psi_{T}} \in\mathcal{H}'$, such that $\rho_{A}=\Tr_{\mathcal{H}_{A'}}(\ket{\psi_T}\bra{\psi_T})$. In analogy to the EoP, see~\eqref{eq.EoPdef}, we define the CoP ${\cal C}_{P}$ as the minimum of the complexity function $\mathcal{C}$ with respect to a reference state $\ket{\psi_{\mathrm{R}}}$ and to all purifications $\ket{\psi_{\mathrm{T}}}$:
\begin{equation}
    {\cal C}_{P}(\rho_A)=\min_{\ket{\psi_{\mathrm{T}}}\in\mathcal{H'}}\mathcal{C}\left(\ket{\psi_{\mathrm{T}}},\ket{\psi_{\mathrm{R}}}\right)\,.\label{eq:CoP}
\end{equation}
CoP inherits the richness of building blocks of complexity for pure states, such as dependence on the choice of a reference state $\ket{\psi_\mathrm{R}}$ as well as on the cost function which evaluates the circuits built from the unitaries generated by the Lie algebra of admissible gates. This is an additional complication with respect to EoP, which requires only minimization over purifications. For a generic definition of complexity underlying CoP one would not only need to optimize over purifications, but also for each purification one would need to solve an intricate optimization problem finding the optimal circuit.

The idea behind the present work is to make use of a particularly natural definitions of cost function for Gaussian states defined in~\cite{Hackl:2018ptj}~(fermions) and~\cite{Chapman:2018hou}~(bosons), which provide closed form and efficient to evaluate expressions for complexity. In this way we make the problem of calculating CoP in free CFTs much more manageable, as similarly to EoP, it requires now only one layer of numerical optimization. Of course, it would be very interesting to explore other cost functions in the CoP context and we leave this rather difficult problem for future studies.

To introduce the relevant cost function, let us recall from section~\ref{sec:purifications} (see also appendix~\ref{app:Gaussian-entanglement}) that bosonic and fermionic Gaussian states $\ket{J}$ can be efficiently characterized by their linear complex structure $J^a{}_b$. The latter can be constructed from their two-point function $C_2^{ab}=\braket{\hat\xi^a\hat\xi^b}$, where we only consider Gaussian states with $\braket{\hat\xi^a}=0$. As shown in~\cite{Hackl:2018ptj,Chapman:2018hou} the geodesic distance between $\ket{J_{\mathrm{R}}}$ and $\ket{J_{\mathrm{T}}}$ \emph{within} the Gaussian state manifold gives rise to a version of complexity based on a $L^2$ cost function
\begin{align}
\label{eq:CompFunct}
    \mathcal{C}(\ket{J_{\mathrm{T}}},\ket{J_{\mathrm{R}}})=
    \sqrt{\frac{\left|\tr\left(\log\left(-J_{\mathrm{T}}J_{\mathrm{R}}\right)^2\right)\right|}{8}}.
\end{align}
To relate with the discussion in the introduction, the above definition of complexity corresponds to optimization with respect to the $L^2$ cost function~\eqref{eq.costL2} with
\begin{equation}
\upeta_{IJ}=\frac{1}{4}\tr(K_I\,G\,K_J^\intercal\,G^{-1})\,\label{eq:eta}
\end{equation}
where $G^{ab}=\braket{J_{\mathrm{R}}|\hat\xi^a\hat\xi^b+\hat\xi^b\hat\xi^a|J_{\mathrm{R}}}$ for the reference state $\ket{\psi_{\mathrm{R}}}$. Due to the canonical anti-commutation relations, this normalization of the Lie algebra elements $K_I$ is independent of the reference state for fermions\ (see~\cite{Hackl:2018ptj}), but for bosons~\eqref{eq:eta} implies that $K_I$ is normalized based on the specific reference state, which was also referred to as equating reference and gate scale (as discussed in~\cite{Jefferson:2017sdb,Chapman:2017rqy}). In the above expression $K_I$ are the respective Lie algebra elements (symplectic for bosons, orthogonal for fermions) associated to the quadratic operators $O_I$ in their fundamental representation acting on the classical phase space.

In the following, we focus on minimal purifications, \ie, purifications whose ancilla have the same number of degrees of freedom as the reduced density matrix of the subsystem. Our focus on minimality comes as a result of a number of numerical computations for the cost function~\eqref{eq:CompFunct} which indicate that purifying the reduced density matrix with a larger number of ancilla does not lead to a lower CoP. It would be very interesting to explore if this feature is special to the cost function and the resulting complexity~\eqref{eq:CompFunct} we considered, however, we leave it for future investigations.

When applying the closed form complexity formula~\eqref{eq:CompFunct} to Gaussian purifications $\ket{J_{\mathrm{T}}}$ of some mixed state $\rho_A$, we need to think about what an appropriate reference state $\ket{J_{\mathrm{R}}}$ can be. The two most immediate applications are thermal states and mixed states resulting from the reduction to spatial subsystem:
\begin{itemize}
    \item \textbf{Thermal states.} Our mixed state could be the thermal state $\rho$ in a system $\mathcal{H}=:\mathcal{H}_A$, which we purify to $\mathcal{H}'=\mathcal{H}_A\otimes\mathcal{H}_{A}$. Here, we can always choose a spatially unentangled and pure reference state, which we can extend to the purifying system as $\ket{J_{\mathrm{R}}}=\ket{J_{\mathrm{R}}}_A\otimes \ket{J_{\mathrm{R}}}_{A}\in\mathcal{H}'$.
    \item \textbf{Subsystems.} We consider a pure Gaussian state $\ket{\psi}\in\mathcal{H}=\mathcal{H}_A\otimes\mathcal{H}_B$, which we reduce to some local subsystem $\rho_A=\Tr_{\mathcal{H}_B}\ket{\psi}\bra{\psi}$. In this subsystem, we have a pure and spatially unentangled Gaussian reference state $\ket{J_{\mathrm{R}}}_A$, which we can extend to the purifying system as $\ket{J_{\mathrm{R}}}=\ket{J_{\mathrm{R}}}_A\otimes \ket{J_{\mathrm{R}}}_{A}\in\mathcal{H}'$.
\end{itemize}
The spatially unentangled character of $\ket{J_{R}}$ is a choice motivated by the fact that such a state is on one hand truly simple and, on the other, in the case of pure state complexity it reduces the kind of divergence encountered in the holographic complexity proposals.

In both scenarios outlined above, only the target state $\ket{J_{\mathrm{T}}}$ is entangled across $\mathcal{H}_A\otimes\mathcal{H}_{A'}$, while the reference state is a product state $\ket{J_{\mathrm{R}}}=\ket{J_{\mathrm{R}}}_A\otimes \ket{J_{\mathrm{R}}}_{A}$. As there is no a priori physical notion of locality in the ancillary system, we only require that $\ket{J_{\mathrm{R}}}_{A}$ is pure and Gaussian. We choose
\begin{subequations}
\begin{align}
    [J_{\mathrm{R}}]&\equiv\bigoplus^{N_A}_{i=1}\begin{pmatrix}
    0 & \mu\\
    -\frac{1}{\mu} & 0
    \end{pmatrix}\,,&& \textbf{(bosons)}\label{eq.refstatebosons}\\
    [J_{\mathrm{R}}]&\equiv\bigoplus^{N_A}_{i=1}\begin{pmatrix}
    0 & 1\\
    -1 & 0
    \end{pmatrix}\,,&& \textbf{(fermions)}
\end{align}
\end{subequations}
over spatially local sites $i$, where we only introduced a reference scale $\mu$ for bosons\footnote{For fermions, the spatially unentangled vacuum is essentially unique if we require it to be translationally invariant over sites and have the same parity as the vacuum of the Ising model.}.

The optimization over all purification in~\eqref{eq:CoP} could therefore be equivalently performed over reference or target state or even both. The minimum would always be the same, which can be seen as follows. By construction, the complexity function is invariant under the action of a single Gaussian unitary $U$ acting on both states, \ie we have
\begin{align}
    \mathcal{C}(\ket{J_{\mathrm{T}}},\ket{J_{\mathrm{R}}})=\mathcal{C}(U\ket{J_{\mathrm{T}}},U\ket{J_{\mathrm{R}}})\,,\label{eq:equivalence}
\end{align}
where $U$ is related to a group transformation $M^a{}_b$ via $U^\dagger\hat{\xi}^a U=M^a{}_b\hat{\xi}^b$. In the case of Gaussian purifications, we optimize over all Gaussian purifications for the target state, \ie if we have found such a purification $\ket{J_{\mathrm{T}}}$, any other purification is given by $\id_{A}\otimes U_{A'}\ket{J_{\mathrm{T}}}$. We thus find
\begin{align}
\begin{split}
    \mathcal{C}_P&=\min_{U_{A'}}\mathcal{C}(\id_A\otimes U_{A'}\ket{J_{\mathrm{T}}},\ket{J_{\mathrm{R}}})\\
    &=\min_{V_{A'}}\mathcal{C}(\ket{J_{\mathrm{T}}},\id_A\otimes V_{A'}\ket{J_{\mathrm{R}}})\\
    &=\min_{U_{A'},V_{A'}}\mathcal{C}(\id_A\otimes U_{A'}\ket{J_{\mathrm{T}}},\id_A\otimes V_{A'}\ket{J_{\mathrm{R}}})\,,
\end{split}
\end{align}
where the equalities follow from~\eqref{eq:equivalence} and where both $U_{A'}$ and $V_{A'}$ are Gaussian unitarites on the system $A'$. In practice, we can therefore start with a basis $\hat{\xi}_A$, such that $[J_{\mathrm{T}}]_A$ takes the mixed standard form~\eqref{eq:restrictedJstandard}. It can then be purified so that the purification takes the standard form with respect to the extended basis $\hat\xi'=(\hat{\xi}_A,\hat{\xi}_{A'})$,
\begin{align}
 J_{\mathrm{T}}\equiv\left(\begin{array}{ccc|ccc}
    \mathbb{C}_1 & \cdots & 0 & \mathbb{S}_1 & \cdots & 0 \\[-2mm]
    \vdots & \ddots & \vdots & \vdots & \ddots & \vdots \\
    0 & \cdots & \mathbb{C}_{N_A} & 0 & \cdots & \mathbb{S}_{N_A}\\[1mm]
    \hline
    \pm\mathbb{S}_1 & \cdots & 0 & \mathbb{C}_1 & \cdots & 0 \\[-2mm]
    \vdots & \ddots & \vdots & \vdots & \ddots & \vdots \\
    0 & \cdots & \pm\mathbb{S}_{N_A} & 0 & \cdots & \mathbb{C}_{N_A}
    \end{array}\right)\,,
\end{align}
as defined in~\eqref{eq:JStandard}. The reference state has the block diagonal form
\begin{align}
    J_{\mathrm{R}}\equiv [J_{\mathrm{R}}]_A\oplus [J_{\mathrm{R}}]_{A'}\equiv\left(\begin{array}{c|c}
    [J_{\mathrm{R}}]_A & 0\\
    \hline
    0 & [J_{\mathrm{R}}]_{A'}
    \end{array}\right)
\end{align}
as it is a product state. We have $(\id_{A}\otimes U_{A'})\ket{J}=\ket{MJM^{-1}}$ with $M=\id_A\oplus M_{A'}$, so the optimization gives
\begin{align}
\label{eq.defourCoP}
    \mathcal{C}_P=\min_{M=\id_{A}\oplus M_{A'}}\sqrt{\frac{\tr{\left(\log(MJ_{\mathrm{T}}M^{-1}J_{\mathrm{R}})^2\right)}}{8}}\,,
\end{align}
which shows explicitly that we can think of the optimization as either applied to target or reference state. When we perform the optimization, it is actually advantageous to optimize over the reference state, as its stabilizer group is larger (\ie there are more group elements that preserve $J_{\mathrm{R}}$ than $J_{\mathrm{T}}$) and so we can identify a fewer number of directions/parameters to optimize over. Our algorithm is described in more detail in appendix~\ref{app:algorithm} and is one of subjects of the companion paper~\cite{Windt:2020tra}.

\subsection{Single interval in the vacuum}\label{sec:single-interval}

In the present and following sections we apply this general framework to the particular case of free CFTs on the lattice, as we did for EoP in section~\ref{sec:EoP}. We first consider the case of a single interval in the vacuum, which appeared earlier in the context of the aforementioned mode-by-mode purifications in~\cite{Caceres:2019pgf}. In the present section we re-address the same problem using the most general purifications, whereas in the later section~\ref{sec:singlemode} we reconsider the same problem using our simplified take on the mode-by-mode purifications to make further contact with~\cite{Caceres:2019pgf}. One expects CoP to diverge in the continuum limit in light of a general physics picture where circuits acting on a spatially disentangled state need to build entanglement at all scales to match features of CFT vacua, as well as from explicit results in~\cite{Caceres:2019pgf}.


For a single interval on a line, fermionic CoP will be a function only of~$\frac{w}{\delta}$ as the system size $N$ becomes large. Bosonic CoP, however, also contains two additional parameters, the reference state scale $\mu$, see~\eqref{eq.refstatebosons}, and the effective mass $m\, \delta$. As changing $\mu \to a \mu$ is equivalent to rescaling the mass and lattice spacing according to $m \to m / a$, $\delta \to a \delta$, we can set $\mu = 1$ in numerical calculations and restore it in analytical formulas containing the (now unit-less and independent) $m$ and $\delta$.

\begin{figure}[tb]
\centering
\includegraphics[height=0.17\textheight]{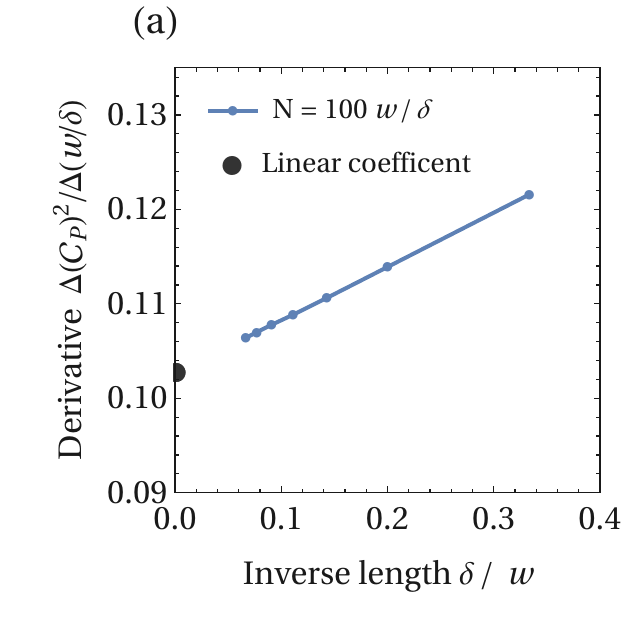} \\
\vspace{0.5cm}
\includegraphics[height=0.17\textheight]{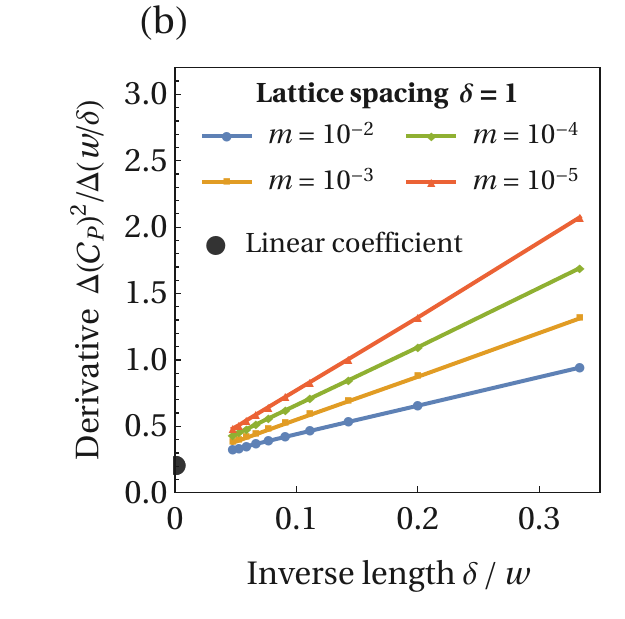}
\includegraphics[height=0.17\textheight]{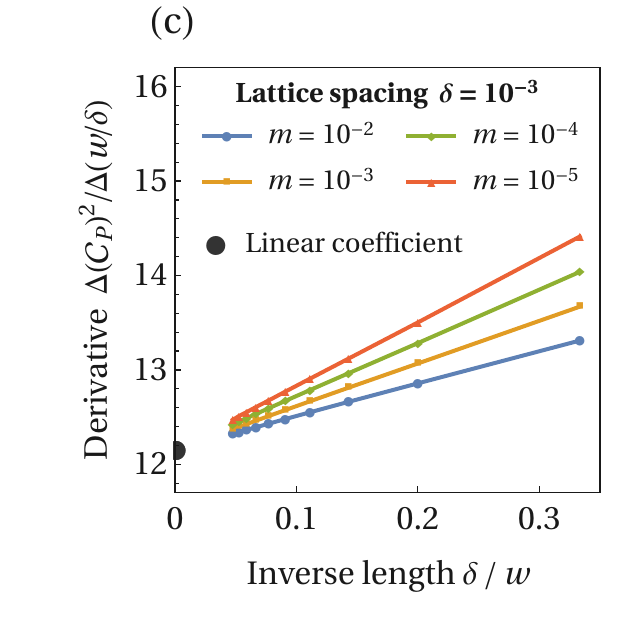}
\caption{Discrete derivative of fermionic (a) and bosonic squared CoP (b, c) of a single interval of length $\ell/\delta$.
$m$ and $\delta$ are given in units where $\mu=1$.
In both cases, the number of total sites is given by $N = 100\, w / \delta$.
}
\label{fig:CoPsingleint}
\end{figure}

We begin with the simpler case of fermionic CoP (at central charge $c=\tfrac{1}{2}$). Here, we find a relation of the form
\begin{align}
\lim_{N\to\infty}{\cal C_{P}}^2 = e_2 \frac{w}{\delta} + e_1 \log\frac{w}{\delta} + e_0 \ .
    \label{EQ_COP_SINGLE_INT_FIT}
\end{align}
Note that we consider the CoP squared. We test this functional form by computing the discrete derivative with respect to $w/\delta$, expected to be described by the expression $e_2 + e_1 \delta/w$. As figure~\ref{fig:CoPsingleint} (top) shows, it is indeed perfectly linear, allowing us to determine
\begin{align}
    e_0 &= 0.0894 \ , &
    e_1 &= 0.0544 \ , &
    e_2 &= 0.103
\end{align}
with the given three significant digits corresponding to the numerical accuracy of the optimization algorithm. 

\begin{figure}[tb]
\centering
\includegraphics[height=0.17\textheight]{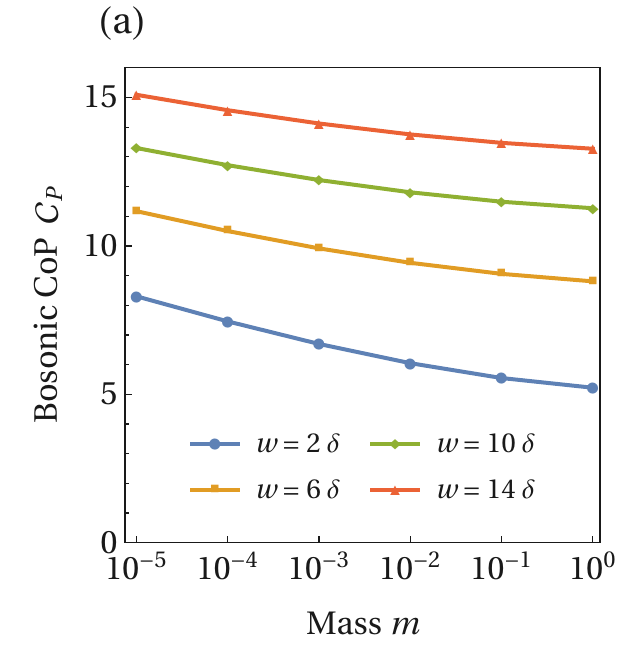}
\hspace{0.2cm}
\includegraphics[height=0.17\textheight]{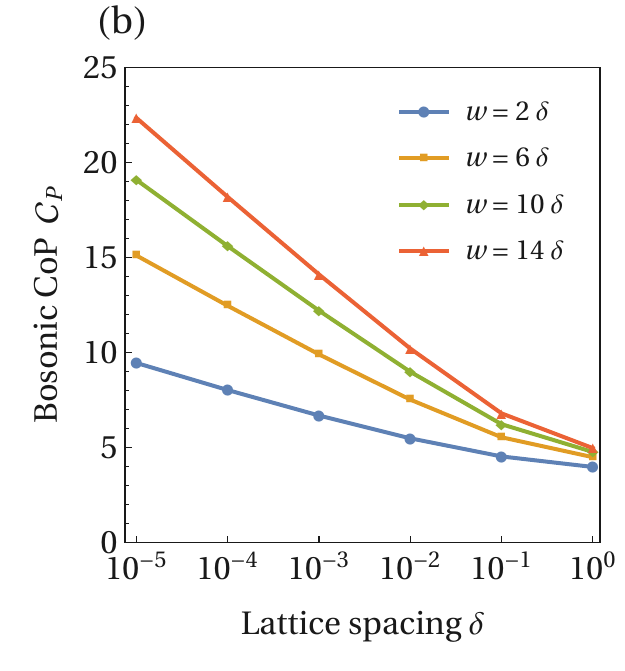}
\caption{Divergences of bosonic CoP of a single interval of size $w$ at fixed $\mu\, \delta = 10^{-3}$ and $\frac{m}{\mu} \to 0$ (a) and at fixed $\frac{m}{\mu} = 10^{-3}$ and $\mu\, \delta \to 0$ (b), in units of $\mu=1$. In both cases, the leading divergence is linear. While the $\mu\, \delta$ divergence is $w$-dependent, the $\frac{m}{\mu}$ one is not.
}
\label{fig:CoPbosDiv}
\end{figure}

\begin{figure}[tb]
\centering
\includegraphics[height=0.17\textheight]{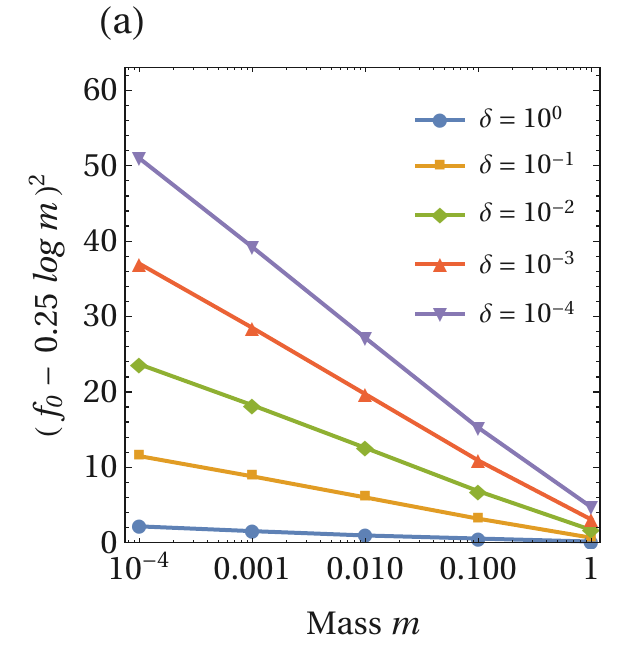}
\hspace{0.2cm}
\includegraphics[height=0.17\textheight]{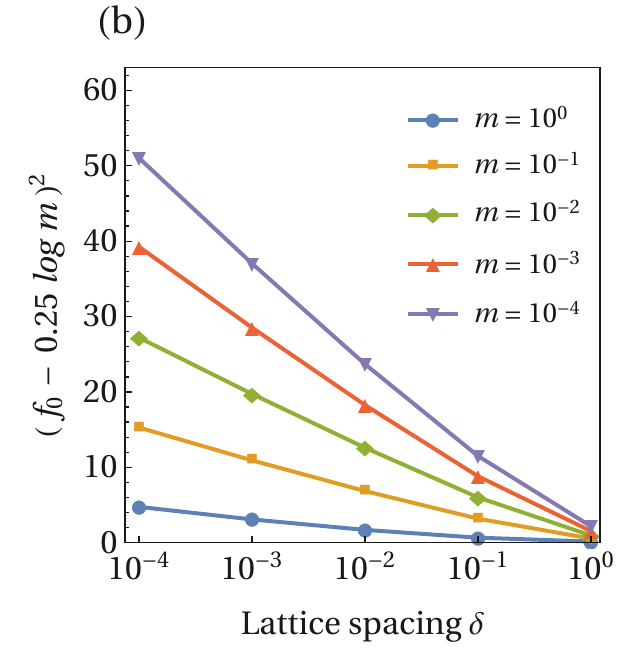}
\caption{Subleading contribution to constant term $f_0$ of bosonic CoP with respect to $m/\mu$ (a) and $\mu\, \delta$ (b), with $\mu=1$. After subtracting the leading contribution $0.25 \log^2\frac{m}{\mu}$ from $f_0$, the square of the remainder is linear in $\log\frac{m}{\mu}$ and $\log(\mu\, \delta)$ when both are small.  
}
\label{fig:CoPsingleconst}
\end{figure}

The bosonic case (with $c=1$) is more subtle, as we must subtract terms in $\tfrac{m}{\mu}$ and $\delta \mu$ that diverge in the continuum limit $\tfrac{m}{\mu}, \delta \mu \to 0$.
However, we still see in figure~\ref{fig:CoPsingleint} (bottom) that the functional form of \eqref{EQ_COP_SINGLE_INT_FIT} still holds, but with dependencies
\begin{align}
   \lim_{N\to\infty}{{\cal C}_{P}}^2= f_2(\mu\, \delta) \tfrac{w}{\delta} + f_1(\tfrac{m}{\mu}, \mu\, \delta) \log\tfrac{w}{\delta} + f_0(\tfrac{m}{\mu}, \mu\, \delta).
    \label{EQ_COP_SINGLE_INT_FIT2}
\end{align}
This form accurately describes the $w$ dependence over a large range of $m / \mu$ and $\mu \delta$.
The functions $f_0$ to $f_2$ are estimated as
\begin{subequations}
\label{eq.CoPf0f1f2}
\begin{eqnarray}
    f_0(\tfrac{m}{\mu}, \mu\, \delta) &=& 0.80 \sqrt{\log(\mu\, \delta) \log\frac{m}{\mu}} + 0.25 \log^2\tfrac{m}{\mu} , \quad \\
    f_1(\tfrac{m}{\mu}, \mu\, \delta) &=& 0.25 - 0.46 \log \frac{m}{\mu} - 0.17 \log(\mu\,\delta) , \\
    f_2(\mu\, \delta) &=& 0.22 + 0.25 \log^2\left( \mu\,\delta \right) ,
\end{eqnarray}
\end{subequations}
\normalsize
for parameters $m/\mu,\mu\,\delta \ll 1$. Because of the increased number of fit parameters relative to the fermionic case, we are unable to produce fit coefficients with more than two significant digits of accuracy.
The leading divergences in $\mu\,\delta$ and $\frac{m}{\mu}$ are visible in figure~\ref{fig:CoPbosDiv}, where we plot ${\cal C}_{P}$ (non-squared) and find linear divergences in $\log (\mu\,\delta)$ and $\log\frac{m}{\mu}$, respectively. The dependence of the slope of the divergence on $w$, given by $\approx\frac{\sqrt{w}}{2}$, is clearly visible in the $\mu\,\delta$ case, while the $\frac{m}{\mu}$ term diverges with a constant slope $\approx\frac{1}{2}$ at $\frac{m}{\mu} \ll 1$, consistent with the appearance of these terms in $f_2$ and $f_0$, respectively.
Note that $f_0$ and $f_1$ are estimated from the setup of two adjacent intervals, analyzed in the next section, where the linear term $f_2$ cancels. In particular, the square root term in $f_0$ can be seen in figure~\ref{fig:CoPsingleconst}, where the leading divergence in $\frac{m}{\mu}$ is subtracted.

To corroborate this discussion, let us also note that the structure of the leading divergences in the two cases matches the result of the vacuum complexity in free CFTs, see~\cite{Chapman:2017rqy,Jefferson:2017sdb} for bosons and~\cite{Khan:2018rzm,Hackl:2018ptj} for fermions. In this pure state case analogy, the role of $w$ is played by the total system size measured in lattice units. The presence of the $\log^2 \mu\,\delta$ contribution in the vacuum case comes from the ratio of the highest momentum frequency of the order of the inverse lattice spacing to the reference state scale and the overall coefficient in front of the whole divergence is $\mu$-independent. The logarithmic divergence is present because the symplectic group is non-compact. For fermions, the group of transformations is compact and there is no logarithmic enhancement of the leading divergence.

\subsection{Two adjacent intervals in the vacuum}
The next case to consider are two adjacent intervals in the vacuum. This is basically the application of the formulas from the previous section with the addition that it allows us to gain a better control over the finite term $f_0(\tfrac{m}{\mu}, \mu\, \delta)$ in the bosonic single interval CoP~\eqref{EQ_COP_SINGLE_INT_FIT2}. 

To this end, we are interested in a better behaved combination of complexities akin to~\eqref{eq:mutualCholo}. We take it to be
\begin{equation}
\label{eq:UVMichalReg}
\Delta {\cal C}_{P}^{(2)} \equiv {\cal C}_{P}(A)^2 + {\cal C}_{P}(B)^2 - {\cal C}_{P}(A\cup B)^2,
\end{equation}
where we put two on the LHS in the brackets to emphasize that it does not denote taking a square. The rational behind this expression is that when one keeps $\mu \, \delta$ fixed, the whole power-law divergent part cancels between the three terms. Similar combinations to~\eqref{eq:UVMichalReg} in the aforementioned context of pure state complexity in thermofield double states appear in~\cite{Chapman:2018hou}. Also, note the difference with respect to holographic mutual complexity~\eqref{eq:mutualCholo}.

\begin{figure}[tb]
\centering
\includegraphics[height=0.17\textheight]{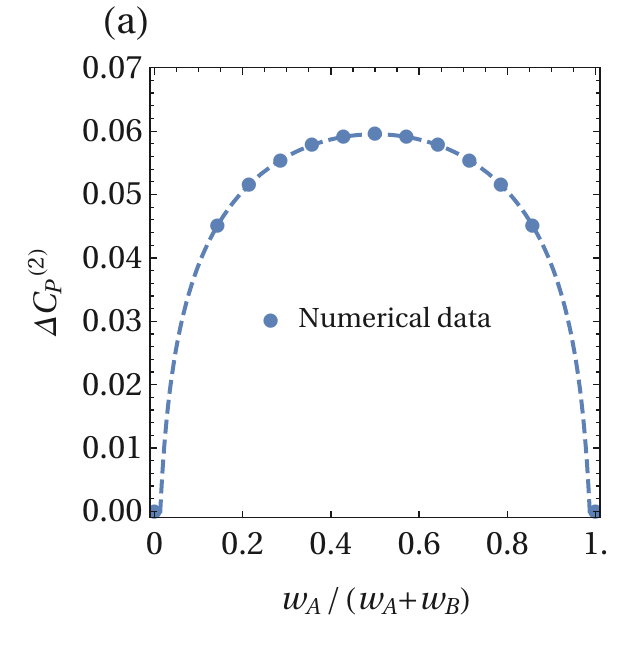}
\hspace{0.2cm}
\includegraphics[height=0.17\textheight]{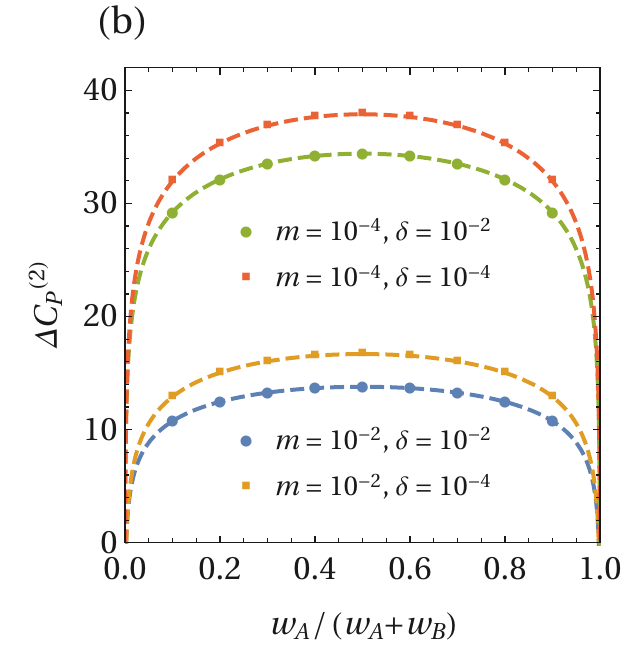}
\caption{Fermionic/Ising (a) and bosonic (b) CoP for two adjacent ($d=0$) subsystems $A$ and $B$, in units of $\mu=1$. The expected analytical forms, \eqref{eq.MCoP.F} for fermions and \eqref{eq.MCoP.B} for bosons, are plotted as dashed curves.
We consider $\frac{w_A + w_B}{\delta} = 14$ total sites for fermions and $20$ sites for bosons. The total system size is set to $N = 100 \frac{w}{\delta}$ for each block width $w$.}
\label{FIG_COP_ADJACENT_COP}
\end{figure}

Simple manipulations of the single interval case lead to the following result for the Ising CFT ($c = \frac{1}{2}$):
\begin{equation}
\label{eq.MCoP.F}
\lim_{N\to\infty}\Delta {\cal C}_{P}^{(2)} =  e_1 \, \log{\frac{w_{A}w_{B}}{(w_{A}+w_{B})\,\delta}} + e_0.
\end{equation}
For the free decompactified boson CFT ($c=1$), we can again use the single-interval result~\eqref{EQ_COP_SINGLE_INT_FIT2} to obtain
\begin{align}
\label{eq.MCoP.B}
\hspace{-2mm}\lim_{N\to\infty}\Delta {\cal C}_{P}^{(2)} = & f_1(\tfrac{m}{\mu},\mu\,\delta) \, \log{\tfrac{w_{A}w_{B}}{(w_{A}+w_{B})\,\delta}}+ f_0(\tfrac{m}{\mu},\mu\,\delta),
\end{align} 
but where the logarithmic coefficient and constant term now depend on the divergences in $\frac{m}{\mu}$ and $\mu \delta$ in precisely the same way as the single-interval expression.
The form of $\Delta {\cal C}_{P}^{(2)}$ for both bosons and fermions in shown in figure~\ref{FIG_COP_ADJACENT_COP} in terms of the ratio $\frac{w_A}{w_A + w_B}$. The qualitative behavior is the same as for the EoP result shown in figure~\ref{FIG_EOP_A1A2t} and also the holographic complexity proposal results encapsulated in table~\ref{TAB_ANALYTIC_RES_COMP}. However, one should bear in mind a different subtraction of complexity with the latter case related to the use of a $L^{2}$ norm in our Gaussian studies. Note also that $\Delta {\cal C}_{P}^{(2)}$ is logarithmically ultraviolet divergent.

Finally, it is interesting to observe that the mutual complexity depicted in figure~\ref{FIG_COP_ADJACENT_COP} is positive, which indicates subadditivity of our CoP definition. This is in line with an earlier observation in the case of two coupled harmonic oscillations in~\cite{Camargo:2018eof}.

\subsection{Single mode optimization for bosons}\label{sec:singlemode}
A natural hope in the context of Gaussian states is to split the problem of finding the CoP for a system with $N_A$ modes into $N_A$ problems for a single mode. In~\cite{Caceres:2019pgf}, the authors use a formula for the $L^1$ norm complexity of a single bosonic degree of freedom for certain Gaussian states derived in~\cite{Jefferson:2017sdb}, where they introduce two types of $L^1$ norm bases (called the physical and the diagonal one). Note that the authors use the geodesic with respect to the $L^2$ norm, but compute its length with respect to the $L^1$, as it is difficult to prove that a path is minimal with respect to the $L^1$ norm (in particular, for several modes). When considering several modes of the free Klein-Gordon field, we need to distinguish two settings reviewed previously:
\begin{itemize}
    \item \textbf{Thermal states.} Here, we can choose a basis $\hat{\xi}_A$, such that both $[J_{\mathrm{T}}]_A$ of a mixed thermal Klein-Gordon state and $[J_{\mathrm{R}}]_A$ of a spatially unentangled vacuum decompose into $2$-by-$2$ blocks, so it is easy to argue that the Gaussian CoP results from optimizing over individual modes.
    \item \textbf{Subsystems.} If we consider a mixed Gaussian state $\rho_A$ resulting from restricting the Klein-Gordon vacuum to a region, it is typically not possible to bring both $[J_{\mathrm{T}}]_A$ and $[J_{\mathrm{R}}]_A$ of a spatially unentangled vacuum into block diagonal form, so it will not suffice to optimize over individual modes.
\end{itemize}
Let us emphasize that in both cases, we have a mode-by-mode purification with respect to the standard form~\eqref{eq:JStandard}, but it is the reference state that will only be a tensor product over \emph{these} modes in the case of thermal states, but not for local subsystems. However, we may encounter situations where the standard decomposition of the mixed target state also approximately decomposes the reference state into individual modes.

We consider a single bosonic mode with a pure Gaussian reference state and a mixed Gaussian target state that both do not have $\varphi\pi$-correlations, \ie $\braket{\hat{\varphi}_i\hat{\pi}_j}=0$. We extend this system to a system of two bosonic modes $\mathcal{H}'=\mathcal{H}_A\otimes\mathcal{H}_{A'}$ to have the extended reference state $\ket{J_{\mathrm{R}}}$ and the purified target state $\ket{J_{\mathrm{T}}}$, such that the respective complex structures are given by
\begin{align}
    J_{\mathrm{T}}&\equiv\left(\begin{array}{cc|cc}
    0 & \lambda  & 0 & \sqrt{\lambda^2-1} \\
    -\lambda  & 0 & \sqrt{\lambda^2-1} &0 \\
    \hline
    \sqrt{\lambda^2-1} & 0 & 0 & \lambda\\
    0 & \sqrt{\lambda^2-1} & -\lambda & 0
    \end{array}\right)\\
    J_{\mathrm{R}}&\equiv\left(\begin{array}{cc|cc}
    0 & \mu  & 0 & 0 \\
    -\tfrac{1}{\mu}  & 0 & 0 &0 \\
    \hline
    0 & 0 & 0 & \nu\\
    0 & 0 & -\tfrac{1}{\nu} & 0
    \end{array}\right)\,,
\end{align}
where $\lambda\in[1,\infty)$ is the same as $c_i$ for several degrees of freedom in~\eqref{eq:restrictedJstandard}, $\mu$ is the reference state frequency for the original single mode, and $\nu$ is a parameter in the reference state, for which we will minimize the complexity functional~\eqref{eq:CompFunct}. The latter is given for us by
\begin{equation} \label{eq:CompFunctOne Mode}
\mathcal{C}(\lambda,\mu,\nu)= \frac{1}{2}\sqrt{\log\left(\frac{\omega_{+}}{\mu}\right)^{2}+\log\left(\frac{\omega_{-}}{\mu}\right)^{2}}\,,
\end{equation}
where we defined the variables
\begin{equation}
    \omega_{\pm}:=\frac{\mu}{2}\left(\lambda(\mu+\nu)\pm \sqrt{\lambda^{2}(\mu+\nu)^{2}-4\mu\nu}\right)\,.
\end{equation}
In order to find the minimum of $\mathcal{C}_P=\min_{\nu}\mathcal{C}(\lambda,\mu,\nu)$, we need to solve a transcendental equation for $\nu$. This can be done numerically for any given value of $c$ and $\mu$ in a very efficient manner. Unfortunately there is no closed analytic expression for $C_P$ of a single mode, but we reduced the problem of a single mode (with vanishing $\varphi\pi$-correlations in reference and target states) to a problem that is much simpler than the optimization over a large manifold. Can we extend this to larger systems?

As derived in~\cite{Hackl:2018ptj,Chapman:2017rqy} for the perspective of the $L^2$ norm, the optimal geodesic from reference to target state gives~\eqref{eq:CompFunct}, so we only need to worry about what goes into this formula. We consider a mixed state of a region $A$ with $N_A$ sites. With respect to the local basis
\begin{align}
    \hat{\xi}_A=(\hat\varphi_1^A,\dots,\hat\varphi_{N_A}^A,\hat\pi_1^A,\dots,\hat\pi_{N_A}^A)\,,
\end{align}
the covariance matrix of reference and target state will have the following forms
\begin{align}
    [G_{\mathrm{T}}]_A\equiv\left(\begin{array}{c|c}
    G_{\varphi\varphi}     &  0\\
    \hline
    0     & G_{\pi\pi}
    \end{array}\right)\,,\quad [G_{\mathrm{R}}]_A\equiv\left(\begin{array}{c|c}
    \mu \id     &  0\\
    \hline
    0     & \frac{1}{\mu}\id
    \end{array}\right)\,.
\end{align}
If the Gaussian target state were pure, it is well-known that we can find symplectic transformation
\begin{align}
    M=\left(\begin{array}{c|c}
    O     &  0\\
    \hline
    0     & O
    \end{array}\right)\,,\label{eq:O}
\end{align}
where $O$ is an orthogonal matrix, such that $\tilde{G}_{\mathrm{T}}=MG_{\mathrm{T}}M^\intercal$ is diagonal, while $G_{\mathrm{R}}=MG_{\mathrm{R}}M^\intercal$ is preserved. As soon as $G_{\mathrm{T}}$ represents a mixed Gaussian state, there still exists a symplectic transformation $M$, such that $\tilde{G}_{\mathrm{T}}=MG_{\mathrm{T}}M^\intercal$ is diagonal, but now $M$ will not be of the form~\eqref{eq:O} anymore and will thus not preserve $G_{\mathrm{R}}$, \ie $MG_{\mathrm{R}}M^\intercal\neq G_{\mathrm{R}}$. However, we could pretend to approximate the true $M$ by only diagonalizing $G_{\varphi\varphi}$ with a respective $O$, such that $\tilde{G}_{\varphi\varphi}=OG_{\varphi\varphi}O^\intercal$ is diagonal. We then apply the respective $M$ of the form~\eqref{eq:O} and pretend that also $\tilde{G}_{\pi\pi}=OG_{\pi\pi}O^\intercal$ is diagonal, \ie we drop the off-diagonal terms which we hope to be sufficiently small. With this assumption, we can apply a mode-by-mode optimization based on~\eqref{eq:CompFunctOne Mode}, such that
\begin{align}
    \mathcal{C}_P(\ket{J_\mathrm{T}},\ket{J_{\mathrm{R}}})\approx\sqrt{\sum_i\min_\nu\mathcal{C}(\lambda_i,\mu,\nu)}\,,\label{eq:CoP-mode-by-mode}
\end{align}
where $\lambda_i$ is extracted from the diagonal entries of $\tilde{G}_{\varphi\varphi}$ and $\tilde{G}_{\pi\pi}$. For a pure target state, \eqref{eq:CoP-mode-by-mode} becomes an equality, where both sides match the regular complexity~\eqref{eq:CompFunct}.

\begin{figure}[tb]
\centering
\includegraphics[height=0.18\textheight]{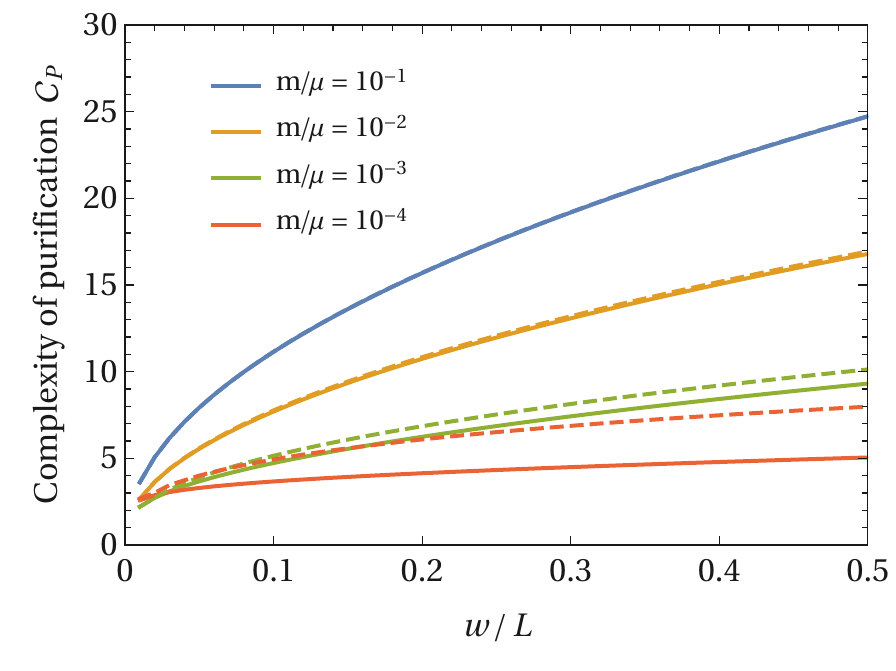}
  \caption{Comparison of bosonic CoP obtained using the Gaussian optimization algorithm (solid) and the Single Mode Optimization approximation (dashed) for the single interval case for mass $m/\mu=10^{-1},10^{-2},10^{-3},10^{-4}$ (top to bottom), $N=100$, and for lattice spacing $\mu\delta=10^{-4}(m/\mu)^{-1}$.}
  \label{fig:ComparisonSingleMode}
\end{figure}

We consider the restriction of the Klein-Gordon vacuum to the subsystem of a single interval as explored in section\  \ref{sec:single-interval}. In this setup, we can compare the approximate single mode optimization with the full optimization. While the full optimization takes several hours on a regular desktop computer, our approximate scheme of optimizing~\eqref{eq:CompFunctOne Mode} over individual modes only takes a few seconds. Figure~\ref{fig:ComparisonSingleMode} shows how the single mode optimization is almost indistinguishable from the full optimization for large $m/\mu$, but the approximation becomes increasingly worse for smaller $m/\mu$. In~\cite{Caceres:2019pgf}, the authors perform a similar calculation for the $L^1$ norm, which they refer to as mode-by-mode purifications\footnote{As pointed out at the beginning of this section, any Gaussian purification is a mode-by-mode purifications, but what the authors of~\cite{Caceres:2019pgf} mean is that they only optimize over the purifications in a specific way, as if reference and target state would decompose into the same individual modes as an approximation.}. The difference between our single mode optimization and what the authors in~\cite{Caceres:2019pgf} do is two-fold: First, we optimize over the $L^2$ norm, while they consider the $L^1$ norm. Second, we change the target state by hand to decompose into a product over modes in the same basis as the reference state and then perform the optimization semi-analytically for individual modes, \ie we optimize for each mode independently. In contrast, the authors~\cite{Caceres:2019pgf} do not change reference or target state, but only consider a subset of possible purifications, \ie they evaluate the full complexity function and optimize over a restricted subset of parameters (one parameter per mode). As they do not change the target state, they cannot evaluate the complexity for individual modes, so would need in principle to optimize over all parameters simultaneously, but find good convergence when optimizing over $\mathcal{O}(1)$ parameters at once. Clearly, the approximation in~\cite{Caceres:2019pgf} and our single mode optimization work because for large $m /\mu$ reference and target states are close to decomposable over individual modes. This will not be the case for generic subsystems (such as two intervals), general states (such as those with $\varphi\pi$-correlations) and fermionic systems (which cannot be decomposed into single mode squeezings), in which case our full optimization algorithm is required.

\subsection{Comparison with the Fisher-Rao distance proposal\label{sec.Fisher-Rao}}
In~\cite{DiGiulio:2020hlz}, the authors propose a measure of bosonic mixed state complexity based on the Fisher-Rao distance, which can be defined on the manifold $\mathbb{P}(N)$ of $2N\times2N$ real and positive definite matrices, of which bosonic covariance matrices are a subset of. Without the need for any purifications, the proposal for the complexity of mixed states is formally equivalent to~\eqref{eq:CompFunct}, where here $G$ and $G_{0}$ are taken to be the covariance matrices of the mixed target and reference state, respectively. The motivation for this definition is that the Fisher-Rao distance function 
\begin{equation}
 d(G,G_{0}):=\sqrt{\tr\left(\log(G_{0}G^{-1})^{2}\right)}/2\sqrt{2}   
\end{equation}
measures the geodesic distance in the manifold of covariance matrices. It is important to highlight that the authors in ~\cite{DiGiulio:2020hlz} focus on bosonic Gaussian states occurring in the Hilbert space of harmonic
lattices, and hence the proposal of the Fisher-Rao distance function should be thought of as applicable in principle only to bosonic states, although one could conjecture that a simliar formula should be applicable to fermionic Gaussian states. Nonetheless, it is interesting to compare the properties of such distance function with the bosonic CoP measure arising from the Gaussian optimization procedure developed in this paper. 

\begin{figure}[tb]
\centering
\includegraphics[height=0.18\textheight]{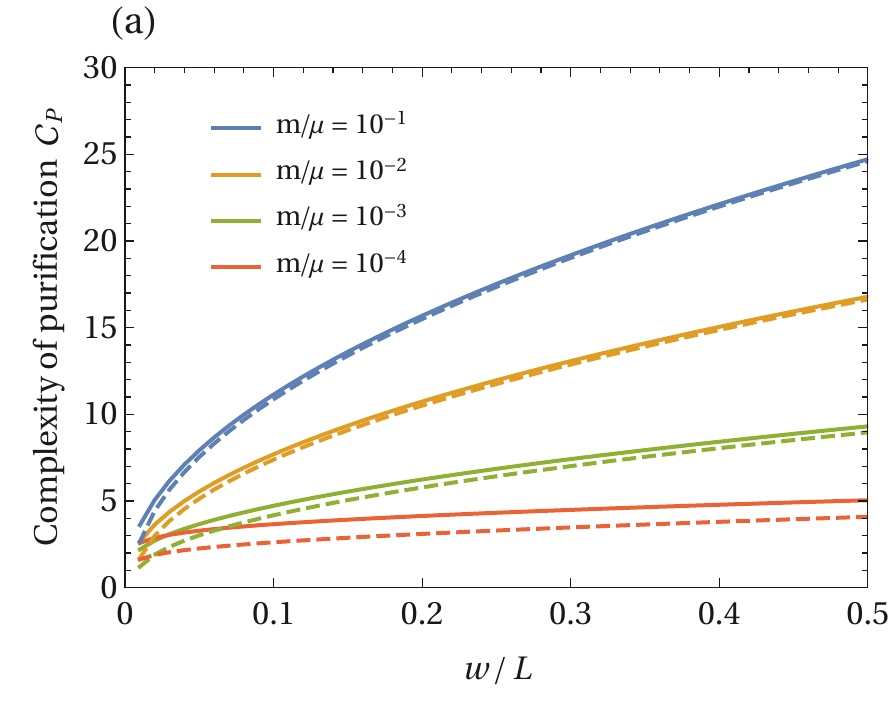} \\
\vspace{0.15cm}
\includegraphics[height=0.18\textheight]{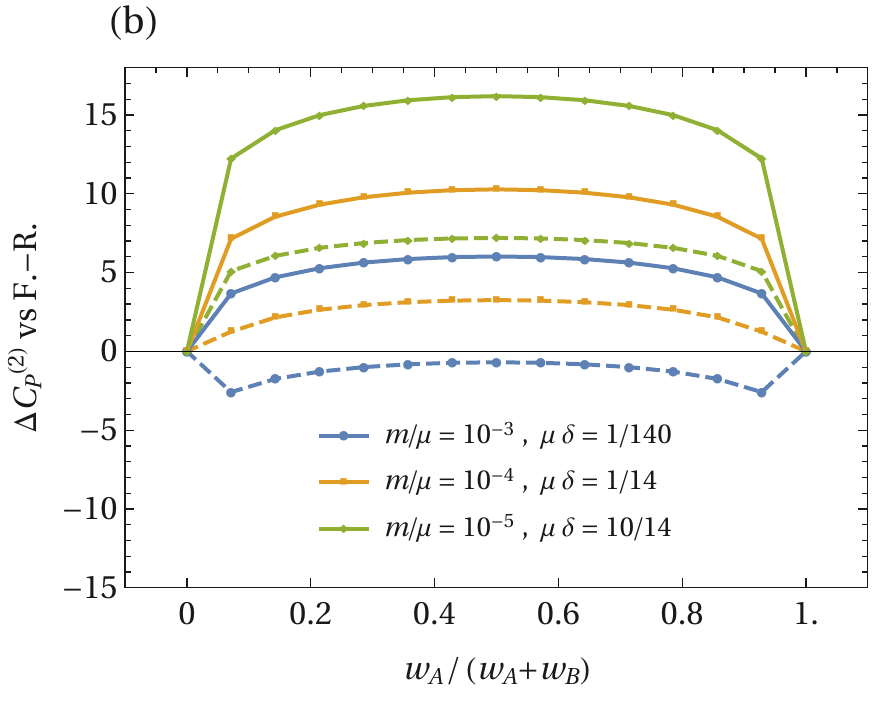}
\caption{Comparison of CoP for a single interval (a) and two adjacent intervals (b) obtained using the Gaussian optimization algorithm (solid) and the Fisher-Rao distance function (dashed). The data for the single interval case (a) are generated for mass $m/\mu=10^{-1},10^{-2},10^{-3},10^{-4}$ (top to bottom), $N=100$, and for lattice spacing $\mu\delta=10^{-4}(m/\mu)^{-1}$, while the data for the adjacent interval case (b) are generated for mass $m/\mu=10^{-3},10^{-4},10^{-5}$, $N=1400$, $(w_{A}+w_{B})=14$, and for lattice spacing $\mu\delta=(1/14)\times10^{-4}(m/\mu)^{-1}$.}
\label{ComparisonErik2}
\end{figure}

For the single interval case, there is in fact a noteworthy qualitative and quantitative agreement between the two, as shown at the top of figure~\ref{ComparisonErik2}. The Fisher-Rao distance function and the CoP measure offer a comparable measure of the complexity of the mixed state associated to the single interval, which is remarkable given the fact that the Fisher-Rao distance function is a geodesic distance being evaluated on the manifold of mixed states on the Hilbert space $\mathcal{H}_{A}$, whereas CoP is a geodesic distance on the manifold of pure states on the larger Hilbert space $\mathcal{H}=\mathcal{H}_{A}\otimes\mathcal{H}_{A'}$ and these two need not be the same or even comparable to one another, as explained on figure~\ref{fig:CoP-sketch}. This comparison seems to work better the larger the mass $m/\mu$, much like in the case of single mode optimization. Our studies also indicate that for two adjacent intervals the distinction between the Fisher-Rao distance and CoP deviate significantly from each other, even though the qualitative behaviour remains comparable, as shown at the bottom of figure~\ref{ComparisonErik2}.



\section{Comments}\label{sec.comment}
In our considerations, two important subtleties of free QFTs, known to the literature, played a key role and influenced the vacuum subregions we could consider to make genuine QFT predictions. They are the zero mode in the case of free boson QFTs and spatial locality of disjoint intervals under the Jordan-Wigner mapping between the Ising model and the free Majorana fermion theory. Below we provide an additional discussion of these two important points.

\subsection{Zero mode for free bosons}\label{sec:zero-mode}
The presence of the zero mode is a known subtlety of the free boson theory in the massless limit, as we discussed in section~\ref{sec:models.KG}. The simplest way to deal with it is to keep the mass term in the Hamiltonian~\eqref{eq:H-KG} nonzero and try to numerically approach the limit $m \rightarrow 0$, which is precisely the strategy we adopted in the calculations of the bosonic EoP and CoP. Given the scarceness of other methods to shed light on EoP and CoP in QFTs, it is important to learn about the role of zero mode in better-understood problems.

To some degree we already explored this issue in section~\ref{sec:models.KG} when understanding what needs to be done in order to reproduce the modular invariant thermal partition function~\eqref{eq.Zmodinv} from the Gaussian calculation in the massive theory~\eqref{eq.Zmassive}. Here we will address another quantity, which is the two interval vacuum MI reviewed in section~\ref{sec:mutualinformation}. 
Fitting a power law for the data on a periodic chain of bosons does not lead to convergence of the estimated power to a nonzero value in the $d/w \to \infty$ limit (see table~\ref{tab:EoP-MI}), implying a decay with slower asymptotic functional dependency.
Indeed, the massless limit of our study can be thought of as a decompactification limit of the free boson theory on a circle in the field space (which can be seen as another way of dealing with the zero mode), as reviewed in section~\ref{sec:models.KG}, and the predicted large distance behaviour in this case is not of a power-law type. 
In our periodic setup we considered the limit of a large number of sites $N$ with the dimensionless scale $m L = m N \delta$ kept constant and small. In this limit, the mass dependence of both MI and EoP is accurately described by an additive $-\frac{1}{2}\log(L m)$ term (as already noted in~\cite{Bhattacharyya:2019tsi}), so that the dependence on $w/d$ can be studied independently of the mass.

\begin{figure}[tb]
\centering
\includegraphics[height=0.18\textheight]{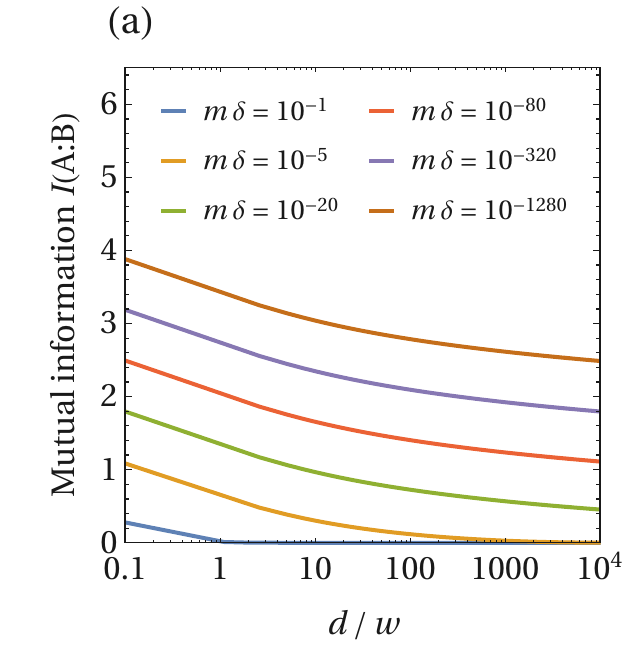}
\hspace{0.15cm}
\includegraphics[height=0.18\textheight]{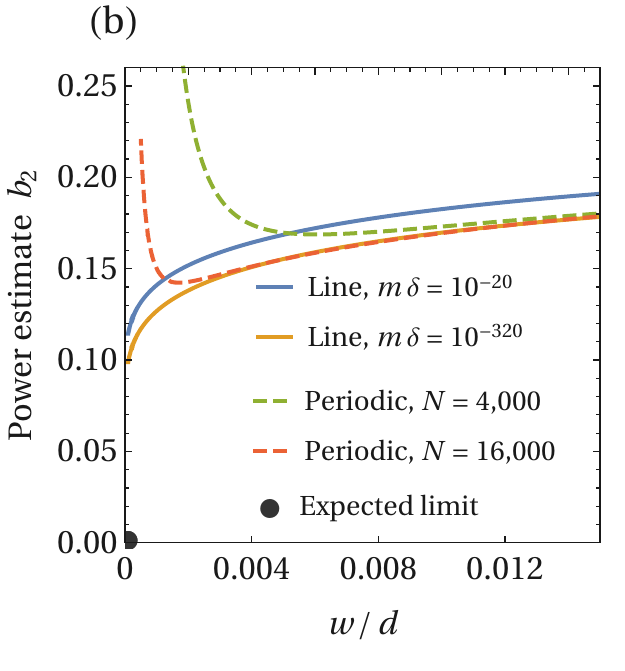}
\caption{MI on an infinite line in the small $m\, \delta$ limit. Shown are the general mass dependence (a) as well as the decay power (b) in the $d/w \to \infty$ limit compared to the periodic system in the large $N$ limit (right). 
}
\label{FIG_BOS_MI_LINE}
\end{figure}

\begin{figure}[tb]
\centering
\includegraphics[height=0.18\textheight]{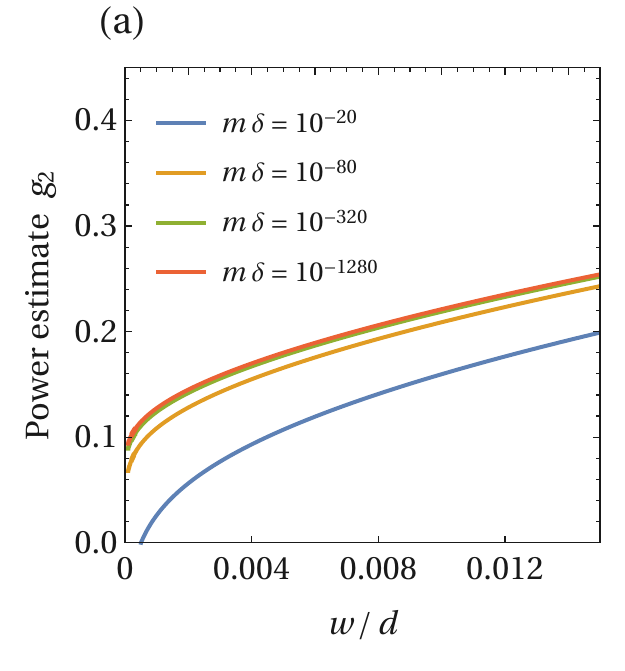}
\hspace{0.15cm}
\includegraphics[height=0.18\textheight]{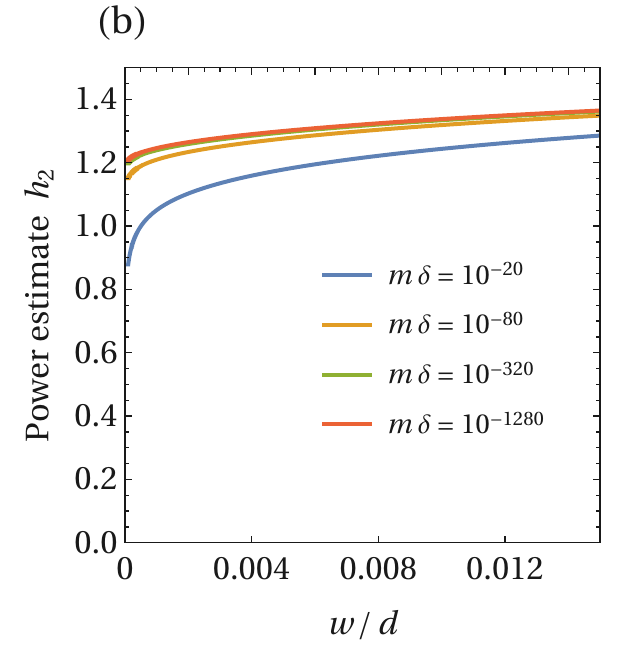}
\caption{Power-logarithmic (a) and power-double-logarithmic (b) fits of bosonic MI on an infinite line at large $d/w$ and small effective mass $m \delta$. Block width set to $w/\delta=10$. Plotted are the estimated coefficients $g_2$ and $h_2$ from~\eqref{EQ_BOS_LINE_POW_LOG} and \eqref{EQ_BOS_LINE_DBL_LOG}.
}
\label{FIG_BOS_POWERLOG}
\end{figure}

However, one may alternatively consider free bosons on an infinite line, \ie taking the limit $m\, \delta \to 0$ only after the limit $N \to \infty$. As the scale $m\,\delta\, N$ formally diverges in the first limit, the resulting mass dependence is qualitatively different from the periodic setup. To investigate the behavior in this case, we adapted our numerical method to free bosons on an infinite line and computed MI at extended precision at small $m \, \delta$. Comparable studies were performed earlier in~\cite{Marcovitch:2008sxc} and the authors reported a power-law fall-off of the form $\left(\frac{w}{d}\right)^{0.05}$, see also~\eqref{eq.IABlarged}. 
The value of $I(A:B)$ on the line changes only very slowly with $m \delta$, as shown in figure~\ref{FIG_BOS_MI_LINE}~(left). For $d/w \ll m \delta$, this mass dependence can be expressed by a constant offset
\begin{equation}
    I(A:B) = f_\text{MI} \left(\frac{d}{w}\right) + \frac{1}{2} \log\left( \log\frac{1}{m \delta} \right) \ ,
\end{equation}
where the factor of $1/2$ is reproduced with a numerical accuracy of four significant digits. This double-logarithmic infrared divergence matches previous results for single-interval entanglement entropies in a similar setup \cite{Casini:2005zv}.
Note that the positivity of $I(A:B)$ prevents this dependence persisting at finite $m\, \delta$, and thus $I(A:B)$ begins to decay exponentially as $d/w \gg m\, \delta$.
Apart from the different mass dependence, the periodic and line setup in their respective limits yield equivalent results. As we show in figure~\ref{FIG_BOS_MI_LINE} (right), the estimated decay power from both limits exactly matches, vanishing as $d/w \to \infty$.

As the line setup is more efficient at probing large values of $d/w$ due to an absence of finite-size effects, we use this setup to test for functional dependencies slower than the previously considered power law and logarithmic functions. As the functional dependence of MI and EoP match in this limit, we expect the results to extend to both measures.
In particular, we consider the power-logarithmic asymptotics
\begin{equation}
    f_\text{MI}(\tfrac{d}{w})\sim g_0 - g_1 \left( \log\frac{d}{w} \right)^{g_2}\quad\text{as}\quad \frac{d}{w}\to\infty \ ,
    \label{EQ_BOS_LINE_POW_LOG}
\end{equation}
as well as a power-double-logarithmic one,
\begin{equation}
    f_\text{MI}(\tfrac{d}{w}) \sim h_0 - h_1 \left( \log \log\frac{d}{w} \right)^{h_2} \quad\text{as}\quad \frac{d}{w}\to\infty\ .
    \label{EQ_BOS_LINE_DBL_LOG}
\end{equation}
In figure~\ref{FIG_BOS_POWERLOG} coefficients from both fits are shown. While the power $e_2$ of the power-logarithmic fit converges to a value $e_2 \lesssim 0.1$ that may be zero in the $d/w \to \infty, m \delta \to 0$ limit, the double-logarithmic power $f_2$ clearly converges to a value $f_2 \lesssim 1.3$ that is visibly larger than zero.
Surprisingly, this clear observation of sub-polynomial functional dependence both in the infinite line and periodic setups contradicts the aforementioned earlier numerical observations of a power law in~\cite{Marcovitch:2008sxc}. These earlier results may be the result of numerical estimates performed only at moderately large $d/w$; while the functional dependence in this range can be well approximated by a power law, as also shown in~\cite{Bhattacharyya:2019tsi}, the apparent power law is not stable as $d/w \to \infty$.
Curiously, the authors of~\cite{Marcovitch:2008sxc} analyze in the same setup another measure of entanglement -- the logarithmic negativity~\cite{Vidal:2002zz} -- and our extended precision calculations in this case reproduce the exponential fall-off as reported in~\cite{Marcovitch:2008sxc} for the same range of masses considered in figure~\ref{FIG_BOS_POWERLOG}. This shows that not all non-local quantities are affected by the zero mode problem.

An additional insight about the expected behaviour of MI in the case of the free boson CFT in the decompactification limit $R \rightarrow \infty$ can be obtained from having another look at the modular invariant partition function on the circle~\eqref{eq.Zmodinv}. The partition function provides the information about the density of states, which, via the state-operator correspondence, describes also the density of operators in the spectrum. The latter quantity, in conjunction with~\eqref{CCDd}, will provide an indication on what to expect from the two interval case in the large separation limit for the decompactified free boson CFT. To this end, the partition function of a CFT with a continuous spectrum of operators (for a decompactified free boson theory vertex operators are labelled with a continuous index) can be written as
\begin{equation}
Z \sim \int_{0}^{\infty} d\Delta\, \rho(\Delta) \, e^{-2 \pi (\beta / L) \, \Delta} \, e^{\frac{\pi}{6} (\beta / L)}, 
\end{equation}
where the second exponent comes from the Casimir energy and $\Delta$ is the scaling dimension of operators in the theory. Note that descendent operators appear in the sum only for $\Delta > 1$. The power law multiplying the Casimir contribution in the low temperature limit of~\eqref{eq.Zmodinv} points to the density of operators behaving in a power law fashion in the vicinity of $\Delta = 0$. In particular, the behavior
\begin{equation}
Z \Big|_{\beta/L \gg 1} \sim (\beta/L)^{-\alpha} \, e^{\frac{\pi}{6} (\beta / L)}
\end{equation}
can be explained by
\begin{equation}
\label{eq.rholight}
\rho(\Delta \approx 0) \sim \Delta^{\alpha-1}.
\end{equation}
Note that for a free decompactified boson
\begin{equation}
\label{eq.alpha05}
\alpha = \frac{1}{2}.    
\end{equation}
In this case, the density of states can be easily understood to be given by $\Delta^{-1/2}$ upon noting that the operators of interest are vertex operators $:e^{i \, \nu \, \phi}:$ specified by a real number $\nu$. The scaling dimension of vertex operators is given by $\Delta \sim \nu^2$. The density of operators is uniform when parametrized by $\nu$ and viewing it as a function of~$\Delta$ brings in the Jacobian $\sim \Delta^{-1/2}$, which gives~\eqref{eq.alpha05}.

Now we can come back to MI at large separation. Since the formula~\eqref{CCDd} incorporates the exchange of a single operator, in the absence of a gap in the spectrum one needs to sum over the continuum of light operators with their density given by~\eqref{eq.rholight}. Following~\cite{Ugajin:2016opf} we can schematically write
\begin{equation}
I(A:B)\Big|_{w \ll d} \sim \int_{0}^{\infty} d\Delta \, \Delta^{\alpha -1} \left( \frac{w}{d} \right)^{4 \, \Delta} \left(c_{{\cal T}{\cal T} O_{\Delta}}\right)^2 + \ldots,
\end{equation}
\normalsize
where $c_{{\cal T}{\cal T} O_{\Delta}}$ is the three-point function coefficient between two twist fields and a primary with dimension~$\Delta$ and the ellipsis denote contributions with higher powers of $\left( \frac{w}{d} \right)^{4 \, \Delta}$. At the present moment we do not have control over the two kinds of contributions. However, neglecting the additional contributions and assuming that $c_{{\cal T}{\cal T} O_{\Delta}}$ has a power-law dependence on $\Delta$ for small scaling dimensions
\begin{equation}
c_{{\cal T}{\cal T} O_{\Delta}} \stackrel{?}{\sim} \Delta^{\kappa},
\end{equation}
the long distance behaviour of MI becomes
\begin{equation}
\label{eq.MIlogscaling}
I(A:B)\Big|_{w \ll d} \stackrel{?}{\sim} \left(\log{\frac{d}{w}}\right)^{-\alpha - 2 \, \kappa}.
\end{equation}
Let us re-stress that the above equation is based on unverified assumptions and the correct answer is likely to be more involved, yet in principle calculable.
However, what~\eqref{eq.MIlogscaling} indicates is that MI may decay much more slowly with distance than a simple power law, as is also shown by our numerical results. In particular, the fitting ansatz corresponding to~\eqref{eq.MIlogscaling} is~\eqref{EQ_BOS_LINE_POW_LOG} consistent with $-\alpha - 2 \, \kappa \lesssim 0.1$. 

Finally, note also that while in the massive theory, the mass scale eventually triggers an exponential decay of MI to zero, the ansatzes~\eqref{EQ_BOS_LINE_POW_LOG} and~\eqref{EQ_BOS_LINE_DBL_LOG} predict divergence when naively extrapolated to $\frac{d}{w} \rightarrow \infty$. It is unclear at the moment if this is a feature of regularizing the zero mode via introducing a non-vanishing mass, or if the behaviour~\eqref{EQ_BOS_LINE_POW_LOG} or~\eqref{EQ_BOS_LINE_DBL_LOG} that we see using our Gaussian numerics persists in the free boson CFT in the decompactification limit.

\subsection{Subsystems in Ising CFT vs. free fermions}\label{sec:fermion-subsystem}
There is a subtle difference for computations of MI and EoP between the XY spin model and the Majorana fermion model. The spin model in the continuum limit becomes the genuine $c=1/2$ Ising CFT, which has a modular invariant partition function. However, the fermion model can only be identified with this CFT in the continuum limit if we impose the correct GSO projections, summing over the sectors of periodic and anti-periodic boundary conditions (also known as the R and NS sector, respectively). This projection, originating from anti-commutative nature of fermions as opposed to spins, maps entanglement entropy and related quantities like MI and EoP for standard choices of subsystems in the spin model to unconventional ones in the fermion theory. This difference due to the projections does not occur in the calculations of MI and EoP when the subsystem $A$ and $B$ are adjacent. However, when $A$ and $B$ are separated, the non-trivial topology in the replica computation of entanglement entropy leads to a difference between the spin and fermion calculation as the simple choice of subsystems does not respect the projections. This is consistent with our numerical results for Majorana fermions and adjacent intervals shown in figure~\ref{FIG_EOP_A1A2t}. We also note that MI and EoP are smaller in the fermion model calculations without projects compared with those in spin model calculations. Refer to appendix~\ref{app:spin-vs-fermion} for more details.

\section{Discussion}\label{sec.discussion}
In this manuscript, we have a presented a systematic and comprehensive analysis of EoP and CoP that characterize mixed states.
We computed the EoP between two blocks of widths $w_A$ and $w_B$ at distance $d$ in one-dimensional periodic systems at large size for both critical bosons and fermions, the latter of which are equivalent to a discretized Ising CFT while $d=0$.
Furthermore, we compared these results with the well-studied MI. At $d=0$, our data shows
\begin{subequations}
\label{eq.IEoPsummary}
\begin{eqnarray}
I(A:B) &=& \frac{c}{3} \log \frac{w_A w_B}{(w_A + w_B)\delta}  , \\
E_P &=& \frac{c}{6} \log \frac{2 w_A w_B}{(w_A + w_B)\delta} \ ,
\end{eqnarray}
\end{subequations}
confirming previous expectations through analytical methods \cite{Holzhey:1994we,Caputa:2018xuf}.
For $d>0$, we considered the symmetric setup $w_A=w_B=w$ in the two limits $d/w \ll 1$ and $d/w \gg 1$. In the former limit, our data is consistent with 
\begin{subequations}
\begin{eqnarray}
I(A:B) &\propto& \frac{1}{3} \log \frac{w}{d} \ ,\\
E_P &\propto& \frac{1}{6} \log \frac{w}{d} \ ,
\end{eqnarray}
\end{subequations}
for our bosonic model.
The latter limit shows a sub-polynomial (logarithmic or double-logarithmic) decay of both $I(A:B)$ and $E_P$ at large $d/w$.
In summary, MI and EoP show the same scaling at large distance $d$.
This result is consistent with the observation in \cite{Bhattacharyya:2019tsi} that EoP appears to weight quantum and classical correlations differently from MI, leading to different qualitative behavior only when both become relevant, \ie at small distances.
Indeed, this is the regime in which our numerical results show new model-dependent features that distinguish both measures.
For both the periodic and infinite line setup, two-interval MI and EoP are divergent in the zero-mass limit; this divergence can be regulated by a $\tfrac{1}{2} \log(m \delta N)$ and $\tfrac{1}{2}\log\log(m \delta)$ term, respectively. Let us also emphasize that the large distance behaviour of MI in the free boson case at small masses is very subtle and is described by the fall-off slower than any power-law, contrary to earlier studies in the literature. We discussed this at length in section~\ref{sec:zero-mode}.

In the studies of CoP, our only guidance were the predictions of holographic complexity proposals for subregions, as summarized in table~\ref{TAB_ANALYTIC_RES_COMP}. It is interesting to note, that our studies reproduce qualitatively terms present in the holographic results. In particular, for appropriately defined mutual complexity, all complexity notions we considered lead to an analogous dependence on the sizes of two adjacent intervals to the one seen in for entanglement measures in~\eqref{eq.IEoPsummary}. What is also worth a separate remark is the intricate dependence of the subleading divergent and the finite term in the bosonic CoP on the reference state scale $\mu$ and the mass $m$ acting as the zero mode regulator, see~\eqref{eq.CoPf0f1f2}.

The other important set of CoP results has to do with a comparison with an earlier study of CoP using a restricted set of Gaussian purifications in~\cite{Caceres:2019pgf}, as well as with the Fisher-Rao distance introduced in a related context in~\cite{DiGiulio:2020hlz}. As we saw respectively in sections~\ref{sec:singlemode} and~\ref{sec.Fisher-Rao}, both of these notions can reproduce qualitative behaviour of the CoP, but they can also exhibit significant deviations from the numerical answer predicted by the full Gaussian optimization. This indicates that in general one indeed needs to optimize over as large sets of states in the Hilbert state as possible to reproduce the true CoP. Of course, even if the reduced density matrix is Gaussian, this does not imply that the optimal circuit is necessarily such. Generalizing our study to non-Gaussian states is an outstanding challenge and in the last paragraph of this discussion, we sketch what we believe might be an interesting and workable example.

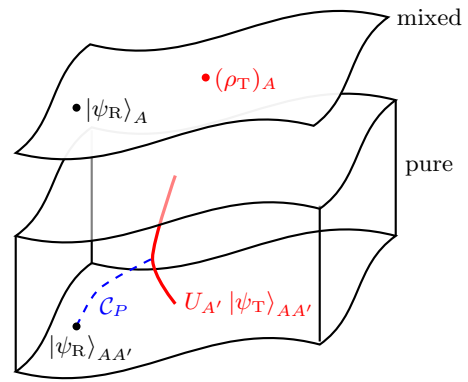
\begin{figure}
    \centering
    \begin{tikzpicture}
    
    \draw[thick] (10.5,2.8) -- node[right]{pure} (10.5,1) (5.5,1) -- (5.5,-.8) (6.5,2.4) -- (6.5,.6);
    \manifold[black,thick,fill=white,fill opacity=0.95]{5.5,-.8}{10.5,1}
    
    \draw[red] (7.6,.1) node[right]{$U_{A'}\ket{\psi_{\mathrm{T}}}_{AA'}$};
    \draw[very thick,red] plot [smooth] coordinates {(7.6,1.8) (7.3,.7) (7.6,.1)};
    
    \draw[thick,blue,dashed] plot [smooth] coordinates {(6.3,-.2) (6.6,.3) (7.3,.7)};
    \draw[blue] (6.6,.3) node[below,xshift=2mm]{$\mathcal{C}_P$};
    
    \draw[thick]  (9.5,1.4) -- (9.5,-.4);
    \manifold[black,thick,fill=white,fill opacity=0.5]{5.5,1}{10.5,2.8}
    \fill (6.3,-.2) node[below,xshift=2mm]{$\ket{\psi_{\mathrm{R}}}_{AA'}$} circle (1.5pt);
    
    \begin{scope}[yshift=-1.1cm]
    \draw (10.4,5) node[right]{mixed};
    \manifold[black,thick,fill=white,fill opacity=0.95]{5.5,3.2}{10.4,5}
    \fill (6.3,3.8) node[right,fill=white,rounded corners=2pt,inner sep=1pt,xshift=2pt,yshift=-1pt]{$\ket{\psi_{\mathrm{R}}}_A$} circle (1.5pt);
    \fill[red] (8,4.2) node[right,fill=white,rounded corners=2pt,inner sep=1pt,xshift=2pt,yshift=-1pt]{$(\rho_{\mathrm{T}})_A$} circle (1.5pt);
    \end{scope}
    \end{tikzpicture}
    \caption{We sketch how the manifold of mixed states on Hilbert space $\mathcal{H}_A$ is related to the manifold of pure states on the larger Hilbert space $\mathcal{H}=\mathcal{H}_A\otimes\mathcal{H}_{A'}$. We indicate the manifold (solid line) of all possible purifications $\ket{\psi_{\mathrm{T}}}_{AA'}$ related by Gaussian unitaries $U_{A'}$. The CoP $\mathcal{C}_P$ is given by the geodesic distance (dashed line) between the purified reference state $\ket{\psi_{\mathrm{R}}}_{AA'}$ and the family of purified target states $U_{A'}\ket{\psi_{\mathrm{T}}}_{AA'}$.}
    \label{fig:CoP-sketch}
\end{figure}

In the context of CoP, we also want to offer an intuitive, yet rigorous interpretation of the setup that we were using. Formula~\eqref{eq:CompFunct} was derived in~\cite{Hackl:2018ptj,Chapman:2018hou} based on a certain metric on the group manifold, which coincided with the geodesic distance on the Gaussian state manifold (Fubini-Study metric), as studied in~\cite{Chapman:2017rqy}. Note that there is an ongoing debate what the most appropriate notion of distance is, both on the manifold of states and the Lie group. While the $L^1$ norm appears to have similar properties as the different holographic complexity proposals, only the $L^2$ norm induced by a certain Riemannian metric can be analytically minimized. This is the key reason why we focused on the $L^2$ norm in the present manuscript, as most other notions of distances are intractable in the setup we are considering (many degrees of freedom, most general Gaussian gates, analytical optimization over all trajectories from reference to target state, numerical optimization of all possible purifications). Figure~\ref{fig:CoP-sketch} provides a visualization of this interpretation: CoP becomes in essence another type of minimization, namely finding the minimal distance between the unique reference state $\ket{\psi_{\mathrm{R}}}$ to the the set of all possible purifications (or all Gaussian purifications) that is fully determined by the single mixed state, whose CoP we are computing.

Let us emphasize that there have been several approaches to define and compute complexity for mixed states, which predominantly focus again on Gaussian states. On the one hand, there are approaches~\cite{Agon:2018zso,Camargo:2018eof,Caceres:2019pgf} based on purifications, to which a notion of pure state complexity is applied. This provides an elegant way to carry any definition of pure state complexity over to arbitrary mixed states. Our present draft uses the same philosophy, but performs the required optimization over all possible (Gaussian) purifications numerically. On the other hand, there are approaches~\cite{DiGiulio:2020hlz} to define mixed state complexity directly on the set of mixed states. Here, one introduces additional non-unitary gates that allow to change the spectrum of the density operator, so that unitarily inequivalent mixed states can be reached. The resulting geodesic distance agrees with the $L^2$ norm when restricted to pure states and the procedure can be understood as measuring the geodesic distance on the manifold of (Gaussian) mixed states equipped with the Fisher information geometry. Remarkably, the resulting analytical formula in terms of covariance matrices for bosons agrees with the one~\eqref{eq:CoP} derived for pure states. One can expect the same result to also hold for fermions. In particular, we can use these gates to transform a pure reference state into a mixed target state. Finally, a complementary approach is based on path integral optimization~\cite{Caputa:2017urj,Caputa:2017yrh}, which uses gates being exponentials of the energy-momentum tensor operators with complex coefficients and so both unitary, as well as Hermitian operators~\cite{Camargo:2019isp}. In the case of free CFTs, these are Gaussian gates, however, in the interacting cases they are not.

An important feature of our works stems from the fact that the Ising model in the spin picture (rather than the fermionic picture) leads to genuinely non-Gaussian mixed states if we restrict to disconnected regions, \ie the spectrum of the reduced density operator to such subsystem cannot be reproduced by bosonic or fermionic mixed Gaussian states. This may open the window to study non-Gaussian circuit complexity (of purification) in a genuine QFT limit. For this, we propose to consider two individual separated sites in the spin picture of the critical Ising model. This leads to a mixed state in a system of two qubits, \ie in a four-dimensional complex Hilbert space. We can now study CoP in this setup as a function of the separation between the two sites\footnote{Of course, one should in principle consider more sites to be closer to continuum, but this will very quickly make the optimization problem intractable.}. While these are ideas for future work, the implemented algorithm to optimize over all possible purifications (in this cases non-Gaussian ones) can be used, once an appropriate notion of non-Gaussian circuit complexity (for generic gates) is defined. In this sense, we believe that our considerations of CoP in the context of the critical Ising model CFT presents a stepping stone to explore genuine non-Gaussian circuit complexity in the context of QFTs.

\begin{acknowledgements}
We thank J.~Eisert, R.~Jefferson, R.~Myers, N.~Shiba, K.~Tamaoka, and K.~Umemoto for helpful conversations and J.~Knaute, L.~Shaposhnik and V.~Svensson for comments on the draft. The Gravity, Quantum Fields and Information group at AEI is supported by the Alexander von Humboldt Foundation and the Federal Ministry for Education and Research through the Sofja Kovalevskaja Award. HC is partially supported by the Konrad-Adenauer-Stiftung through their Sponsorship Program for Foreign Students and by the International Max Planck Research School for Mathematical and Physical Aspects of Gravitation, Cosmology and Quantum Field Theory. AJ has been supported by the FQXi as well as the Perimeter Institute for Theoretical Physics. Research at Perimeter Institute is supported by the Government of Canada through the Department of Innovation, Science, and Economic Development, and by the Province of Ontario through the Ministry of Research and Innovation. TT is supported by Inamori Research Institute for Science and World Premier International Research Center Initiative (WPI Initiative) from the Japan Ministry of Education, Culture, Sports, Science and Technology (MEXT). TT is also supported by JSPS Grant-in-Aid for Scientific Research (A) No.16H02182 and by JSPS Grant-in-Aid for Challenging Research (Exploratory) 18K18766. We are also grateful to the long term workshop Quantum Information and String Theory (YITP-T-19-03) held at Yukawa Institute for Theoretical Physics, Kyoto University, where this work was initiated.
\end{acknowledgements}

\appendix

\section{Transverse field Ising model}\label{sec:Ising-model}
We review the construction of fermionic Hamiltonian describing the transverse field Ising model. In particular, we explain how we decompose the fermionic Hamiltonian into two quadratic parts and how we compute its bipartite entanglement entropy using standard Gaussian techniques.

\subsection{Definition}
We consider the transverse field Ising model in one dimension with $N$ sites arranged in circle. We denote sites as $i=1,\ldots, N$, with the site $N+1$ is identified with the site $1$. For simplicity, we will assume that $N$ is an even integer. The Hamiltonian is given by
\begin{align}
\hat{H}=-\sum^N_{i=1} (2J\hat{S}^{\X}_i\hat{S}^{\X}_{i+1}+h\hat{S}^{\Z}_i) \,,
\end{align}
where $\hat{S}^{\X}_i $, $\hat{S}^{\Y}_i $ and $\hat{S}^{\Z}_i$ are the standard Pauli operators. This model can be solved by applying the following steps. We follow the conventions introduced in~\cite{vidmar2016generalized,vidmar2018volume,hackl2019average}.

\textbf{(1) Jordan-Wigner transformation.} We express the spin operators $\hat{S}_i^{\X}$ and $\hat{S}_i^{\Y}$ in terms of ladder operators $\hat{S}^\pm_i=\hat{S}_i^{\X}\pm\ii\hat{S}_i^{\Y}$. We then perform a Jordan-Wigner transformation by expressing all spin operators in terms of fermionic creation and annihilation operators $\hat{f}_i^\dagger$ and $\hat{f}_i$, namely
	\begin{align}
	\hat{S}_i^{+}=\hat{f}_i^\dagger \,\exp\left(\ii \pi\sum^{i-1}_{j=1}\hat{f}^\dagger_j\hat{f}_j\right)\,.\label{eq:Jordan-Wigner}
	\end{align}
The resulting Hamiltonian is then given by
	\begin{align}
	\begin{split}
	\hat{H}&=-\sum^N_{i=1}\left[\frac{J}{2}\left(\hat{f}_i^\dagger(\hat{f}_{i+1}+\hat{f}_{i+1}^\dagger)+\text{h.c.}\right)+h\hat{f}_i^\dagger \hat{f}_i\right]\\
	&\qquad+\frac{J}{2}\left[\hat{f}_N^\dagger(\hat{f}_{1}+\hat{f}_{1}^\dagger)+\text{h.c.}\right](\hat{P}+1)+\frac{hN}{2}\,,
	\label{eq:HP}
	\end{split}
	\end{align}
where $\hat{P}_{\mathrm{tot}}=e^{\ii \pi\hat{\mathcal{N}}}$ is the parity operator with $\hat{\mathcal{N}}=\sum^N_{i=1}\hat{f}^\dagger_i\hat{f}_i$. The last term is a boundary term.

\textbf{(2) Quadratic Hamiltonians.} Because of the operator $\hat{P}$ in~\eqref{eq:HP}, the Hamiltonian is not quadratic in $\hat{f}_i^\dagger$ and $\hat{f}_i$. The non-quadratic term containing $\hat{P}$  distinguishes the sectors of even and odd eigenvalues of the number operator $\hat{\mathcal{N}}$. The Hilbert space can be decomposed as direct sum $\mathcal{H}=\mathcal{H}^+\oplus\mathcal{H}^-$ where $\mathcal{H}^+$ and $\mathcal{H}^-$ are the eigenspaces of the number parity operator $\hat{P}=e^{\ii\pi\hat{\mathcal{N}}}$ with eigenvalues $\pm 1$. The projectors onto these eigenspaces are given by
	\begin{align}
	\hat{\mathcal{P}}^\pm=\frac{1}{2}(\mathbb{1}\pm \hat{P})\,.
	\end{align}
	We can diagonalize $\hat{H}_{\mathrm{XY}}$ over $\mathcal{H}^\pm$ individually by applying the Fourier transformations
	\begin{align}
	\hat{c}_\kappa=\frac{1}{\sqrt{N}}\sum^N_{j=1}e^{\ii \kappa j}\hat{f}_j\,,
	\end{align}
where $\kappa\in\mathcal{K}^\pm$ with
	\begin{align}
	\mathcal{K}^+&=\left\{\tfrac{\pi}{N}+\tfrac{2\pi k}{N}\,\big|\,k\in\mathbb{Z}\,,-\tfrac{N}{2}\leq k<\tfrac{N}{2}\right\}\,,\\
	\mathcal{K}^-&=\left\{\tfrac{2\pi k}{N}\,\big|\,k\in\mathbb{Z}\,,-\tfrac{N}{2}\leq k<\tfrac{N}{2}\right\}\,.
	\end{align}
The resulting Hamiltonian $\hat{H}=\hat{H}_+\hat{\mathcal{P}}^++\hat{H}_-\hat{\mathcal{P}}^-$ is composed of the quadratic pieces
	\begin{align}
	\begin{split}
	\hat{H}_\pm&=\sum_{\kappa\in\mathcal{K}^\pm>0}\left[a_\kappa\left(\hat{c}_\kappa^\dagger\hat{c}_\kappa+\hat{c}_{-\kappa}^\dagger\hat{c}_{-\kappa}-1\right)\right.\\
	&\hspace{7mm}\qquad\left.-b_\kappa\left(\ii \,\hat{c}_\kappa^\dagger\hat{c}_{-\kappa}^\dagger-\ii \,\hat{c}_{-\kappa}\hat{c}_{\kappa}\right)\right]
	\end{split}
	\end{align}	
with the parameters defined as
	\begin{align}
		a_\kappa=-J\cos(\kappa)-h\,,\qquad b_\kappa= J\sin(\kappa)\,.
	\end{align}

\textbf{(3) Diagonalizing Hamiltonian.} At this point, we only need to perform individual fermionic two-mode squeezing transformations mixing the mode pair $(\kappa,-\kappa)$
\begin{align}\label{eq:eta-from-c}
	\hat{\eta}_\kappa= u_\kappa \hat{c}_\kappa - v_\kappa^* \hat{c}_{-\kappa}^\dagger\,.
\end{align}
with transformation coefficients explicitly given by
\begin{align}
	\begin{split}
	\epsilon_\kappa&=\sqrt{h^2+2hJ\cos(\kappa)+J^2}\,,\\
	u_\kappa&=\frac{\epsilon_\kappa+a_\kappa}{\sqrt{2\epsilon_\kappa(\epsilon_\kappa+a_\kappa)}}\,,\quad v_\kappa=\frac{ib_\kappa}{\sqrt{2\epsilon_\kappa(\epsilon_\kappa+a_\kappa)}}\,.
	\end{split}
\end{align}
As fermionic Bogoliubov coefficients, $u_\kappa$ and $v_\kappa$ satisfy $|u_\kappa|^2+|v_\kappa|^2=1$. The cases $\kappa\in\{0,\pi\}$, for which the Hamiltonian is already diagonal, can be treated by defining the special coefficients
	\begin{align}
	u_0=v_{\pi}=v_{-\pi}=0\,,\, u_{\pi}=u_{-\pi}=1\,,\,v_{0}=\ii\,,
	\end{align}
which leads to the identification $\hat{\eta}_0=\hat{c}_0^\dagger$ and $\hat{\eta}_\pi=\hat{c}_{-\pi}^\dagger$. After performing this last transformation, the quadratic pieces take the diagonal form 
\begin{align}
\hat{H}_\pm&=\sum_{\kappa\in\mathcal{K}^\pm}\epsilon_\kappa\,\left(\hat{\eta}_{\kappa}^\dagger\hat{\eta}_{\kappa}-\tfrac{1}{2}\right)\,,
\end{align}
With this in hand, we can analyze efficiently the entanglement structure of eigenstates. The relevant information is fully contained in the transformation from the (local) fermionic operators $(\hat{f}_i,\hat{f}^\dagger_i)$ to the (non-local) operators $(\hat{\eta}_\kappa ,\hat{\eta}_\kappa ^\dagger)$. This allows us to define and compute the covariance matrix of an eigenstate $\ket{\{N_\kappa\}}$ as~\cite{hackl2019average}
\begin{align}
\begin{split}
    \Omega^{ab}_{ij}&\equiv\braket{0|\hat{\xi}^a_i\hat{\xi}^b_j-\hat{\xi}^b_j\hat{\xi}^a_i|0}=\frac{1}{N}\sum_{{\kappa}\in\mathcal{K}^\pm}\begin{pmatrix}
    0 & -N_{\kappa}\,C^{i-j}_{\kappa}\\
    N_{\kappa}\,C^{j-i}_{\kappa} &0
    \end{pmatrix}
\end{split}
\end{align}
with the functions $C^{i-j}_{\kappa}$ given by
\begin{align}
\begin{split}
    C^{d}_{\kappa}&=(|v_{\kappa}|^2-|u_{\kappa}|^2)\cos(\kappa d)+2\mathrm{Im}(u_{\kappa}v_{\kappa})\sin(\kappa d)\,.
\end{split}
\end{align}
We find $C^d_{\kappa}=\cos{\kappa(\frac{1}{2}+d)}$ for $J=h$. The numbers $N_\kappa\in\{-1,1\}$ are given by $N_{\kappa}=(-1)^{n_{\kappa}}$, where we have the regular occupation numbers $n_\kappa$ as eigenvalues of $\hat{\eta}_{\kappa}^\dagger\hat{\eta}_{\kappa}$. For states with an even number of excitations, \ie where $\sum_{\kappa} N_\kappa$ is even, we need to use $\kappa\in\mathcal{K}^+$, while for an odd number of excitations, we use $\kappa\in\mathcal{K}^-$.

\subsection{Notions of locality}
First, let us note that the transverse field Ising model is invariant under lattice translations, which implies that we can choose the reference point $i=1$ of our Jordan-Wigner transformation without loss of generality. It therefore suffices to consider intervals that start at $i=1$.

The Jordan-Wigner transformation~\eqref{eq:Jordan-Wigner} does not only provide an isomorphism between operators, but it also preserves bipartite entanglement of connected interval, \ie the entanglement entropy associated to a region consisting of adjacent sites $R=(1,\ldots,\frac{w}{\delta})$ is the same regardless if we use the tensor product structure induced by the spin operators or the fermionic creation and annihilation operators. This is a consequence of the remarkable fact that, despite the Jordan-Wigner transformation being non-local, the operator $\hat{S}^\pm_i$ only depends on creation and annihilation operators $\hat{f}^\dagger_j$ and $\hat{f}_j$ in the range $1\leq j\leq i$. Therefore, the subalgebras generated either by $\hat{S}_i^{\sigma}$ (with $\sigma\in\{\mathrm{x},\mathrm{y},\mathrm{z}\}$) in the spin formulation or by $\hat{f}_i$ and $\hat{f}^\dagger_i$ in the fermionic formulation will both probe the same observables.

This is no longer true if we consider regions consisting two non-adjacent intervals, such as $R=(1,\dots,\frac{w}{\delta},\frac{d+w}{\delta},\dots,\frac{d+2w}{\delta})$ for $d>0$ and $w>0$. In this case, the spin operators $\hat{S}_i^\sigma$ on sites $(\frac{d+w}{\delta},\dots,\frac{d+2w}{\delta})$ will depend on the fermionic operators associated to all sites $(1,\dots,\frac{d+2w}{\delta})$ as seen from~\eqref{eq:Jordan-Wigner}, which in particular includes the sites $(\frac{w}{\delta}+1,\dots,\frac{w+d}{\delta})$. Consequently, computing the bipartite entanglement entropy associated to the region $R$ will be different depending on if we define the subsystem in the spin picture vs. the fermionic picture. We will discuss this issue in more detail in appendix~\ref{app:spin-vs-fermion}.

\section{Spin vs. Majorana Fermion}\label{app:spin-vs-fermion}
In this section, we highlight the known subtlety due to the different notion of locality in the fermionic model compared to the spin model.

\subsection{Partial traces and subsystems}
The inequivalence between spin and Majorana entanglement can also be seen in the behavior of partial traces under an explicit mapping between both models, the Jordan-Wigner transformation.
Explicitly, it relates spin and fermionic operators via
\begin{align}
\fe_j &= \left(\prod_{i=1}^{j-1} Z_i \right) \frac{X_j - \ii\, Y_j}{2} \ , \\
\fd_j &= \left(\prod_{i=1}^{j-1} Z_i \right) \frac{X_j + \ii\, Y_j}{2} \ ,
\end{align}
where we defined the $k$-site Pauli operators on $N$ total sites as
\begin{align}
X_j &= (\id_2)^{\otimes (j-1)} \otimes \sigma_x \otimes (\id_2)^{\otimes (N-j)} \ , \\
Y_j &= (\id_2)^{\otimes (j-1)} \otimes \sigma_y \otimes (\id_2)^{\otimes (N-j)} \ , \\
Z_j &= (\id_2)^{\otimes (j-1)} \otimes \sigma_z \otimes (\id_2)^{\otimes (N-j)} \ .
\end{align}
Alternatively, we can write the Jordan-Wigner transformation in terms of $2N$ real Majorana operators $\m_j$, related to the standard fermionic operators by $\fe_j = (\m_{2j-1} + \ii \m_{2j})/2$. It then takes the form
\begin{align}
\m_{2j-1} &=\left(\prod_{i=1}^{j-1} Z_i \right) X_j \ , \\
\m_{2j} &=\left(\prod_{i=1}^{j-1} Z_i \right) Y_j \ .
\end{align}
Consider the mapping between basis states under this transformation. The basis decomposition of a pure state,
\begin{align}
    \ket{\psi} &= \sum_{n \in \{0,1\}^{\times N}} T_n\, \ket{n_1}_1 \ket{n_2}_2 \dots \ket{n_N}_N \ ,
\end{align}
differs between fermions and spins: For the former, the basis states $\ket{n_k}_k = (\fd_k)^{n_k} \ket{0}_k$ (with a local Fock vacuum $\ket{0}_k$) only commute for $b_k=0$, while for the latter $(\ket{0},\ket{1}) = (\ket{\uparrow},\ket{\downarrow})$ are simply commuting spin states.
Thus, choosing a different ordering under the Jordan-Wigner transformation leads to different fermionic states.
Its entanglement entropies $S(A) = - \Tr_A (\rho_A \log \rho_A)$, computed from the spectrum of the reduced density matrix $\rho_A = \Tr_{\bar{A}} \rho$, are then generally different as well.
We can easily see that a partial trace over fermionic sites, unlike spin degrees of freedom, is not commuting:
\begin{align}
    \Tr_j \Tr_k \rho &= \bra{0} (1+\fe_j)(1+\fe_k) \,\rho\, (1+\fd_k)(1+\fd_j) \ket{0} \nonumber\\
    \neq \Tr_k \Tr_j \rho &= \bra{0} (1+\fe_k)(1+\fe_j) \,\rho\, (1+\fd_j)(1+\fd_k) \ket{0}
    \ . 
\end{align}
Now consider a total system of $N$ sites separated into $|A|$ sites in $A$ and $|\bar{A}|$ sites in the complement region $\bar{A}$.
In the simplest case, $A$ and $\bar{A}$ are connected regions and $\bar{A}$ begins at the first site of the given ordering.
By defining the trace over a subsystem as 
\begin{equation}
\Tr_B = \Tr_{B_{|B|}} \Tr_{B_{|B|-1}} \dots \Tr_{B_2} \Tr_{B_1} \ ,
\end{equation}
\ie tracing out sites in their reverse order, the reduced density matrix $\rho_A$ is equivalent in both fermions and spins, \ie
\begin{align}
    \rho_A &= \Tr_{\bar{A}} \ket\psi\bra\psi \nonumber\\
    &= (\bra{0}_{\bar{A}_N} {+} \bra{1}_{\bar{A}_N}) \dots (\bra{0}_{\bar{A}_1} {+} \bra{1}_{\bar{A}_1}) \,\ket\psi+ \mathrm{H.c.}   \ .
\end{align}
We call this setup, where creation operators appear in products with increasing site index and annihilation operators in reverse order, the \emph{canonical ordering}.
All other subsystem choices can be brought into this form by permutation of indices.
While a reordering of modes leaves spin states invariant due to their commuting basis, fermionic permutations change the equivalence under Jordan-Wigner transformations: Entanglement entropies $S(A)$ are only equivalent between spins and fermions for regions $A$ in the canonical ordering.

This equivalence can be extended to connected regions $A$ of fermionic Gaussian states. 
Consider a first Jordan-Wigner transformation where $A$ is in the canonical ordering, and a second one where spin indices are cyclically permuted as $i \to (i+1) \mod N$. We denote the Majorana operators for both transformations as $\m_k$ and $\mt_k$, respectively. They  are related as follows:
\begin{equation}
\begin{aligned}
\m_1 &\equiv X_1  \;\to\;  X_2 \equiv -\ii \mt_1 \mt_2 \mt_3 \ , \\
\m_2 &\equiv Y_1   \;\to\; Y_2 \equiv -\ii \mt_1 \mt_2 \mt_4 \ , \\
\m_3 &\equiv Z_1 X_2  \;\to\; Z_2 X_3 \equiv -\ii \mt_1 \mt_2 \mt_5 \ , \\
     &\dots   \\
\m_{2N-2} &\equiv Z_1 \dots Z_{N-2} Y_{N-1} \\
    &\;\to\;  Z_2 \dots Z_{N-1} Y_N \equiv -\ii \mt_1 \mt_2 \mt_N \ , \\
\m_{2N-1} &\equiv Z_1 \dots Z_{N-1} X_{N} \\
    &\;\to\;  X_1 Z_2 \dots Z_{N} \equiv -\ii \mt_1 \mt_2 \hat{P}_{\mathrm{tot}} \mt_1 \ , \\
\m_{2N} &\equiv Z_1 \dots Z_{N-1} Y_{N} \\
    &\;\to\;  Y_1 Z_2 \dots Z_{N} \equiv  -\ii \mt_1 \mt_2 \hat{P}_{\mathrm{tot}} \mt_2 \ . \\
\end{aligned}
\end{equation}
Here, we defined the total parity operator $\hat{P}_{\mathrm{tot}}=\prod_i Z_i \equiv (-\ii)^N \prod_k \mt_k$. 
As Gaussian states are fully characterized by their covariance matrix entries $\Gamma_{j,k}$, we consider how they change between both Jordan-Wigner transformations.
For parity-even states, $\m_k \to -\ii \mt_1 \mt_2 \mt_{\tilde{k}}$ with $\tilde{k}=(k+1)\mod 2N$, so we find
\begin{align}
\Gamma^+_{ij} = \frac{\ii}{2} \bra{\psi^+} [ \m_i,\m_j ] \ket{\psi^+} \;\to\; \tilde{\Gamma}^+_{i,j} = \frac{\ii}{2} \bra{\psi^+} [ \mt_{i},\mt_{j} ] \ket{\psi^+} \ ,
\end{align}
where $\ket{\psi^+}$ is a Gaussian state vector with $\hat{P}_{\mathrm{tot}} \ket{\psi^+} = \ket{\psi^+}$. In matrix form, this can be written as
\begin{align}
\Gamma^+ \;\to\; \tilde{\Gamma}^+ = 
\begin{pmatrix}
0 & \Gamma^+_{[1,2],[3,2N]} \\
\Gamma^+_{[3,2N],[1,2]} & \Gamma^+_{[3,2N],[3,2N]}
\end{pmatrix}\,,
\end{align}
where $\Gamma^+_{[i,j],[k,l]}$ is the sub matrix consisting of the rows from $i$, $i+1$ up to row $j$ and columns from $k$, $k+1$ up to $l$. For a parity-odd Gaussian state vector $\ket{\psi^-}$ with $P_\text{tot} \ket{\psi^-} = - \ket{\psi^-}$, however, the rows and columns corresponding to the first two Majorana modes are sign-flipped:
\begin{align}
\Gamma^- \;\to\; \tilde{\Gamma}^- = 
\begin{pmatrix}
0 & -\Gamma^-_{[1,2],[3,2N]} \\
-\Gamma^-_{[3,2N],[1,2]} & \Gamma^-_{[3,2N],[3,2N]}
\end{pmatrix} \ .
\end{align}
In consequence, only parity-even fermionic Gaussian states are equivalent under different Jordan-Wigner transformations related by cyclic permutations, while parity-odd ones acquire a sign flip in the two-point correlations related to modes that are moved from the end to the beginning of the fermionic ordering. Fortunately, this sign flip does not affect fermionic entanglement entropies $S(A)$, which are computed from the  spectrum of the submatrix of $\Gamma$ corresponding to sites in $A$, denoted $\Gamma_A$.

As the eigenvalue spectrum is not affected by change of sign across entire rows and columns of a matrix, $\tilde{\Gamma}_{|\tilde{A}}$ and $\Gamma_{|A}$ have the same eigenvalues, and hence, $S(A) = S(\tilde{A})$ independent of parity.  Note that once we consider index transpositions more general than cyclic permutations, the relative fermionic ordering is broken and even Gaussian states are no longer equivalent under different Jordan-Wigner transformations.

\subsection{Ising CFT representations}\label{app:spin-cft}
Here, we would like to discuss subtle differences between the spin model and Majorana fermion model when we calculate MI and EoP. Refer also to \cite{Coser:2015dvp} for an earlier detailed analysis on this problem, which is essentially equivalent to our argument below.

First let us remember that the Ising spin CFT is defined from Majorana fermion CFT by the GSO projection \cite{Ginsparg:1988ui}. Explicitly the modular invariant torus partition function of the Ising CFT is schematically written as
\begin{eqnarray}
Z_{\mathrm{Ising}}=\Tr_{NS}\left[\frac{1+(-1)^F}{2}\right]+\Tr_{R}\left[\frac{1-(-1)^F}{2}\right], \nonumber \\ 
\label{torus}
\end{eqnarray}
where the first trace is taken for the NS sector (${\cal H}_+$ sector), \ie the anti-periodic boundary condition is imposed for the free fermion on a circle. Also $\frac{1+(-1)^F}{2}$ is the restriction to the even fermion number state. The second one is for the R sector (${\cal H}_-$ sector), \ie the periodic boundary condition is imposed for the free fermion on a circle. Also $\frac{1-(-1)^F}{2}$ is the restriction to the odd fermion number state.  Note that the spin operator 
$\sigma$ is included in the R sector of the Majorana fermion model. 

\begin{figure}[tb]
\centering
\includegraphics[height=0.22\textheight]{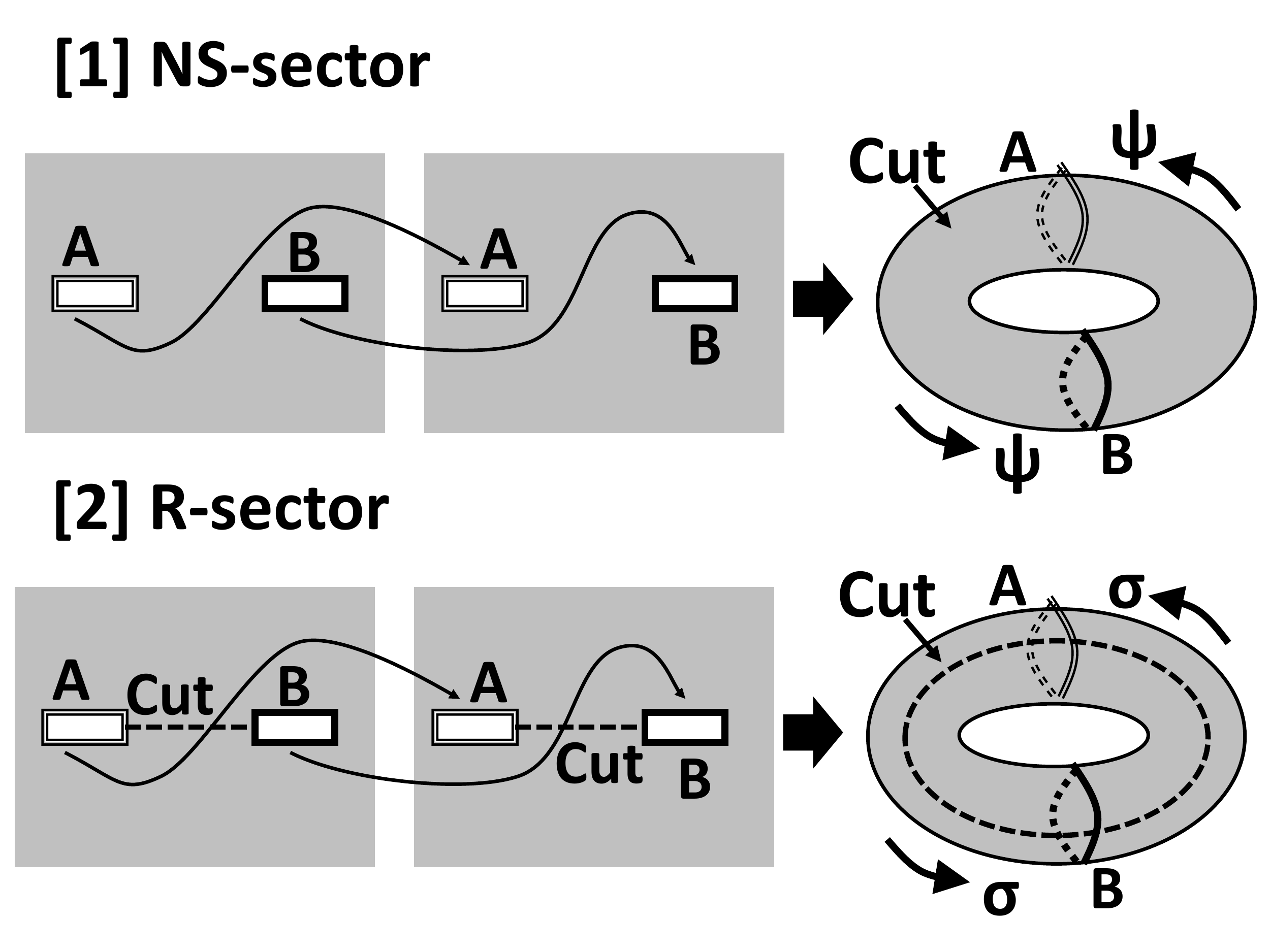}\\
\includegraphics[height=0.22\textheight]{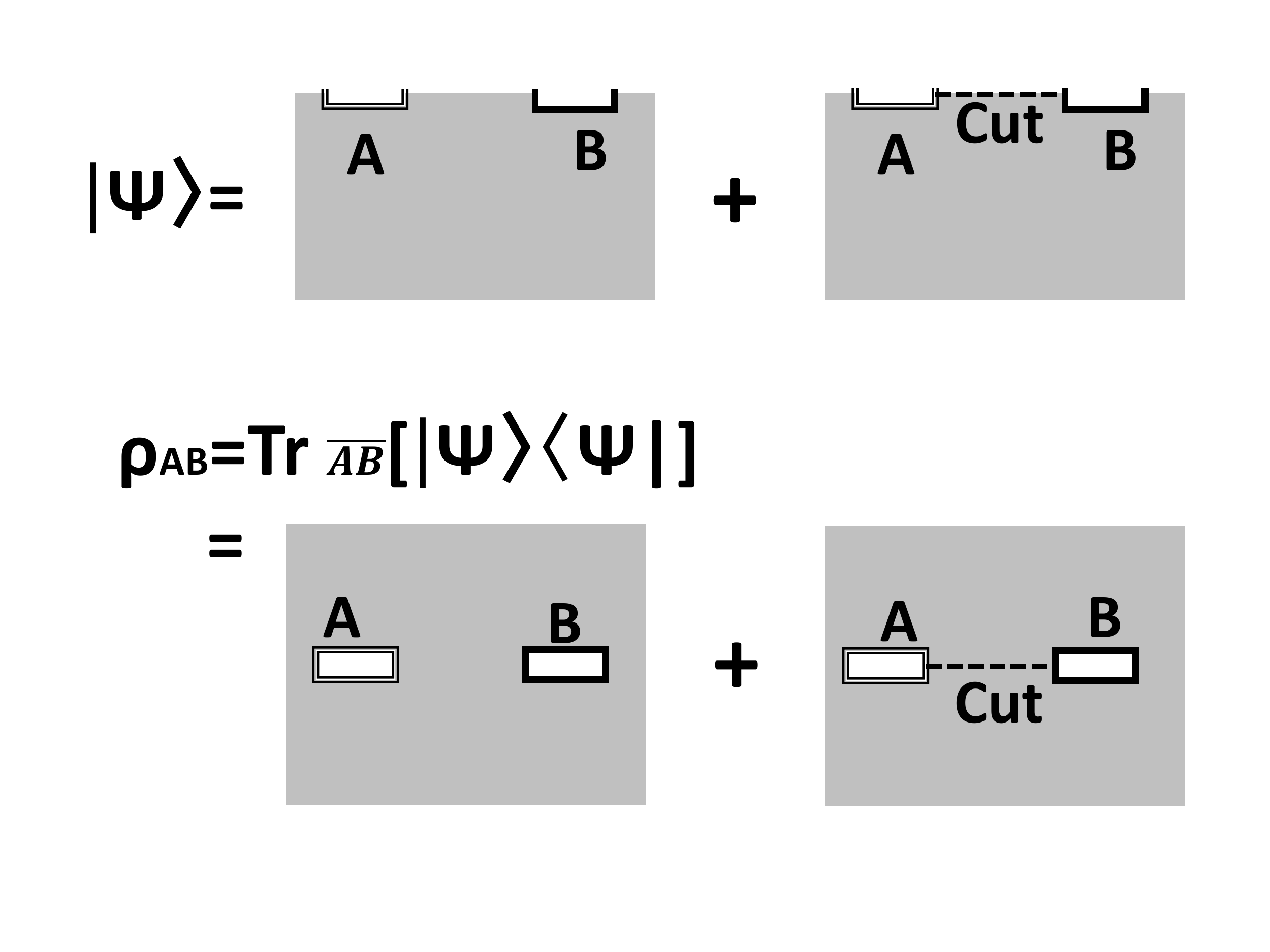}
\caption{The replica calculation of $\Tr[\rho_{AB}^2]$ and the resulting wave function for the GSO projected Majorana fermion CFT which is equivalent to the Ising CFT.}
\label{GSOfig}
\end{figure}

In the calculation of entanglement entropy $S(A\cup B)$ or its Renyi entropy in Ising CFT, we need to perform this GSO projection of the Majorana fermion along the interval $C$ between $A$ and $B$. This is illustrated in the upper pictures in figure~\ref{GSOfig} by having in mind the calculation of $\Tr[\rho_{AB}^2]$ as an example, which is essentially the 2nd Renyi entropy and which is equivalent to a torus partition function (\ref{torus}). If we write the fermion number on this interval $C$ as $F_C$ (\ie $F_C=\sum_{i\in C}\hat{f}^\dagger_i \hat{f}_i$), then the wave function for the calculation of Renyi entropy should be 
\begin{align}
\ket{\Psi}_{\mathrm{Ising}}=\ket{\Psi}_{\mathrm{Majorana}}+(-1)^{F_C}\ket{\Psi}_{\mathrm{Majorana}}\,, \label{majgso}
\end{align}
which is depicted in the lower pictures in  figure \ref{GSOfig} This procedure leads to the spin operator in the spectrum and leads to the expected behavior of MI $I(A:B)\sim (w/d)^{1/2}$ when 
$d\gg w$. However, this form (\ref{majgso}) is not included in our Gaussian ansatz of the numerical calculation. The Gaussian ansatz only takes into account the first term which ignores the twisted sector of Majorana fermion namely the spin operator sector (\ie R sector). This causes the absence of the spin operator excitation and explains our numerical result $I(A,B)\sim (w/d)^2$ for the Majorana fermion model when $d\gg w$.

Moreover, this difference between the XY spin model and the Majorana fermion model
leads to a significantly different behavior of MI when $d\ll w$. 
To see this, as a toy model which mimics the calculation of MI in XY spin model  at $d=1$, we would like to analyze a three qubit system, whose spins are called $A, C$ and $B$, from the left to right. We consider the following spin state for this $ABC$ system:
\begin{eqnarray}
&& |\Psi(x,y,\theta)\rangle_{ACB} \nonumber\\
&&=\frac{1}{\sqrt{2}}U(x)_{AC}\cdot U(y)_{CB}\cdot 
(|0\rangle_A|\theta\rangle_C|0\rangle_B+|1\rangle_A|\theta\rangle_C|1\rangle_B),\nonumber\\
\end{eqnarray}
where $|\theta\rangle=\cos\theta|0\rangle+\sin\theta|1\rangle$. The $4\times 4$ unitary matrix $U(x)$ is defined by 
\begin{eqnarray}
&& |00\rangle\to \cos x |00\rangle -\sin x |11\rangle,\nonumber\\
&& |11\rangle\to -\sin x |00\rangle +\cos x |11\rangle,\nonumber\\
&& |10\rangle\to \cos x |10\rangle -\sin x |01\rangle,\nonumber\\
&& |01\rangle\to -\sin x |10\rangle +\cos x |01\rangle.
\end{eqnarray}
Note that $U(x)_{AC}$ and $U(y)_{CB}$ mimic the entanglement between nearest neighbor sites due to the standard interactions of spin system.

If we consider this state from the viewpoint of Majorana fermions, then we need to take into account that 
the fermion in $C$ and the fermion in  $B$ do anti-commute.  This happens only if there is a fermion in both $B$ and $C$ site. Therefore we have the following rule
\begin{eqnarray}
|p\rangle_C |q\rangle_B= (-1)^{ab} |q\rangle_B |p\rangle_C,
\end{eqnarray}
where $p=0,1$ and $q=0,1$.  We act this  rule to  $|\Psi(x,y,\theta)\rangle_{ACB}$ 
and obtain the wave function for  the Majorana fermion, which is written as 
$|\Psi^f(x,y,\theta)\rangle_{ABC}$. 

\begin{figure}[tb]
\centering
\hspace{-0.15cm}
\includegraphics[height=0.18\textheight]{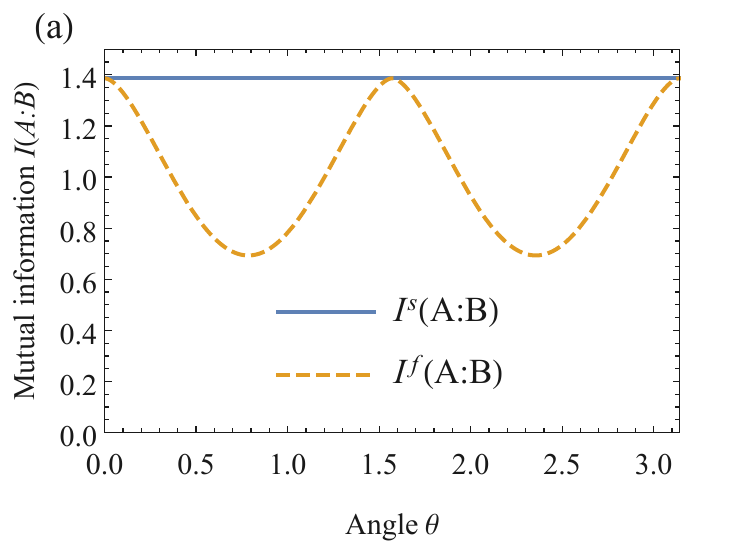}\\[0.25cm]
\includegraphics[height=0.18\textheight]{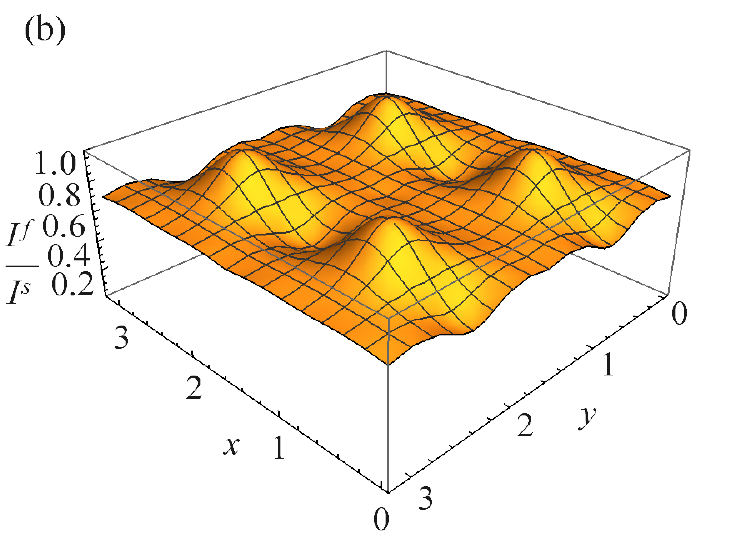}\\[0.25cm]
\includegraphics[height=0.18\textheight]{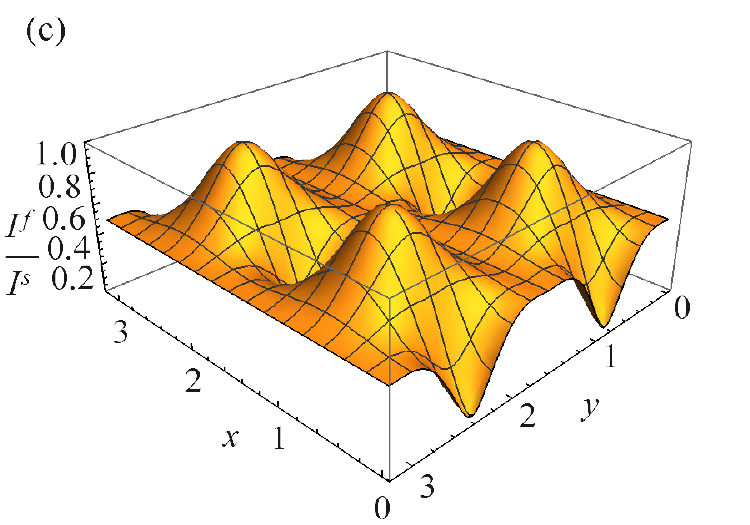}
\caption{(a) The value of $I^s(A:B)$ (solid line) and $I^f(A:B)$ (dashed line) as a function of $\theta$ at $x=y=0$. (b) A 3D plot of the ratio $I^f(A:B)/I^s(A:B)$ as a function $x$ (horizontal) and 
$y$ (depth) at $\theta=\pi/8$. (c) A 3D plot of the ratio $I^f(A:B)/I^s(A:B)$ as a function $x$ (horizontal) and  $y$ (depth) at $\theta=\pi/4$.}
\label{fig:MIcompare}
\end{figure}

We compare MI $I^s(A:B)$ obtained from the spin wave function $|\Psi(x,y,\theta)\rangle_{ACB}$, assuming the spin at $B$ and that at $C$ do commute, with the fermionic one $I^f(A,B)$ calculated from $|\Psi^f(x,y,\theta)\rangle_{ABC}$ in figure \ref{fig:MIcompare}. In the next subsection we will explain the same difference from the viewpoint of orderings of traces.

First of all, even at $x=y=0$ (no nearest neighbor entanglement), we find that $I^f(A:B)$ can be reduced from
 $I^s(A:B)$. The reduction is maximized at $\theta=\pi/4$, where we have $I^f(A:B)/I^s(A:B)=1/2$. By turning on the neighbor entanglement, we can reduce the fermion MI more. In particular, we find at $x=y=\frac{\pi}{4}$ and $\theta=\frac{\pi}{2}$ we obtain  $I^f(A,B)=0$. In summary, we always have
$I^f(A:B)/I^s(A:B)\leq 1$ \ie the fermion MI is smaller than that of spin.

Let us consider this difference between the spin calculation and fermion calculation from the viewpoint of 
two dimensional CFTs. Remember again the computation of Tr$\rho_{AB}^2$. In the $c=1/2$ Majorana fermion CFT, which is equivalent to the critical spin system, we need to impose the GSO projection to fermions in each of two cycles of the torus to obtain the modular invariant partition function. In particular,
we need to impose the projection on the horizontal cycle (the dotted circle,  called 'cut') in the upper two tori in figure~\ref{GSOfig}. This cycle corresponds to the subsystem $C$.  This GSO projection on $C$ picks up the 
periodic or anti-periodic boundary condition for the fermion $\hat{f}_i^\dagger$  depending on the even or odd value of the total fermion number in $C$. This projection is automatically performed in the spin system calculation. However if we just focus on the Majorana fermion system and ignore this phase factor, we do not find the correct GSO projection or equally the boundary condition to define the correct CFT partition function. Indeed, in the limit $d\ll w$ we are interested, the entropy $S(A \cup B)$ significantly depends on this boundary condition of the fermion on $C$ because the size of $C$ is very small. In summary, calculations in the spin system correspond to the CFT calculations of entanglement entropy and related quantities such as MI and EoP for standard choices of subsystems, while those in the naive Majorana fermion calculations without the GSO projection do not.

\section{Gaussian entanglement entropy}\label{app:Gaussian-entanglement}
As explained in section~\ref{sec:mutualinformation} and~\ref{sec:EoP}, important measures of quantum correlations, such as MI and EoP, are constructed from the bipartite entanglement entropy. While this is hard to evaluate for general quantum states $\ket{\psi}$, it is well-known that the entanglement entropy of a Gaussian state $\ket{\psi}$ can be computed analytically based on the entanglement entropy. Let us therefore briefly review the respective formulas.

A bosonic or fermionic system with $N$ degrees of freedom is characterized by the linear observables $\hat{\xi}^a\equiv(\hat{q}_1,\hat{p}_1,\dots,)$, as introduced in section~\ref{sec:models}. Mathematically, we refer $\hat{\xi}^a$ forms a basis of the classical phase space $V$. They satisfy the canonical commutation or anti-commutation relations, namely
\begin{align}
    [\hat{\xi}^a,\hat{\xi}^b]&=\ii\Omega^{ab}\,,\label{eq:CCR}\\
    \{\hat{\xi}^a,\hat{\xi}^b\}&=G^{ab}\,,\label{eq:CAR}
\end{align}
where $\Omega^{ab}$ is a symplectic form and $G^{ab}$ a positive definite metric. For a normalized quantum state $\ket{\psi}$ with $\braket{\psi|\hat{\xi}^a|\psi}=0$, we compute its two-point function as
\begin{align}
    C^{ab}_2=\braket{\psi|\hat{\xi}^a\hat{\xi}^b|\psi}\,.
\end{align}
We can always decompose $C^{ab}_2$ into
\begin{align}
    C^{ab}_2=\frac{1}{2}(G^{ab}+\ii\Omega^{ab})\,,
\end{align}
where $G^{ab}$ is a symmetric positive definite metric and $\Omega^{ab}$ is non-degenerate antisymmetric symplectic form. Here, $\Omega^{ab}$ is fixed by~\eqref{eq:CCR} for bosons, while $G^{ab}$ is similarly fixed by~\eqref{eq:CAR} for fermions. Consequently, only the respective other piece, \ie $G^{ab}$ for bosons and $\Omega^{ab}$ for fermions, which are often called the bosonic and fermionic \emph{covariance matrix}, respectively,  will depend on the state $\ket{\psi}$. We can define the linear map
\begin{align}
    J^a{}_b=G^{ac}\Omega^{-1}_{cb}\,.
\end{align}
We refer to the state $\ket{\psi}$ as Gaussian if and only if $J$ squares to minus identity, \ie
\begin{align}
    \ket{\psi}\text{  is pure Gaussian state}\quad\Longleftrightarrow\quad J^2=-\id\,,
\end{align}
in which case $J$ is called \emph{linear complex structure}.

The entanglement entropy $S(A)$ of a Gaussian state $\ket{\psi}$ with complex structure $J$ can be efficiently computed from the restriction $J_A$ of $J$ to the respective subsystem $A$. More precisely, if we have $\mathcal{H}=\mathcal{H}_A\otimes\mathcal{H}_B$ with phase space decomposition $V=A\oplus B$. The restriction
\begin{align}
    J_A=J_{[1,2N_A],[1,2N_A]}\label{eq:JA}
\end{align}
represents then the $2N_A$-by-$2N_A$ sub matrix of $J$ associated to the subspace $A\subset V$. While $J$ associated to a pure Gaussian state has eigenvalues $\pm \ii$, its restriction $J_A$ will have eigenvalues $\pm\lambda_i$ with $\lambda_i\in[0,\infty)$ for bosons and $\lambda_i\in[0,1]$ for fermions. The entanglement entropy can be computed from these eigenvalues using the formulas
\begin{align}
    S(A)=\left\{\begin{array}{rl}
    \tr\left(\frac{\id_A-\ii J_A}{2}\log\left|\frac{\id_A-\ii J_A}{2}\right|\right)    &  \textbf{(bosons)}\\[2mm]
    -\tr\left(\frac{\id_A-\ii J_A}{2}\log\color{white}\left|\color{black}\frac{\id_A-\ii J_A}{2}\color{white}\right|\color{black}\right)     & \textbf{(fermions)}
    \end{array}\right.,\label{eq:SA-derivation}
\end{align}
where we wrote covariant matrix equations, which are equivalent to replacing $J_A$ by its eigenvalues and the trace by a sum over them. Such analytical formulas for Gaussian entanglement entropy were first derived in~\cite{sorkin1983entropy,peschel2003calculation} based on the covariance matrices and later also phrased in terms of linear complex structures~ \cite{bianchi2015entanglement,vidmar2017entanglement,hackl2018aspects}. A unified framework to describe bosonic and fermionic Gaussian states in terms of Kähler structures (and in terms of complex structures, in particular) is presented in\ \cite{hackl2020bosonic}.

Note that the definition of entanglement is a subtle issue. Usually, one decomposes the Hilbert space into a regular tensor product $\mathcal{H}=\mathcal{H}_A\otimes\mathcal{H}_B$, such that operators $\mathcal{O}_A$ and $\mathcal{O}_B$ that are only supported in one of the subsystems are given as $\mathcal{O}_A=\hat{A}\otimes\id$ and $\mathcal{O}_B=\id\otimes\hat{B}$, leading to $[\mathcal{O}_A,\mathcal{O}_B]=0$. For fermions, we would like to define a similar structure, but independent operators should anticommute, \ie $\{\mathcal{O}_A,\mathcal{O}_B\}=0$. This requires a slightly different definition of ``fermionic tensor product'', as recognized and discussed in~\cite{caban2005entanglement,banuls2007entanglement,friis2013fermionic,szalay2020fermionic}. We discuss some of the resulting subtleties in the next section.

\section{Algorithm implementation}\label{app:algorithm}
Our minimization algorithm has the goal to minimize a function $f(J)$, where $J$ is the linear complex structure of a pure Gaussian state. Intuitively, we would like to apply gradient descent, \ie solve the differential equation
\begin{align}
    \dot{x}^\mu=-\sum_j\bm{G}^{\mu\nu}(x)\frac{\partial f}{\partial x^\nu}(x)\,,
\end{align}
where $\bm{G}^{\mu\nu}$ is the matrix representation of the Riemannian metric on the manifold of states. Using coordinate has two disadvantages: First, in general it will be difficult or even impossible to find a global coordinate system depending on the topology of manifold (in particular, for fermions the topology is non-trivial). Second, the matrix $\bm{G}^{\mu\nu}(x)$ will depend on the position and would need to be evaluated at every step, which will slow down the computation. Our approach based on the natural Lie group parametrization avoids both of these disadvantages.

Our parametrization is defined relative to an initial state $\ket{J_0}$. We then choose a basis of Lie algebra generators $(\Xi_\mu)^a{}_b$ satisfying the conditions $\{\Xi_\mu,J_0\}=0$ and
\begin{equation}
\begin{aligned}
    \tfrac{1}{8}\tr(\Xi_\mu G \Xi_\nu^{\intercal}g)&=\delta_{\mu\nu}\,,
    &K_\mu\Omega&=-\Omega K^\intercal_\mu\,,\quad\textbf{(bosons)}\\
    \tfrac{1}{8}\tr(\Xi_\mu G \Xi_\nu^{\intercal}g)&=-\delta_{ij}\,,
    &K_iG&=-G K^\intercal_i\,,\quad\textbf{(fermions)}
\end{aligned}
\end{equation}
where $G^{ab}=\braket{J_0|\{\hat{\xi}^a,\hat{\xi}^b\}|J_0}$ and $\Omega^{ab}=\braket{J_0|[\hat{\xi}^a,\hat{\xi}^b]|J_0}$. One can show~\cite{hackl2020geometry} that the dimension of this space is $N(N+1)$-dimensional for bosons and $N(N-1)$-dimensional for fermions. Rather than using coordinates $x^i$, we use the matrix $M^a{}_b$ to parametrize all Gaussian states $J=MJ_0M^{-1}$ connected to $J_0$. The gradient descent equation for $M$ reduces then to
\begin{align}
    \frac{dM}{dt}=-M \left(\sum_iX^\mu(M)\Xi_\mu\right)\,.\label{eq:Lie-gradient-descent}
\end{align}
where the gradient vector is computed as~\cite{hackl2020geometry}
\begin{align}
    \mathcal{F}^\mu(M)=-\frac{d}{ds}\bigg|_{s=0}\hspace{-4mm}f\left(Me^{s\Xi_\mu}J_0e^{-s\Xi_\mu}M^{-1}\right)\,.
\end{align}
In our algorithm, we discretize~\eqref{eq:Lie-gradient-descent}. Starting from the identity $M_{0}=\id$, we perform individual steps as
\begin{align}
    M_{n+1}=M_n\, e^{\epsilon K_{n}}\approx M_n\,\left(\frac{\id+\tfrac{\epsilon}{2} \,K_{n}}{\id-\tfrac{\epsilon}{2} \,K_{n} }\right)\,,
\end{align}
where $K_{n}=\sum_\mu \mathcal{F}^\mu(M_{n})\Xi_\mu/\lVert \mathcal{F}\rVert^2$. Here, we approximate the exponential for small $\epsilon$ in such a way that $M_{n+1}$ is ensured to lie in the symplectic or orthogonal group for bosons or fermions, respectively, which $e^{\epsilon K_n}\approx\id+\epsilon K_n$ would not achieve. At each step, we choose $0<\epsilon<1$ in such a way that $f(M_{n+1})<f(M_n)$, which is always possible to achieve for sufficiently small $\epsilon$.

Our algorithm circumvents the disadvantages of a coordinate parametrization by distinguishing between state and tangent space. While we use $M$ to parametrize our state $\ket{J}$ with $J=MJ_0M^{-1}$, we construct an orthonormal basis of Lie algebra generators $K_i$ which can be identified with the respective tangent vector of the curve $\gamma(s)=Me^{s \Xi_\mu}$ at the point $M$. This allows us in particular that our Riemannian metric on the manifold of Gaussian states is left-invariant, such that the so constructed tangent vectors are orthonormal at every point. We therefore do not need to compute its matrix representation $\bm{G}^{\mu\nu}(x)$, but can instead work with orthonormal frames at every point $M$.

\end{document}